\documentclass[twocolumn]{aastex6}
\usepackage{hyperref}
\usepackage{color}
\usepackage{appendix}
\usepackage{subfigure}
\definecolor{purple}{rgb}{0.62,0.12,0.94}
\newcommand\degrees{$^{\circ}$}

\begin{document}

\sloppy

\title{Search for new cosmic-ray acceleration sites within the 4FGL catalog Galactic plane sources}

\author{
S.~Abdollahi\altaffilmark{1}, 
F.~Acero\altaffilmark{2}, 
M.~Ackermann\altaffilmark{3}, 
L.~Baldini\altaffilmark{4}, 
J.~Ballet\altaffilmark{2*}, 
G.~Barbiellini\altaffilmark{6,7}, 
D.~Bastieri\altaffilmark{8,9,10}, 
R.~Bellazzini\altaffilmark{11}, 
B.~Berenji\altaffilmark{12}, 
A.~Berretta\altaffilmark{13}
% ...
}
\altaffiltext{1}{IRAP, Universit\'e de Toulouse, CNRS, UPS, CNES, F-31028 Toulouse, France}
\altaffiltext{2}{Université Paris Saclay and Université Paris Cité, CEA, CNRS, AIM, F-91191 Gif-sur-Yvette, France}
\altaffiltext{3}{Deutsches Elektronen Synchrotron DESY, D-15738 Zeuthen, Germany}
\altaffiltext{4}{Universit\`a di Pisa and Istituto Nazionale di Fisica Nucleare, Sezione di Pisa I-56127 Pisa, Italy}
% ...
\altaffiltext{6}{Istituto Nazionale di Fisica Nucleare, Sezione di Trieste, I-34127 Trieste, Italy}
\altaffiltext{7}{Dipartimento di Fisica, Universit\`a di Trieste, I-34127 Trieste, Italy}
\altaffiltext{8}{Istituto Nazionale di Fisica Nucleare, Sezione di Padova, I-35131 Padova, Italy}
\altaffiltext{9}{Dipartimento di Fisica e Astronomia ``G. Galilei'', Universit\`a di Padova, I-35131 Padova, Italy}
\altaffiltext{10}{Center for Space Studies and Activities ``G. Colombo", University of Padova, Via Venezia 15, I-35131 Padova, Italy}
\altaffiltext{11}{Istituto Nazionale di Fisica Nucleare, Sezione di Pisa, I-56127 Pisa, Italy}
\altaffiltext{12}{California State University, Los Angeles, Department of Physics and Astronomy, Los Angeles, CA 90032, USA}
\altaffiltext{13}{Dipartimento di Fisica, Universit\`a degli Studi di Perugia, I-06123 Perugia, Italy}
% ...

%%%%%%%
%% Mark off the abstract in the ``abstract'' environment. 

\begin{abstract}
Cosmic rays are mostly composed of protons accelerated to relativistic speeds. When those protons encounter interstellar material, they produce neutral pions which in turn decay into gamma rays. This offers a compelling way to identify the acceleration sites of protons. A characteristic hadronic spectrum, with a low-energy break around 200 MeV, was detected in the gamma-ray spectra of four Supernova Remnants (SNRs), IC 443, W44, W49B and W51C, with the \emph{Fermi} Large Area Telescope. This detection provided direct evidence that cosmic-ray protons are (re-)accelerated in SNRs. Here, we present a comprehensive search for low-energy spectral breaks among 311 4FGL catalog sources located within 5\degrees\ from the Galactic plane. Using 8 years of data from the \emph{Fermi} Large Area Telescope between 50 MeV and 1 GeV, we find and present the spectral characteristics of 56 sources with a spectral break confirmed by a thorough study of systematic uncertainty. Our population of sources includes 13 SNRs for which the proton-proton interaction is enhanced by the dense target material; the high-mass $\gamma$-ray binary LS~I +61 303; the colliding wind binary $\eta$ Carinae; and the Cygnus star-forming region. This analysis better constrains the origin of the $\gamma$-ray emission and enlarges our view to potential new cosmic-ray acceleration sites. 
\end{abstract}
%% Keywords should appear after the \end{abstract} command. 
%% See the online documentation for the full list of available subject
%% keywords and the rules for their use

\section{Introduction} \label{sec:intro}
The acceleration site of protons, the main components of cosmic rays, is one of the most fundamental topics of high energy astrophysics. The strong shocks associated with supernova remnants (SNRs) are widely believed to accelerate the bulk of Galactic cosmic rays (E $< 10^{15}$~eV) through the diffusive shock acceleration mechanism \citep[e.g.][]{1983RPPh...46..973D}. Indeed, accelerated cosmic rays interact with surrounding matter and produce $\pi^0$ mesons which usually quickly decay into two gamma rays, each having an energy of $67.5$~MeV in the rest frame of the neutral pion. In turn, the gamma-ray number spectrum $F(E)$ is symmetric at this same energy in log-log representation \citep{1971NASSP.249.....S} which then leads to a $\gamma$-ray spectrum in the usual $E^2F(E)$ representation rising below 200~MeV and approximately tracing the energy distribution of parent protons at energies greater than a few GeV. This characteristic spectral feature, often referred to as the "pion-decay bump", uniquely identifies proton acceleration since leptonic $\gamma$-ray production mechanisms such as bremsstrahlung and inverse Compton (IC) emission require fine tuning to produce a similar feature. \cite{esposito} explored this hypothesis by studying the $\gamma$-ray emission from SNRs, and potential associations of $\gamma$-ray sources with five radio-bright shell-type SNRs were reported using data taken by the EGRET instrument on board the \emph{Compton} Gamma Ray Observatory. More recently, this signature of protons was detected in five SNRs interacting with molecular clouds (MCs) and detected at gamma-ray energies by \emph{Fermi}-LAT: IC 443 and W44 \citep{2011ApJ...742L..30G, 2013Sci...339..807A}, W49B \citep{2018A&A...612A...5H}, W51C \citep{2016ApJ...816..100J} and HB~21 \citep{2019A&A...623A..86A}, although in this last source both the leptonic and hadronic processes are able to reproduce the $\gamma$-ray emission. Finally, the young SNR Cassiopeia A was also analyzed at low energy and \cite{2013ApJ...779..117Y} derived an energy break at $1.72^{+1.35}_{-0.89}$~GeV which is better reproduced by a hadronic scenario. More details on this characteristic feature observed in the gamma-ray emission are provided in Appendix \ref{appen:pion}, showing a stronger signature for a soft proton injection index ($\Gamma = 2.5$) than for a hard index ($\Gamma = 1.5$).\\

Electrons can also radiate at gamma-ray energies via the inverse Compton scattering and bremsstrahlung processes. It has been demonstrated, for the supernova remnants interacting with molecular clouds cited above, that the large gamma-ray luminosity is difficult to explain via inverse Compton scattering. In addition, the steep gamma-ray spectrum detected at low energy requires additional breaks in the electron spectrum if we consider a model in which electron bremsstrahlung is dominant. Accurate estimation of the spectral characteristics of a $\gamma$-ray source at low energy is therefore crucial since it probes the nature of the particles (electrons or protons) emitting these gamma rays. However, the analysis of sources below 100 MeV is complicated due to large uncertainties in the arrival directions of the gamma rays, which lead to confusion among point sources and difficulties in separating point sources from diffuse emission. Thus, catalogs released by the \emph{Fermi}-LAT Collaboration have focused on energies greater than 100\,MeV until the 4FGL catalog \citep{2020ApJS..247...33A} expanded the lower bound to 50\,MeV.  This allows better constraint of low-energy spectra, but since the 4FGL upper energy bound is 1 TeV, the spectral model for most sources is dominated by data with energies above a few hundred MeV. In addition, the spectral representation of sources in the 4FGL catalog considered three spectral models: power law (PL), power law with sub-exponential cutoff, and log-normal (or log-parabola, hereafter called LP). This means that any source presenting a spectral break will be represented by a log-normal shape which may not adequately represent the low-energy behavior. Similarly, sources presenting two spectral breaks, as it is the case for W49B \citep{2018A&A...612A...5H} will be represented with a log-normal shape that better describes the high-energy interval due to the better angular resolution and increased effective area at these high energies. This directly implies that the description of the low-energy spectral parameters of a source requires a dedicated spectral analysis.\\

In this paper we use 8 years of Pass~8 data to analyse 311 Galactic sources detected in the 4FGL catalog and search for significant spectral breaks between 50\,MeV and 1\,GeV. The paper is organized as follows: Section~\ref{section:Description} describes the LAT and the observations used, Section~\ref{subsection:Analysis} presents our systematic methods for analyzing LAT sources in the plane at low energy, Section~\ref{section:results} discusses the main results and a summary is provided in Section~\ref{sec:summary}.
\section{\emph{Fermi}-LAT description and observations} 
\label{section:Description}

\subsection{\emph{Fermi}-LAT}
The \emph{Fermi}-LAT is a $\gamma$-ray telescope which detects photons with energies from 20 MeV to more than 500\,GeV by conversion into electron-positron pairs, as described in \cite{2009ApJ...697.1071A}. The LAT is composed of three primary detector subsystems: a high-resolution converter/tracker (for direction measurement of the incident $\gamma$ rays), a CsI(Tl) crystal calorimeter (for energy measurement), and an anti-coincidence detector to identify the background of charged particles. Since the launch of the spacecraft in June 2008, the LAT event-level analysis has been upgraded several times to take advantage of the increasing knowledge of how the \emph{Fermi}-LAT functions as well as the environment in which it operates. Following the Pass 7 data set, released in August 2011, Pass 8 is the latest version of the \emph{Fermi}-LAT data. Its development is the result of a long-term effort aimed at a comprehensive revision of the entire event-level analysis and comes closer to realizing the full scientific potential of the LAT \citep{2013arXiv1303.3514A}. The current version of the LAT data is Pass 8 P8R3 \citep{2013arXiv1303.3514A, 2018arXiv181011394B}. It offers 20\% more acceptance than P7REP \citep{2013arXiv1304.5456B}. We used the SOURCE class event selection, with the Instrument Response Functions (IRFs) P8R3\_SOURCE\_V3. \\

\subsection{Data selection and reduction}\label{subsection:DataSelection}
We used exactly the same dataset as that used to derive the 4FGL catalog of sources, namely 8 years (2008 August 4 to 2016 August 2) of Pass 8 SOURCE class photons. This means that similarly to the 4FGL dataset, our data were filtered removing time periods when the rocking angle was greater than 90$^\circ$ and intervals around solar flares and bright GRBs were excised.\\  
Pass~8 introduced a new partition of the events, called PSF event types, based on the quality of the angular reconstruction, with approximately equal effective area in each event type at all energies. Due to the very low signal to noise ratio at low energy, the angular resolution is critical to distinguish point sources from the background and we decided to use only PSF3 events (the best-quality events) below 100 MeV. We add PSF2 events between 100 MeV and 1 GeV. This high energy bound was selected since middle-aged SNRs commonly exhibit a high energy spectral break at around 1–10 GeV which would then bias our low energy analysis~\citep{2010ApJ...723L.122U}. For both PSF3 and PSF2 events, we excised photons detected with zenith angles larger than 80$^\circ$ to limit the contamination from $\gamma$ rays generated by cosmic-ray interactions in the upper layers of the atmosphere. That procedure eliminates the need for a specific Earth limb component in the model.\\ 
The data reduction and exposure calculations are performed using the LAT $fermitools$ version 1.2.23 and $fermipy$ \citep{2017ICRC...35..824W} version 0.19.0. We used only binned likelihood analysis because unbinned mode is much more CPU intensive and does not support energy dispersion.\\
We accounted for the effect of energy dispersion (reconstructed event energy not equal to the true energy of the incoming $\gamma$ ray) which becomes significant at low energies (see below). To do so, we used edisp\_bins=-3 which means that the energy dispersion correction operates on the spectra with three extra bins below and above the threshold of the analysis\footnote{\url{https://fermi.gsfc.nasa.gov/ssc/data/analysis/documentation/Pass8\_edisp\_usage.html}}.\\ 
Our binned analysis includes three logarithmically spaced energy bins between 50 MeV and 100 MeV, and 9 energy bins between 100 MeV and 1 GeV. The Galactic diffuse emission was modeled by the standard file gll\_iem\_v07.fits and the residual background and extragalactic radiation were described by an isotropic component (depending on the PSF event type) with the spectral shape in the tabulated model iso\_P8R3\_SOURCE\_V3\_PSF(3/2)\_v1.txt. The models are available from the \emph{Fermi} Science Support Center (FSSC)\footnote{\url{http://fermi.gsfc.nasa.gov/ssc/}}. In the following, we fit the normalizations of the Galactic diffuse and the isotropic components.

\begin{figure*}[!ht]
\begin{centering} 
\includegraphics[width=0.32\textwidth]{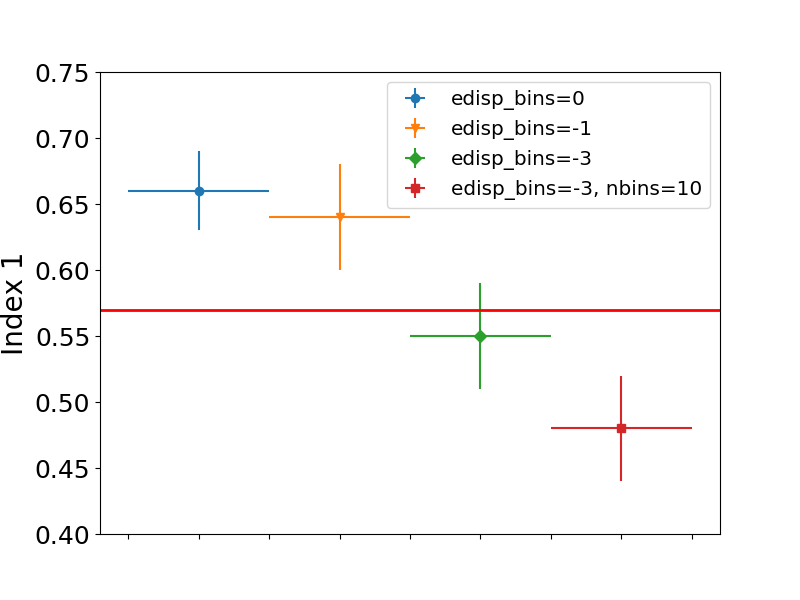}
\includegraphics[width=0.32\textwidth]{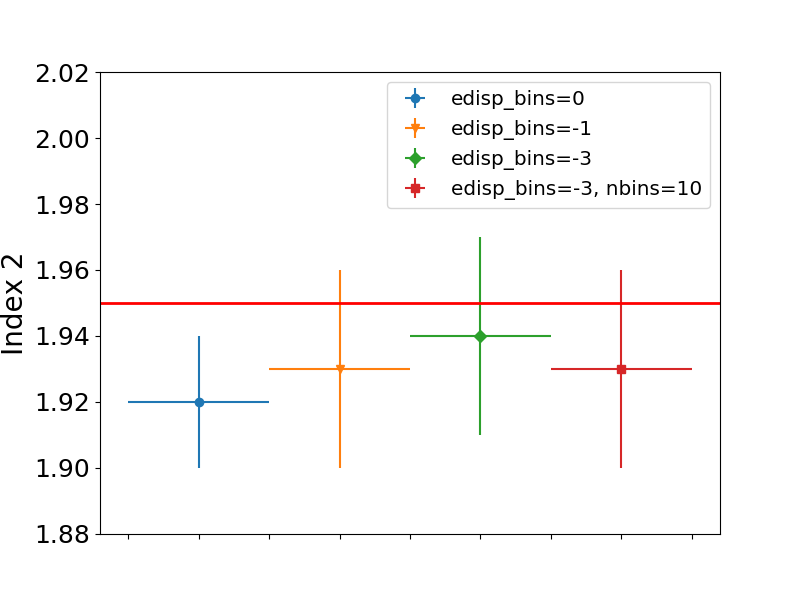}
\includegraphics[width=0.32\textwidth]{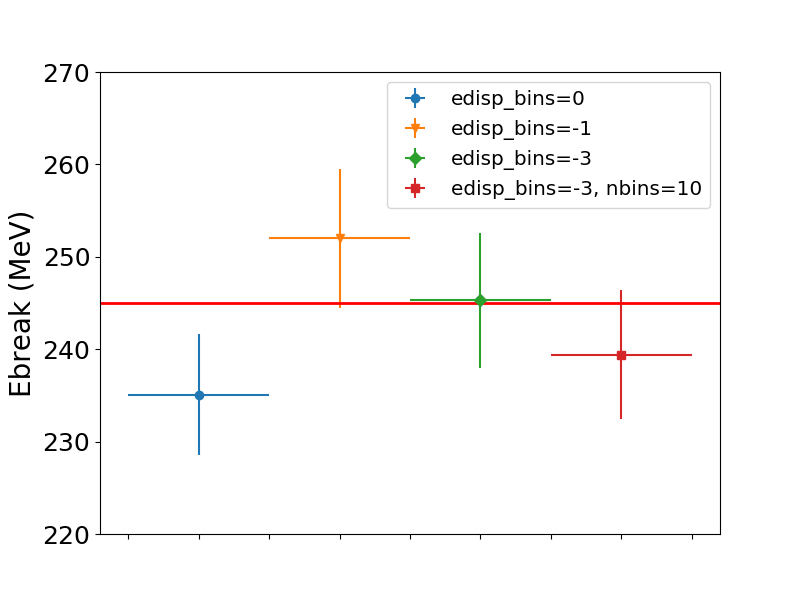}
\caption{Effect of the number of energy bins and value of edisp\_bins on the reconstructed values of the spectral index $\Gamma_1$ (left), $\Gamma_2$ (middle) and the break energy (right) of the broken power-law model of IC~443. Four configurations are tested: 12 energy bins and  edisp\_bins = 0 (blue circle), 12 energy bins and edisp\_bins = $-1$ (orange triangle), 12 energy bins and edisp\_bins = $-3$ (green diamond), 10 energy bins and edisp\_bins = $-3$ (red square). 
%For each case, 200 simulations of the spectrum of IC 443 were analyzed to measure the energy of the break and the spectral indices $\Gamma_1$ and $\Gamma_2$. The Figure reports the centroid and the sigma of the Gaussian fit performed on the distribution of these three spectral parameters. In each panel, the red line indicates the value of the spectral parameter set in the simulations.
    \label{fig:eres}}
\end{centering}
\end{figure*}

\begin{figure*}[!ht]
\begin{centering} 
\includegraphics[width=0.98\textwidth]{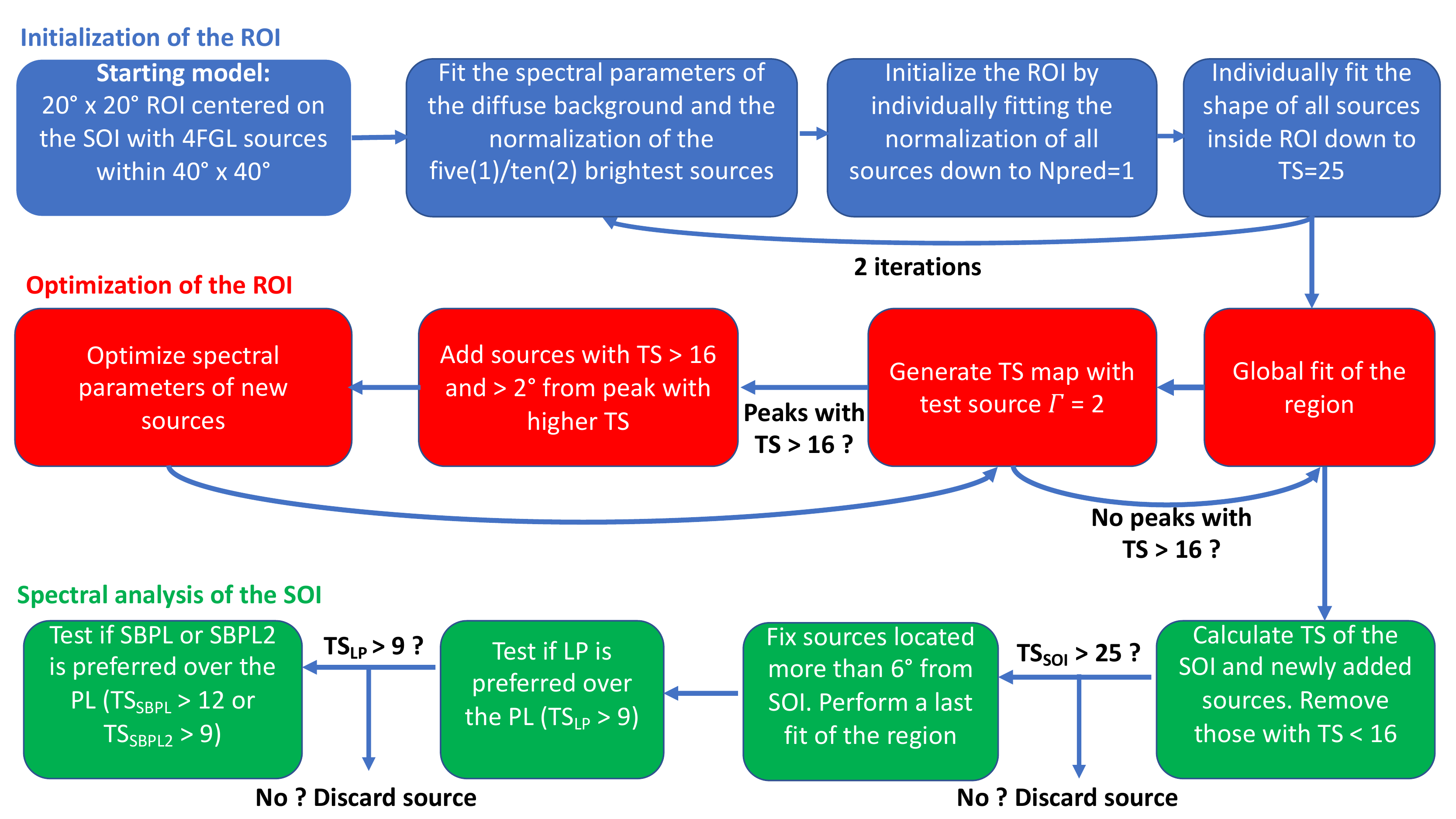}
\caption{Flow chart illustrating the individual analysis procedure of each source of interest (SOI) located in a $20^{\circ} \times 20^{\circ}$ region of interest. See text for further details.
    \label{fig:diag}}
\end{centering}
\end{figure*}

\subsection{Effect of the energy dispersion}\label{subsection:eres}
A crucial point that needs to be considered when analyzing LAT data at low energies is the effect of energy dispersion. For Pass 8, the energy resolution is $<10$\% between 1 GeV and 100 GeV but it worsens below 1 GeV. It is $\sim20$\% at 100 MeV and $\sim28$\% at 30 MeV. The combination of energy dispersion and the rapidly changing effective area below 100 MeV could result in biased measurements of flux and spectral index of the source under study. In order to quantify the effects of energy dispersion, 200 simulations of the spectrum of IC~443 as published by \cite{2013Sci...339..807A} were performed for a 8-year observation time using the $gtobssim$ tool included in the LAT $fermitools$. For these simulations, we assumed a point source spatial model located at (RA, Dec (J2000): 94.51$^\circ$, 22.66$^\circ$) and a smooth broken power-law spectral model of the form:
\begin{equation}
\frac{dN}{dE} = N_0 (E/E_0) ^{-\Gamma_1} (1 + (E/E_{\rm br})^{(\Gamma_2 - \Gamma_1)/\alpha})^{-\alpha},
\label{eq:sbpl}
\end{equation}
where $\alpha =0.1$, the break energy $E_{\rm br} = 245$ MeV and the spectral indices $\Gamma_1 = 0.57$, $\Gamma_2 = 1.95$. These simulations include the effect of energy dispersion. The analysis of these simulations was performed with the exact same configuration (region size, PSF components used, spatial bin size, energy interval) as the one used for real data. The only two parameters that have been varied in this study are the number of energy intervals and the value of the parameter edisp\_bins as discussed in Section~\ref{subsection:DataSelection}. For each combination of (energy bins, edisp\_bins), we analyzed the 200 simulations, plotted the distributions of the reconstructed values of the break energy, $\Gamma_1$ and $\Gamma_2$ and fitted a Gaussian on each distribution.\\ 
The centroid of the Gaussian fit together with their size are reported in Figure~\ref{fig:eres} for the four tests performed: (12 energy bins, edisp\_bins = 0), (12 energy bins, edisp\_bins = $-1$), (12 energy bins, edisp\_bins = $-3$), (10 energy bins, edisp\_bins = $-3$). As can be seen on this Figure, the main effect is on $\Gamma_1$, as expected. If the energy dispersion is not taken into account (edisp\_bins = 0), the spectrum is less steeply falling at low energy and the spectral index $\Gamma_1$ is reconstructed with a value 0.1 higher than the simulated value set in the simulations. This is also true if the energy dispersion is taken into account with only one extra bin (edisp\_bins = -1) which is not sufficient to properly take into account the effect of energy dispersion at these low energies even if this configuration has the advantage to reproduce slightly more accurately the value of the break energy. Even with a configuration using edisp\_bins=-3, if the number of bins is too small, the reconstructed value of $\Gamma_1$ will be biased towards lower value which will create artificially a stronger break at low energy. This is directly due to the fact that the energy resolution varies with energy. This imposes to choose an energy binning that is fine enough to capture this energy dependence.\\ 
The best compromise that was found between good reconstruction and computation time (since higher values of edisp\_bins or of the number of energy bins increase the CPU time) was obtained for a configuration using edisp\_bins=-3 and 12 energy bins between 50 MeV and 1 GeV. This configuration was used for all results presented in the following.

\section{Detection of spectral breaks}
\label{subsection:Analysis}
\subsection{List of candidates}\label{subsection:list}
This analysis intends to find new cosmic-ray acceleration sites in our Galaxy. When cosmic-ray protons accelerated by a source penetrate into high density clouds, the gamma-ray emission is expected to be enhanced relative to the interstellar medium because of the more frequent proton-proton interactions. Targeting the presence of such clouds, we restricted our search to sources within $5^{\circ}$ from the Galactic plane. In addition, we removed from our list all identified pulsars and active galactic nuclei. For AGNs, we removed all subclasses, namely flat-spectrum radio quasars (FSRQs), BL Lac-type objects (BLLs), blazar candidates of uncertain type (BCUs), radio galaxies (RDGs), narrow-line Seyfert 1 (NLSY1s), steep spectrum radio quasars (SSRQs), Seyfert galaxies (SEYs) or simply AGNs. Finally, to ensure that the source is significant in the low energy domain covered by our analysis, we removed all sources with a significance below $3\sigma$ between 300 MeV and 1 GeV as reported in the 4FGL catalog. In the end, these selection criteria provide us with a list of 311 candidates reported in the Appendix \ref{appen:sources}.

\subsection{Input source model construction}\label{subsection:ROI}
We perform an independent analysis of the 311 candidates selected in Section~\ref{subsection:list}. The procedure followed is inspired by the \emph{Fermi} High-Latitude Extended Sources Catalog \citep{2018ApJS..237...32A}, which already used the functions provided by $fermipy$.\\
For each source of interest, we define a $20^{\circ} \times 20^{\circ}$ region and include in our baseline model all 4FGL sources located in a $40^{\circ} \times 40^{\circ}$ region centered on our source of interest. We model each 4FGL source using the same spectral parameterization as used in the 4FGL. For extended sources, we use the spatial models from the 4FGL and keep them as fixed parameters since the angular resolution between 50 MeV and 1 GeV does not allow us to perform a morphological analysis. Similarly the positions of all point sources are fixed at their 4FGL values.\\ 
Starting from the baseline model, we proceed to optimize the model using the optimize function provided by $fermipy$. In this optimization step, we first fit the spectral parameters of the Galactic interstellar emission model and residual background together with the normalization of the five brightest sources.\\ 
Then, we individually fit the normalizations of all sources inside the region of interest (ROI) that were not included in the first step in the order of their total predicted counts in the model (Npred) down to Npred = 1. The optimization is concluded by individually fitting the index and normalization parameters of all sources with a test statistic (${\rm TS}$) value above 25 starting from the highest TS sources. This ${\rm TS}$ value is determined from the first two steps of the ROI optimization by ${\rm TS}=2(\ln \mathcal{L}_1 - \ln \mathcal{L}_0)$, where $\mathcal{L}_0$ and $\mathcal{L}_1$ are the likelihoods of the background (null hypothesis) and the hypothesis being tested (source plus background). This optimization is followed by a second one where the number of bright sources fit together with the diffuse backgrounds is increased to 10. This allows a better convergence for complex regions containing a large number of bright sources.\\
After optimizing the parameters of the baseline model components, we then perform a fit of the region by leaving free the normalization of all sources within $2^{\circ}$ of the source of interest, their spectral shape if their ${\rm TS}$ value is above 16, the normalization of all sources with ${\rm TS} > 100$ in the ROI and the spectral shape of all sources in the ROI with ${\rm TS} > 200$. If the number of degrees of freedom (Ndof) is above 100, we increase the two last ${\rm TS}$ criteria by 100 until Ndof becomes smaller than 100.\\
Once this complete fit of the ROI is performed, we further refine the model by identifying and adding new point source candidates. We identify candidates by generating a TS map for a point source that has a PL spectrum with an index $\Gamma = 2$. When generating the TS map, we fix the parameters of the background sources and fit only the amplitude of the test source. We add a source at every peak in the TS map with ${\rm TS} > 16$ that is at least $2^{\circ}$ from a peak with higher TS due to the poor angular resolution at these low energies. New source candidates are modeled with a PL whose normalization and index are fit in this procedure. We then generate a new TS map after adding the point sources to the model and repeat the procedure until no candidates with ${\rm TS} > 16$ are found. Though we do not expect to find a large number of additional sources, this step is crucial since {we are not using the weights, first introduced for the 4FGL Catalog \citep{2020ApJS..247...33A}. Indeed,} the generation of the candidate list for the Catalog is done above 100 MeV (instead of 50 MeV for our analysis) and each candidate is kept if the TS value obtained via a weighted maximum likelihood fit is above 25. These weights mitigate the effect of systematic errors due to our imperfect knowledge of the Galactic diffuse emission. As a consequence, the TS value of soft sources decreases.\\ 
In the final pass of the analysis, a second general fit of the ROI is performed using the same criteria as above to free the spectral parameters of all sources. If sources added previously by using the TS map fall down below ${\rm TS} > 16$, they are removed from the model. If their ${\rm TS}$ value is above 16 and they are located at a distance smaller than $5^{\circ}$ from our source of interest, we test iteratively for each of them the improvement of the log-normal representation with respect to the power-law model. The log parabola model is defined as: 
\begin{equation}
\frac{dN}{dE} = N_0 \left( \frac{E}{E_0} \right)^{-\left(\alpha + \beta \log\left(E/E_0\right) \right)},
\label{eq:lp}
\end{equation}
where $N_0$ is the overall normalisation factor to scale the observed brightness of a source, $E_0$ is a fixed scale energy (kept at 300 MeV in our analysis), and $\alpha$, $\beta$ are left free, which adds one degree of freedom with respect to the power-law representation. The improvement of the log-parabola model (LP) with respect to the power law one (PL) is performed by determining ${\rm TS_{LP}} = 2(\ln \mathcal{L}_{\rm LP} - \ln \mathcal{L}_{\rm PL})$. If ${\rm TS_{LP}} $ is above 9 (which corresponds to a $3\sigma$ improvement for one additional degree of freedom), we switch to the log-normal representation. The spectral parameters of all added sources located within $5^{\circ}$ of a candidate are reported in Table~\ref{tab:syst2}. As can be seen, the curvature index $\beta$ is hard to constrain for the additional faint sources, even if the log-parabola model significantly improves the fit. It is also clear that several added sources are located within the Vela and Cygnus regions for which the morphological templates used for the Vela-X PWN or the Cygnus Cocoon are not precise enough to characterize properly the region. Because the 4FGL Catalog rejects most point sources found inside extended sources, this leaves many residuals which translate into sources.

% Table 3: Added srcs
\begin{deluxetable*}{ccccccc}
\tablecaption{Localization and TS value of added sources localized
  within 5$^{\circ}$ of a confirmed candidate with a significant
  spectral break (the reference energy $E_0$ is fixed at 300 MeV in
  all cases) \label{tab:syst2}}
\tablehead{
\colhead{Name} &
\colhead{RA, Dec} &
\colhead{TS value} &
\colhead{${\rm TS_{LP}}$} &
\colhead{Prefactor} &
\colhead{Index} &
\colhead{$\beta$}  \\
\colhead{} &
\colhead{($^{\circ}$), ($^{\circ}$)} &
\colhead{} &
\colhead{} &
\colhead{($10^{-11}$ cm$^{-2}$s$^{-1}$MeV$^{-1}$)}&
\colhead{} &
\colhead{} 
}
\startdata
PS J0216.4+6213 & 34.12, 62.23 & 31 & 1 & 2.1 $\pm$ 0.6 & 1.8 $\pm$ 0.3 &  \\ 
PS J0327.6+5329 & 51.92, 53.49 & 30 & 5 & 1.0 $\pm$ 0.3 & 1.7 $\pm$ 0.3 &  \\ 
PS J0533.7+2501 & 83.45, 25.03 & 72 & 3 & 0.8 $\pm$ 0.2 & 4.2 $\pm$ 0.2 &  \\ 
PS J0845.8$-$4448 & 131.46, $-$44.81 & 30 & 4 & 3.0 $\pm$ 0.7 & 2.3 $\pm$ 0.2 &  \\ 
PS J0838.1$-$4212 & 129.55, $-$42.21 & 66 & 16 & 5.2 $\pm$ 0.9 & 2.0 $\pm$ 0.3 & 1.0 \\ 
PS J0856.8$-$4245 & 134.21, $-$42.76 & 59 & 5 & 2.6 $\pm$ 0.2 & 1.9 $\pm$ 0.1 &  \\ 
PS J0900.7$-$4438 & 135.20, $-$44.64 & 92 & 5 & 2.8 $\pm$ 0.2 & 1.5 $\pm$ 0.1 &  \\ 
PS J1558.2$-$5029 & 239.56, $-$50.50 & 51 & 28 & 5.5 $\pm$ 0.7 & 2.7 $\pm$ 0.2 & 1.0 \\ 
PS J1603.6$-$4621 & 240.92, $-$46.35 & 35 & 10 & 2.5 $\pm$ 0.4 & 2.0 $\pm$ 0.2 & 1.0 \\ 
PS J1632.5$-$4221 & 248.14, $-$42.35 & 39 & 7 & 1.7 $\pm$ 0.3 & 1.9 $\pm$ 0.1 &  \\ 
PS J1642.0$-$4802 & 250.50, $-$48.05 & 32 & 2 & 3.6 $\pm$ 0.9 & 2.2 $\pm$ 0.2 &  \\ 
PS J1816.9$-$1619 & 274.23, $-$16.32 & 46 & 8 & 4.9 $\pm$ 0.9 & 1.8 $\pm$ 0.2 &  \\ 
PS J2026.4+4004 & 306.62, 40.07 & 88 & 0 & 5.7 $\pm$ 0.1 & 1.8 $\pm$ 0.2 &  \\ 
PS J2032.0+3935 & 308.02, 39.59 & 131 & 7 & 6.6 $\pm$ 0.7 & 1.9 $\pm$ 0.1 &  \\ 
PS J2038.7+4114 & 309.70, 41.24 & 87 & 10 & 7.4 $\pm$ 0.7 & 1.8 $\pm$ 0.2 & 0.7 $\pm$ 0.2 \\ 
PS J2035.7+4242 & 308.94, 42.71 & 50 & 0 & 3.5 $\pm$ 0.4 & 1.7 $\pm$ 0.1 &  \\ 
PS J2018.8+4112 & 304.70, 41.20 & 74 & 12 & 8.9 $\pm$ 1.0 & 3.0 $\pm$ 0.2 & 0.5 $\pm$ 0.1 \\ 
PS J2045.5+4205 & 311.38, 42.10 & 48 & 9 & 4.9 $\pm$ 0.8 & 2.0 $\pm$ 0.3 & 1.0 \\ 
PS J2045.9+5044 & 311.48, 50.74 & 32 & 4 & 2.5 $\pm$ 0.5 & 1.8 $\pm$ 0.2 &  \\ 
PS J2047.9+4456 & 311.99, 44.94 & 47 & 6 & 2.4 $\pm$ 0.4 & 1.8 $\pm$ 0.2 & \\ 
\enddata
\tablecomments{Columns 2 and 3 provide the Right Ascension,
  Declination of the added source and its ${\rm TS}$ value. Column 4 gives
  the improvement of the lognormal representation with respect to the
  power-law model ${\rm TS_{LP}}$ as defined in
Section~\ref{subsection:ROI}. If ${\rm TS_{LP}} > 9$, a lognormal
representation is favoured and the Index provided in Column 5
corresponds to the spectral parameter $\alpha$ in
Equation~\ref{eq:lp}, while $\beta$ is indicated in Column 6 in such cases. No errors on $\beta$  are reported when it hits the boundary of 1.0.}
\end{deluxetable*}

\subsection{Spectral energy breaks}\label{subsection:break}
Once the ROI is well characterized, we first test the ${\rm TS}$ value of our source of interest in our energy range (50 MeV to 1 GeV). If it is below 25, we stop the analysis for this source since it is not significantly detected in our pipeline. It is the case for 64 sources among the 311 selected and their ${\rm TS}$ value is reported in Table~\ref{tab:candidates}. If the ${\rm TS}$ of the source of interest is above 25, we move on and we fix all sources located more than $5^{\circ}$ from the source of interest and we test the spectral curvature of our source of interest.\\ 
To ensure that the curvature is real and affects several energy bins as it would be the case for a ``pion-decay bump" signature, we first test a log-normal representation for the source as defined in Equation~\ref{eq:lp}. If ${\rm TS_{LP}} $ is below 9, we consider that no significant curvature is detected by our pipeline, we report this value in Table~\ref{tab:candidates} and we stop the analysis of this source. It is the case for 167 sources in our sample.\\ If the value is above 9, we then test a smoothly broken power law following Equation~\ref{eq:sbpl} where $N_0$ is the differential flux at $E_0 = 300$ MeV and $\alpha =0.1$, as was done previously for the cases of IC~443 and W44 \citep{2013Sci...339..807A}. This adds two additional degrees of freedom with respect to the power-law model (the break energy $E_{\rm br}$ and a second spectral index $\Gamma_2$). The improvement with respect to the power law one is determined by ${\rm TS_{SBPL}} = 2(\ln \mathcal{L}_{\rm SBPL} - \ln \mathcal{L}_{\rm PL})$. Since this test requires the addition of two degrees of freedom to the fit and diffusive shock acceleration predicts $\Gamma_2 \sim 2$, we also test the improvement of the smooth broken power law with the second index fixed at 2 with respect to the power law one ${\rm TS_{SBPL2}} = 2(\ln \mathcal{L}_{\rm SBPL2} - \ln \mathcal{L}_{\rm PL})$. We require ${\rm TS_{SBPL}} > 12$ or ${\rm TS_{SBPL2}} > 9$ (implying a $3\sigma$ improvement for 2 and 1 additional degrees of freedom respectively) to keep the source in the significant energy break list reported in Table~\ref{tab:syst}. We switch to the SBPL parameterization for all sources detected in this list. This means that, when a source located within $5^{\circ}$ shows a significant energy break, we re-optimize the ROI and we re-do the whole process as illustrated in the flow chart in Figure~\ref{fig:diag}. This procedure allowed the detection of 77 sources presenting a significant energy break in their low-energy spectrum. The values of ${\rm TS_{LP}}$, ${\rm TS_{SBPL}}$, ${\rm TS_{SBPL2}}$ for each of them is reported in Table~\ref{tab:candidates}.

\subsection{Diffuse and IRF Systematics}\label{subsection:sys}

The primary source of systematic error in this low energy analysis is the Galactic interstellar emission model (IEM). Our nominal Galactic IEM is the recommended one for PASS~8 source analysis. It represents the first major update to the LAT Collaboration’s  IEM since the model for the 3FGL catalog analysis gll\_iem\_v05.fits, developed for Pass~7 Source class, and later rescaled for Pass 8 Source as gll\_iem\_v06.fits. The development of the new model is described in more detail (including illustrations of the templates and residuals) online\footnote{\url{https://fermi.gsfc.nasa.gov/ssc/data/analysis/software/aux/4fgl/Galactic\_Diffuse\_Emission\_Model\_for\_the\_4FGL\_Catalog\_Analysis.pdf}}. The new model has higher resolution and correspondingly greater contrast but some shortcomings in the new Galactic IEM have been recognized when producing the 4FGL catalog.\\ 
To quantify the impact of diffuse systematics, we repeated our analysis for the 77 sources listed in Table~\ref{tab:syst} using the old diffuse rescaled for Pass 8 Source gll\_iem\_v06.fits. This alternative analysis means that the whole flow chart in Figure~\ref{fig:diag} was performed again, from the optimization of the ROI to the source finding algorithm, up to the determination of the spectral curvature. Performing the same complete analysis with the eight alternate diffuse models from \cite{2016ApJS..224....8A} would have become extremely CPU time consuming and this is why the old diffuse model only is tested. Here, since we already know that the source presents a break with the new model, we directly tested if ${\rm TS_{SBPL}} > 12$ or ${\rm TS_{SBPL2}} > 9$. If it was not the case, then this source was discarded from the final list of sources presenting significant energy break. \\

% Table 2 : Systematics
\begin{deluxetable*}{cccccccccc}
\tablecaption{Results of the systematic studies\label{tab:syst}}
%\tabletypesize{\footnotesize}
%\label{tab:syst}
\tablehead{
\colhead{4FGL Name} &
\colhead{$\rm TS_{SBPL}$} &
\colhead{$\rm TS_{SBPL2}$} & 
\colhead{$\rm TS_{SBPL}$} &
\colhead{$\rm TS_{SBPL2}$} & 
\colhead{$\rm TS_{SBPL}$} &
\colhead{$\rm TS_{SBPL2}$} \\ 
\colhead{} &
\colhead{diffuse} &
\colhead{diffuse} & 
\colhead{Aeff min} &
\colhead{Aeff min} & 
\colhead{Aeff max} &
\colhead{Aeff max} \\ 
}
\startdata
$\star$4FGL J0222.4+6156e & 24.3 & 16.5 & 34.9 & 27.5 & 35.1 & 27.7 \\ 
$\star$4FGL J0240.5+6113 & 170.7 & 143.5 & 127.5 & 123.9 & 124.0 & 123.2 \\ 
$\star$4FGL J0330.7+5845 & 13.8 & 7.7 & 15.8 & 12.5 & 16.0 & 12.5 \\ 
$\star$4FGL J0340.4+5302 & 81.6 & 67.5 & 147.0 & 143.5 & 140.1 & 138.5 \\ 
$\star$4FGL J0426.5+5434 & 13.0 & 6.9 & 26.3 & 21.0 & 25.6 & 20.4 \\ 
$\star$4FGL J0500.3+4639e & 12.2 & 8.1 & 15.2 & 14.5 & 14.9 & 14.2 \\ 
$\star$4FGL J0540.3+2756e & 12.7 & 8.0 & 10.8 & 10.6 & 10.7 & 10.6 \\ 
$\star$4FGL J0609.0+2006 & 14.7 & 6.6 & 17.6 & 14.0 & 17.1 & 14.1 \\ 
$\star$4FGL J0617.2+2234e & 103.7 & 81.1 & 96.5 & 79.3 & 95.2 & 79.5 \\ 
4FGL J0618.7+1211 & 10.4 & 5.7 & 16.5 & 9.3 & 15.2 & 9.5 \\ 
$\star$4FGL J0620.4+1445 & 13.5 & 6.0 & 14.0 & 9.2 & 14.2 & 9.3 \\ 
$\star$4FGL J0634.2+0436e & 26.3 & 21.3 & 17.6 & 17.6 & 10.5 & 10.6 \\ 
$\star$4FGL J0639.4+0655e & 33.3 & 28.9 & 44.8 & 39.3 & 45.0 & 39.4 \\ 
$\star$4FGL J0709.1$-$1034 & 26.5 & 14.6 & 19.5 & 13.0 & 19.4 & 13.0 \\ 
4FGL J0722.7$-$2309 & 11.1 & 5.5 & 21.5 & 10.6 & 21.6 & 10.7 \\ 
4FGL J0731.5$-$1910 & 9.4 & 5.1 & 16.4 & 9.4 & 16.3 & 9.4 \\ 
$\star$4FGL J0844.1$-$4330 & 27.1 & 13.2 & 38.9 & 13.2 & 32.7 & 11.2 \\ 
$\star$4FGL J0850.8$-$4239 & 15.2 & 9.0 & 27.4 & 19.7 & 27.7 & 19.9 \\ 
$\star$4FGL J0904.7$-$4908c & 15.8 & 9.4 & 11.8 & 9.7 & 12.5 & 10.0 \\ 
4FGL J0911.6$-$4738 & 11.8 & 7.1 & 11.8 & 9.7 & 10.8 & 8.2 \\ 
4FGL J0924.1$-$5202 & 11.9 & 6.6 & 16.8 & 9.2 & 18.5 & 9.3 \\ 
$\star$4FGL J1008.1$-$5706c & 21.5 & 9.9 & 25.7 & 20.6 & 25.7 & 20.8 \\ 
$\star$4FGL J1018.9$-$5856 & 14.7 & 5.2 & 28.9 & 27.9 & 28.5 & 28.2 \\ 
$\star$4FGL J1045.1$-$5940 & 25.6 & 19.8 & 15.2 & 15.0 & 17.0 & 16.9 \\ 
4FGL J1244.3$-$6233 & 11.4 & 8.0 & 30.9 & 10.8 & 31.2 & 11.1 \\ 
$\star$4FGL J1351.6$-$6142 & 13.6 & 5.7 & 15.7 & 14.1 & 17.9 & 16.2 \\ 
$\star$4FGL J1358.3$-$6026 & 21.7 & 6.1 & 22.1 & 12.8 & 22.5 & 13.1 \\ 
$\star$4FGL J1405.1$-$6119 & 23.6 & 16.2 & 25.1 & 20.1 & 25.1 & 20.1 \\ 
4FGL J1408.9$-$5845 & 10.3 & 5.4 & 11.5 & 9.0 & 11.9 & 9.1 \\ 
$\star$4FGL J1442.2$-$6005 & 15.1 & 6.5 & 16.3 & 12.0 & 16.6 & 12.2 \\ 
$\star$4FGL J1447.4$-$5757 & 15.4 & 7.7 & 18.1 & 14.4 & 18.3 & 14.6 \\ 
4FGL J1501.0$-$6310e & 7.9 & 3.2 & 17.8 & 10.0 & 18.4 & 10.1 \\ 
$\star$4FGL J1514.2$-$5909e & 14.0 & 11.3 & 34.1 & 27.8 & 32.9 & 29.1 \\ 
$\star$4FGL J1534.0$-$5232 & 12.2 & 5.3 & 13.6 & 8.5 & 13.9 & 8.4 \\ 
$\star$4FGL J1547.5$-$5130 & 17.0 & 12.9 & 32.8 & 16.3 & 30.8 & 18.3 \\ 
$\star$4FGL J1552.9$-$5607e & 12.0 & 7.9 & 11.5 & 10.9 & 11.9 & 10.9 \\ 
4FGL J1553.8$-$5325e & 7.9 & 4.1 & 73.5 & 63.2 & 74.0 & 64.6 \\ 
4FGL J1556.0$-$4713 & 9.8 & 5.8 & 11.4 & 4.8 & 10.2 & 4.6 \\ 
$\star$4FGL J1601.3$-$5224 & 34.2 & 21.9 & 42.9 & 36.9 & 44.6 & 36.5 \\ 
$\star$4FGL J1608.8$-$4803 & 20.4 & 13.7 &  30.8 &  13.2 &  30.7 & 13.4 \\ 
$\star$4FGL J1626.6$-$4251 & 20.8 & 13.9 & 18.2 & 8.7 & 15.2 & 8.8 \\ 
$\star$4FGL J1633.0$-$4746e & 12.8 & 8.5 & 37.2 & 36.4 & 38.1 & 37.1 \\ 
4FGL J1639.3$-$5146 & 7.2 & 4.0 & 17.2 & 6.4 & 18.1 & 7.1 \\ 
4FGL J1645.8$-$4533 & 8.9 & 6.6 & 28.9 & 10.5 & 30.9 & 14.6 \\ 
4FGL J1708.6$-$4312 & 10.4 & 6.8 & 19.6 & 9.2 & 19.9 & 9.4 \\ 
4FGL J1730.1$-$3422 & 9.8 & 1.5 & 35.6 & 13.8 & 35.9 & 14.1 \\ 
4FGL J1734.5$-$2818 & 8.7 & 4.0 & 29.9 & 14.7 & 35.9 & 14.1 \\ 
$\star$4FGL J1742.8$-$2246 & 18.3 & 6.8 & 15.2 & 8.4 & 19.5 & 10.1 \\ 
4FGL J1743.4$-$2406 & 7.0 & 3.8 & 14.7 & 5.3 & 14.9 & 6.5 \\ 
4FGL J1759.7$-$2141 & 10.2 & 6.0 & 17.7 & 6.9 & 18.0 & 7.1 \\ 
$\star$4FGL J1801.3$-$2326e & 89.1 & 83.5 & 173.9 & 146.6 & 175.5 & 147.6 \\ 
$\star$4FGL J1808.2$-$1055 & 13.3 & 7.9 & 14.6 & 10.0 & 14.6 & 10.1 \\ 
$\star$4FGL J1812.2$-$0856 & 13.8 & 7.4 & 15.8 & 7.8 & 16.0 & 13.6 \\ 
$\star$4FGL J1813.1$-$1737e & 17.7 & 12.9 & 25.0 & 18.4 & 27.5 & 14.9 \\ 
$\star$4FGL J1814.2$-$1012 & 17.7 & 7.8 & 20.2 & 11.1 & 17.4 & 11.0 \\ 
$\star$4FGL J1839.4$-$0553 & 14.0 & 9.7 & 22.2 & 20.4 & 22.6 & 20.9 \\ 
$\star$4FGL J1852.4+0037e & 14.1 & 4.5 & 20.3 & 19.4 & 22.1 & 19.9 \\ 
$\star$4FGL J1855.2+0456 & 20.8 & 9.7 & 31.7 & 12.2 & 31.9 & 12.0 \\ 
$\star$4FGL J1855.9+0121e & 90.0 & 82.3 & 91.1 & 91.5 & 94.3 & 94.8 \\ 
4FGL J1856.2+0749 & 8.5 & 4.5 & 21.1 & 19.5 & 19.3 & 14.8 \\ 
$\star$4FGL J1857.7+0246e & 12.0 & 5.6 & 24.4 & 20.5 & 24.9 & 19.1 \\ 
$\star$4FGL J1906.9+0712 & 11.3 & 10.9 & 28.1 & 18.7 & 28.0 & 19.8 \\ 
$\star$4FGL J1908.7+0812 & 15.8 & 10.3 & 62.3 & 41.8 & 62.9 & 42.1 \\ 
$\star$4FGL J1911.0+0905 & 14.4 & 10.4 & 27.8 & 27.6 & 27.8 & 27.4 \\ 
4FGL J1912.5+1320 & 7.9 & 4.0 & 21.2 & 14.0 & 21.8 & 14.1 \\ 
$\star$4FGL J1923.2+1408e & 23.0 & 17.7 & 20.8 & 20.7 & 22.3 & 22.1 \\ 
$\star$4FGL J1931.1+1656 & 13.5 & 7.0 & 23.1 & 17.0 & 23.3 & 17.3 \\ 
$\star$4FGL J1934.3+1859 & 28.4 & 12.5 & 31.1 & 15.6 & 30.5 & 14.5 \\ 
4FGL J1952.8+2924 & 8.0 & 4.0 & 20.6 & 12.4 & 21.0 & 12.6 \\ 
4FGL J2002.3+3246 & 8.3 & 4.1 & 14.3 & 11.2 & 14.3 & 10.4 \\ 
$\star$4FGL J2021.0+4031e & 31.6 & 14.6 & 25.6 & 10.2 & 25.8 & 10.2 \\ 
$\star$4FGL J2028.6+4110e & 49.2 & 34.3 & 94.5 & 91.8 & 132.9 & 129.9 \\ 
$\star$4FGL J2032.6+4053 & 13.6 & 15.2 & 21.3 & 19.0 & 22.2 & 19.2 \\ 
$\star$4FGL J2038.4+4212 & 17.0 & 9.4 & 14.4 & 10.3 & 14.5 & 10.3 \\ 
$\star$4FGL J2045.2+5026e & 24.6 & 15.4 & 37.4 & 25.7 & 37.3 & 26.0 \\ 
$\star$4FGL J2056.4+4351c & 17.2 & 11.0 & 18.9 & 10.0 & 18.1 & 10.2 \\ 
$\star$4FGL J2108.0+5155 & 13.5 & 7.1 & 18.2 & 12.3 & 18.3 & 12.4 \\ 
\enddata
\tablecomments{Columns 2 and 3 are obtained with the galactic diffuse
  background rescaled for Pass 8 Source (gll\_iem\_v06.fits) and
  provide values of the improvement of the smooth broken power-law
  representation with respect to the power-law model ${\rm TS_{SBPL}}$
  and the improvement of the smooth broken power-law representation
  when fixing $\Gamma_2 = 2$ called ${\rm TS_{SBPL2}}$  as defined in
  Section~\ref{subsection:ROI}. Columns 4, 5, 6 and 7 provide the same
  values of ${\rm TS_{SBPL}}$ and ${\rm TS_{SBPL2}}$ for the two
  bracketing IRFs. Stars $\star$ denote spectral breaks that are
  robust to all tests. See Section~\ref{subsection:sys} for more details.}
\end{deluxetable*}

The second source of systematic error in our analysis is the instrument response functions (IRFs) and especially the inaccuracies in the effective area. Following the standard method (Ackermann et al. 2012a), we estimated the systematic error associated with the effective area by calculating uncertainties in the IRFs which symmetrically bracket the standard effective area and flip from one extrema to the other at the measured value of the break energy. Here we started from the best fit model obtained with the standard IRF which is optimized before running the final spectral fit of each candidate with each of the two bracketing IRFs. The source finding algorithm was not relaunched in this case since these changes mainly affect the spectral parameters of the source and will not produce extra sources in the field of view.\\
A third source of systematic error that can affect the presence or absence of a spectral break for the source of interest is related to the inaccuracy of the emission models of nearby point sources. A thorough investigation of this effect is beyond the scope of this paper but we included in Table~\ref{tab:next}, for each candidate, the distance of the nearest source as well as the relative contribution of photons from the neighboring sources and from the diffuse backgrounds. Those values show that the diffuse background impacts the sources much more than their neighbors, with the exception of 4FGL J2021.0+4031e around the bright PSR J2021+4026.

% Table 4 : Spectral Parameters
\begin{deluxetable*}{cccc}
\tablecaption{Fractions of photons from neighboring sources and diffuse background affecting all confirmed sources showing a significant break\label{tab:next}}
\tablehead{
\colhead{4FGL Name} &
\colhead{Distance ($^{\circ}$)} &
\colhead{$\rm N_{soi} / N_{diff}$} &
\colhead{$\rm N_{soi} / N_{srcs}$} \\ 
}
\startdata
4FGL J0222.4+6156e & 0.76 & 0.85 & 3.61 \\ 
4FGL J0240.5+6113 & 1.28 & 7.34 & 112.58 \\ 
4FGL J0330.7+5845 & 2.51 & 0.14 & 22.24 \\ 
4FGL J0340.4+5302 & 1.39 & 0.86 & 38.35 \\ 
4FGL J0426.5+5434 & 0.99 & 0.68 & 311.20 \\ 
4FGL J0500.3+4639e & 1.31 & 0.17 & 7.70 \\ 
4FGL J0540.3+2756e & 1.35 & 0.08 & 1.99 \\ 
4FGL J0609.0+2006 & 0.48 & 0.20 & 1.78 \\ 
4FGL J0617.2+2234e & 0.40 & 5.40 & 28.79 \\ 
4FGL J0620.4+1445 & 1.03 & 0.16 & 1.56 \\ 
4FGL J0634.2+0436e & 1.29 & 0.22 & 2.74 \\ 
4FGL J0639.4+0655e & 1.47 & 0.09 & 0.85 \\ 
4FGL J0709.1$-$1034 & 1.42 & 0.25 & 10.44 \\ 
4FGL J0844.1$-$4330 & 0.85 & 0.25 & 0.22 \\ 
4FGL J0850.8$-$4239 & 0.68 & 0.29 & 0.80 \\ 
4FGL J0904.7$-$4908c & 0.66 & 0.18 & 1.60 \\ 
4FGL J1008.1$-$5706c & 0.59 & 0.19 & 2.76 \\ 
4FGL J1018.9$-$5856 & 0.33 & 2.28 & 3.39 \\ 
4FGL J1045.1$-$5940 & 0.52 & 1.39 & 2.76 \\ 
4FGL J1351.6$-$6142 & 0.72 & 0.25 & 1.78 \\ 
4FGL J1358.3$-$6026 & 0.48 & 0.27 & 1.32 \\ 
4FGL J1405.1$-$6119 & 0.48 & 0.51 & 1.51 \\ 
4FGL J1442.2$-$6005 & 0.24 & 0.17 & 0.95 \\  
4FGL J1447.4$-$5757 & 1.28 & 0.28 & 2.97 \\ 
4FGL J1514.2$-$5909e & 0.69 & 0.24 & 1.06 \\ 
4FGL J1534.0$-$5232 & 1.23 & 0.12 & 2.98 \\ 
4FGL J1547.5$-$5130 & 0.66 & 0.17 & 2.28 \\ 
4FGL J1552.9$-$5607e & 2.25 & 0.15 & 6.69 \\ 
4FGL J1601.3$-$5224 & 1.46 & 0.14 & 3.64 \\ 
4FGL J1608.8$-$4803 & 1.30 & 0.15 & 1.92 \\ 
4FGL J1626.6$-$4251 & 0.75 & 0.12 & 1.33 \\ 
4FGL J1633.0$-$4746e & 0.28 & 0.32 & 2.44 \\ 
4FGL J1742.8$-$2246 & 1.00 & 0.20 & 1.22 \\ 
4FGL J1801.3$-$2326e & 0.08 & 0.70 & 2.83 \\ 
4FGL J1808.2$-$1055 & 1.14 & 0.17 & 1.38 \\ 
4FGL J1812.2$-$0856 & 1.36 & 0.21 & 3.33 \\ 
4FGL J1813.1$-$1737e & 0.50 & 0.20 & 2.43 \\ 
4FGL J1814.2$-$1012 & 1.29 & 0.16 & 1.20 \\ 
4FGL J1839.4$-$0553 & 0.28 & 0.45 & 0.87 \\ 
4FGL J1852.4+0037e & 0.76 & 0.15 & 0.87 \\ 
4FGL J1855.2+0456 & 1.25 & 0.16 & 2.01 \\ 
4FGL J1855.9+0121e & 0.44 & 1.25 & 5.17 \\ 
4FGL J1857.7+0246e & 0.45 & 0.22 & 1.34 \\ 
4FGL J1906.9+0712 & 0.23 & 0.26 & 0.65 \\ 
4FGL J1908.7+0812 & 0.94 & 0.20 & 1.54 \\ 
4FGL J1911.0+0905 & 0.21 & 0.49 & 2.81 \\ 
4FGL J1923.2+1408e & 0.35 & 0.91 & 3.27 \\ 
4FGL J1931.1+1656 & 0.74 & 0.22 & 1.92 \\ 
4FGL J1934.3+1859 & 0.55 & 0.21 & 0.72 \\ 
4FGL J2021.0+4031e & 0.12 & 0.79 & 0.19 \\ 
4FGL J2028.6+4110e & 0.73 & 0.16 & 0.43 \\ 
4FGL J2032.6+4053 & 0.57 & 0.22 & 0.46 \\ 
4FGL J2038.4+4212 & 0.71 & 0.24 & 1.09 \\ 
4FGL J2045.2+5026e & 0.32 & 0.32 & 1.77 \\ 
4FGL J2056.4+4351c & 1.07 & 0.18 & 3.63 \\ 
4FGL J2108.0+5155 & 1.14 & 0.19 & 15.44 \\ 
\enddata
\tablecomments{Column 1 indicates the distance (in degrees) of the
  nearest neighboring source. Columns 2 and 3 report the ratio, in the pixel at the source position, between the predicted number of photons from the source of interest with respect to those of the galactic and isotropic diffuse background ($\rm N_{soi} / N_{diff}$), and to those of all neighboring sources $\rm N_{soi} / N_{srcs}$, respectively.}
\end{deluxetable*}

Overall, 56 sources among the 77 sources detected with the standard IEM and IRFs are confirmed with our systematic studies. The 21 candidates rejected are all sources that do not meet the ${\rm TS_{SBPL}}$ or ${\rm TS_{SBPL2}}$ criteria when using the old diffuse model, while the inaccuracy in the effective area has a minor effect in our analysis as can be seen in Table~\ref{tab:syst}. The spectral parameters of the confirmed sources are reported in Table~\ref{tab:spectra}. As can be seen in this Table, even if the old diffuse background detects a significant energy break, the energy of this break can be significantly different than with the standard IEM, leading to large systematics as well on $\Gamma_1$. 
%However, the values of $\Gamma_2$ and the integral flux between 50 MeV and 1 GeV are much more robust. In addition to performing a spectral fit over the entire energy range, we computed an SED by fitting the flux of the source independently in 10 energy bins spaced uniformly in log space from 50 MeV to 1 GeV. 
However, the value of $\Gamma_2$ is much more robust.\\ 
In addition to performing a spectral fit over the entire energy range, we computed an SED by fitting the flux of the source independently in 10 energy bins spaced uniformly in log space from 50 MeV to 1 GeV.  During this fit, we fixed the spectral index of the source at 2 as well as the model of background sources to the best fit obtained in the whole energy range except the normalizations of the Galactic diffuse and isotropic backgrounds. We determined the flux in an energy bin when TS $\ge 1$ and otherwise computed a 95\% confidence level Bayesian flux upper limit, assuming a uniform prior on flux following \cite{1983NIMPR.212..319H}. The systematic studies with the old diffuse and bracketing IRFs were also computed on all SED points for the 56 confirmed sources and the two uncertainties were added in quadrature. When an upper limit was derived, the maximal and minimal upper limits derived in this energy interval are plotted to indicate the systematics related to this data point.

% Table 4 : Spectral Parameters
\begin{deluxetable*}{cccccccccc}
\tablecaption{Spectral parameters of all confirmed sources showing a significant break\label{tab:spectra}}
\tablehead{
\colhead{4FGL Name} &
\colhead{$\rm I(50 - 1000)$} &
\colhead{$\rm \Delta I(50 - 1000)$} & 
\colhead{$\rm E_{break}$} &
\colhead{$\rm \Delta E_{break}$} & 
\colhead{$\rm \Gamma_1$} &
\colhead{$\rm \Delta \Gamma_1$} & 
\colhead{$\rm \Gamma_2$} &
\colhead{$\rm \Delta \Gamma_2$} \\ 
\colhead{} &
\colhead{$10^{-6}$ (MeV/cm$^2$/s)} &
\colhead{stat/syst} & 
\colhead{(MeV)} &
\colhead{stat/syst} & 
\colhead{} &
\colhead{stat/syst} & 
\colhead{} &
\colhead{stat/syst} 
}
\startdata
4FGL J0222.4+6156e & 47.8 & 2.7/0.6 & 465 & 78/40 & 1.35 & 0.14/0.03 & 2.34 & 0.21/0.14 \\ 
4FGL J0240.5+6113 & 237.6 & 1.9/6.6 & 142 & 10/74 & 1.63 & 0.03/0.36 & 2.10 & 0.02/0.10 \\ 
4FGL J0330.7+5845 & 3.2 & 0.5/0.3 & 367 & 38/52 & $-$0.68 & 0.75/0.81 & 3.42 & 0.64/0.21 \\ 
4FGL J0340.4+5302 & 34.1 & 1.3/5.8 & 284 & 43/116 & 1.60 & 0.14/0.38 & 3.27 & 0.23/0.35 \\ 
4FGL J0426.5+5434 & 15.1 & 0.8/0.9 & 338 & 47/80 & 1.25 & 0.16/0.35 & 2.50 & 0.18/0.07 \\ 
4FGL J0500.3+4639e & 11.6 & 1.0/1.6 & 252 & 43/107 & 0.14 & 0.61/1.06 & 2.17 & 0.19/0.08 \\ 
4FGL J0540.3+2756e & 14.8 & 1.5/4.8 & 493 & 82/146 & 0.90 & 0.25/0.54 & 2.64 & 0.52/0.37 \\ 
4FGL J0609.0+2006 & 4.7 & 0.7/0.8 & 499 & 134/59 & 0.11 & 0.67/0.56 & 3.52 & 0.66/0.35 \\ 
4FGL J0617.2+2234e & 122.5 & 2.4/1.1 & 276 & 19/3 & 1.06 & 0.05/0.03 & 1.75 & 0.03/0.03 \\ 
4FGL J0620.4+1445 & 3.2 & 0.6/0.4 & 355 & 36/55 & 0.26 & 0.44/0.36 & 4.03 & 0.71/0.63 \\ 
4FGL J0634.2+0436e & 24.1 & 1.4/15.5 & 243 & 41/121 & 1.07 & 0.13/0.50 & 2.00 & 0.13/0.26 \\ 
4FGL J0639.4+0655e & 36.6 & 3.3/19.2 & 233 & 31/167 & $-$0.13 & 0.66/0.95 & 2.51 & 0.23/0.59 \\ 
4FGL J0709.1$-$1034 & 5.1 & 0.8/2.2 & 351 & 57/23 & 0.06 & 0.90/0.25 & 3.40 & 0.56/0.36 \\ 
4FGL J0844.1$-$4330 & 15.2 & 2.6/2.4 & 159 & 28/76 & 0.35 & 0.19/0.46 & 3.28 & 0.20/0.41 \\ 
4FGL J0850.8$-$4239 & 10.8 & 1.4/1.7 & 424 & 83/26 & 1.24 & 0.12/0.11 & 3.71 & 0.30/0.03 \\ 
4FGL J0904.7$-$4908c & 10.6 & 0.7/1.4 & 402 & 12/173 & 1.10 & 0.07/1.19 & 2.99 & 0.16/0.71 \\ 
4FGL J1008.1$-$5706c & 12.3 & 1.6/5.1 & 409 & 76/37 & 0.96 & 0.43/0.55 & 3.40 & 0.64/0.33 \\ 
4FGL J1018.9$-$5856 & 130.0 & 3.4/11.9 & 73 & 1/24 & 0.32 & 0.02/0.31 & 1.98 & 0.02/0.05 \\ 
4FGL J1045.1$-$5940 & 49.8 & 2.3/6.0 & 525 & 26/178 & 1.12 & 0.05/0.17 & 2.12 & 0.11/0.14 \\ 
4FGL J1351.6$-$6142 & 26.9 & 2.7/12.5 & 125 & 8/22 & $-$0.87 & 0.17/0.59 & 2.37 & 0.12/0.30 \\ 
4FGL J1358.3$-$6026 & 20.8 & 1.5/2.3 & 131 & 4/28 & $-$0.63 & 0.05/0.52 & 2.55 & 0.07/0.13 \\ 
4FGL J1405.1$-$6119 & 61.9 & 2.7/9.2 & 110 & 2/14 & 0.06 & 0.02/0.44 & 2.14 & 0.03/0.05 \\ 
4FGL J1442.2$-$6005 & 21.3 & 1.7/6.9 & 126 & 2/21 & $-$1.10 & 0.03/0.73 & 2.58 & 0.07/0.44 \\  
4FGL J1447.4$-$5757 & 12.2 & 1.4/9.1 & 303 & 42/164 & 0.72 & 0.27/0.71 & 2.56 & 0.24/0.41 \\ 
4FGL J1514.2$-$5909e & 38.4 & 3.2/10.4 & 116 & 9/27 & 1.08 & 0.10/0.69 & 2.92 & 0.10/0.05 \\ 
4FGL J1534.0$-$5232 & 4.5 & 0.9/3.3 & 375 & 30/161 & 0.68 & 0.29/0.47 & 3.95 & 0.24/0.79 \\ 
4FGL J1547.5$-$5130 & 12.8 & 2.8/1.1 & 349 & 331/47 & 1.31 & 0.09/0.49 & 4.68 & 0.14/0.18 \\ 
4FGL J1552.9$-$5607e & 8.9 & 0.8/8.9 & 386 & 38/87 & 0.04 & 0.09/1.15 & 2.15 & 0.26/0.09 \\ 
4FGL J1601.3$-$5224 & 26.1 & 2.4/3.2 & 356 & 23/177 & 1.19 & 0.17/0.77 & 3.78 & 0.32/0.89 \\ 
4FGL J1608.8$-$4803 & 11.3 & 4.0/1.3 & 346 & 112/188 & 1.51 & 0.95/2.20 & 3.36 & 0.22/0.52 \\ 
4FGL J1626.6$-$4251 & 4.5 & 0.7/1.0 & 354 & 16/32 & 0.63 & 0.31/0.28 & 4.57 & 0.15/0.58 \\ 
4FGL J1633.0$-$4746e & 78.1 & 1.9/21.9 & 517 & 18/152 & 1.19 & 0.04/2.28 & 2.11 & 0.15/0.12 \\ 
4FGL J1742.8$-$2246 & 5.7 & 0.7/0.8 & 364 & 22/44 & 0.28 & 0.17/0.32 & 3.40 & 0.15/0.30 \\ 
4FGL J1801.3$-$2326e & 135.2 & 11.8/2.6 & 401 & 138/150 & 1.33 & 0.06/0.40 & 2.14 & 0.79/0.28 \\ 
4FGL J1808.2$-$1055 & 3.5 & 1.4/1.9 & 354 & 6/39 & 0.22 & 0.51/0.67 & 2.81 & 0.75/0.31 \\ 
4FGL J1812.2$-$0856 & 8.2 & 0.7/0.8 & 284 & 7/107 & 0.55 & 0.05/0.88 & 3.11 & 0.11/0.30 \\ 
4FGL J1813.1$-$1737e & 56.0 & 3.1/12.4 & 154 & 3/84 & 0.22 & 0.41/0.25 & 2.17 & 0.03/0.42 \\ 
4FGL J1814.2$-$1012 & 5.5 & 0.7/0.5 & 471 & 50/10 & 0.19 & 0.42/0.53 & 4.25 & 0.17/0.34 \\ 
4FGL J1839.4$-$0553 & 62.4 & 3.8/8.4 & 86 & 3/30 & $-$0.29 & 0.33/0.30 & 1.94 & 0.04/0.10 \\ 
4FGL J1852.4+0037e & 43.4 & 2.5/7.9 & 119 & 2/18 & $-$1.19 & 0.51/0.91 & 2.41 & 0.05/0.33 \\ 
4FGL J1855.2+0456 & 13.5 & 3.1/0.1 & 379 & 157/56 & 0.53 & 0.12/0.44 & 3.76 & 0.25/0.44 \\ 
4FGL J1855.9+0121e & 184.1 & 2.5/7.7 & 347 & 5/62 & 1.03 & 0.04/0.05 & 1.91 & 0.02/0.07 \\ 
4FGL J1857.7+0246e & 37.7 & 0.8/17.7 & 615 & 20/284 & 1.51 & 0.04/1.58 & 2.45 & 0.12/0.24 \\ 
4FGL J1906.9+0712 & 28.6 & 2.0/8.2 & 134 & 3/21 & $-$0.69 & 0.06/0.70 & 2.44 & 0.07/0.15 \\ 
4FGL J1908.7+0812 & 30.6 & 1.1/17.5 & 137 & 3/170 & $-$1.19 & 0.05/1.54 & 2.75 & 0.08/0.88 \\ 
4FGL J1911.0+0905 & 38.8 & 1.9/12.0 & 364 & 11/73 & 0.51 & 0.16/0.19 & 2.01 & 0.06/0.17 \\ 
4FGL J1923.2+1408e & 93.6 & 2.1/3.9 & 381 & 14/131 & 1.39 & 0.01/0.51 & 2.11 & 0.04/0.11 \\ 
4FGL J1931.1+1656 & 17.1 & 2.1/9.9 & 203 & 8/19 & $-$0.60 & 0.10/0.59 & 2.64 & 0.10/0.04 \\ 
4FGL J1934.3+1859 & 15.9 & 2.0/3.5 & 211 & 23/11 & 0.17 & 0.38/0.23 & 3.13 & 0.27/0.12 \\ 
4FGL J2021.0+4031e & 119.8 & 4.3/15.9 & 147 & 7/31 & 1.64 & 0.05/0.18 & 2.55 & 0.05/0.05 \\ 
4FGL J2028.6+4110e & 201.5 & 5.2/77.9 & 383 & 13/138 & 1.00 & 0.02/0.37 & 2.23 & 0.06/0.24 \\ 
4FGL J2032.6+4053 & 22.6 & 4.9/0.9 & 561 & 217/21 & 1.90 & 0.16/0.07 & 4.48 & 0.47/0.23 \\ 
4FGL J2038.4+4212 & 20.2 & 2.0/4.3 & 152 & 22/187 & 0.65 & 0.23/0.29 & 2.29 & 0.14/0.31 \\ 
4FGL J2045.2+5026e & 35.6 & 1.9/13.0 & 397 & 24/155 & 1.09 & 0.09/0.29 & 2.44 & 0.13/0.38 \\ 
4FGL J2056.4+4351c & 9.0 & 1.1/5.2 & 183 & 5/65 & 0.02 & 0.04/0.29 & 2.52 & 0.07/0.22 \\ 
4FGL J2108.0+5155 & 9.8 & 1.7/0.4 & 451 & 77/247 & 1.09 & 0.30/0.18 & 2.68 & 0.68/0.70 \\ 
\enddata
\tablecomments{Results of the maximum likelihood spectral fits for
  sources showing significant breaks confirmed by the systematic
  studies. These results are obtained using a smooth broken power law
  representation. Columns 2, 4, 6 and 8 report the integrated flux,
  the break energy and the photon indices $\Gamma_1$ and $\Gamma_2$  of the source fit in the energy range from 50 MeV to
  1 GeV following Equation~\ref{eq:sbpl}. Columns 3, 5, 7 and 9 report the statistic and systematic
  uncertainties on these spectral parameters.}
\end{deluxetable*}

\begin{figure*}[ht]
\begin{center}
\begin{tabular}{ll}
\includegraphics[width=0.98\textwidth]{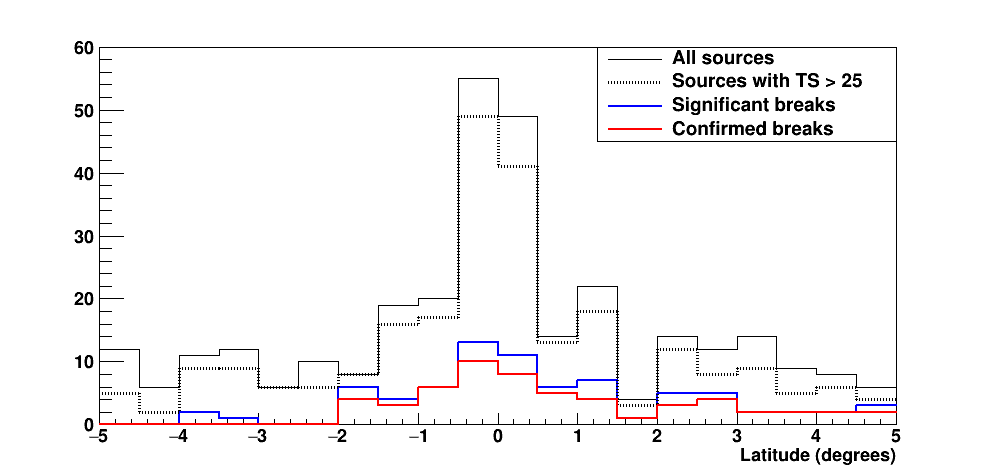}\\
\includegraphics[width=0.98\textwidth]{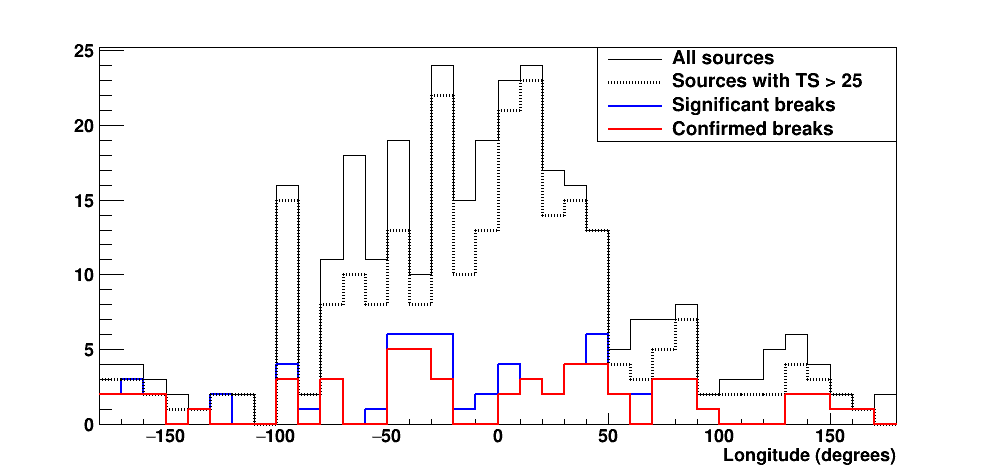}\\
\end{tabular}
\end{center}
\caption{
\label{fig:lonlat}Latitude (top) and longitude (bottom) distributions of the 311 sources selected (black line), the 247 sources with $TS > 25$ in our pipeline (black dashed line), the 77 sources with significant breaks (blue line) and the 56 confirmed cases by our studies of systematics (red line). 
}
\end{figure*}

% Table : Source population
\begin{deluxetable*}{ccc}
\tablecaption{Summary of source classes\label{tab:class}}
\tablehead{
\colhead{Source class} &
\colhead{Analyzed} &
\colhead{Confirmed} 
}
\startdata
Supernova remnant (SNR) & 23 & 13  \\ 
Pulsar wind nebulae (PWN) & 4 & 2  \\ 
Supernova remnant / Pulsar wind nebula (SPP) & 37 & 6  \\ 
Star-forming region (SFR) & 1 & 1  \\ 
Unknown (UNK) & 31 & 4  \\ 
Binary/High-mass binary (BIN/HMB) & 5 & 4  \\ 
Unidentified (UNID) & 210 & 26  \\ 
\enddata
\tablecomments{For the source classes SNR, PWN, SPP, SFR, BIN and HMB, we add both the firm identifications reported in the 4FGL catalog as well as the associations (capital and lower case letters as seen in Column 6 of Table \ref{tab:candidates})}
\end{deluxetable*}

\section{Discussion}\label{section:results}
\subsection{Population study}\label{subsection:pop}
\begin{figure}[ht]
\begin{center}
\begin{tabular}{ll}
\includegraphics[width=0.48\textwidth]{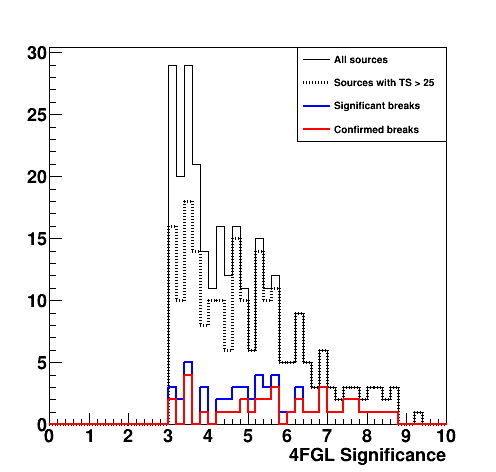}\\
\end{tabular}
\end{center}
\caption{
\label{fig:signi}Distribution of the 4FGL significance between 300 MeV and 1 GeV for the 311 sources selected (black line), the 247 sources with $TS > 25$ in our pipeline (black dashed line), the 77 sources with significant breaks (blue line) and the 56 confirmed cases by our studies of systematics (red line). 
}
\end{figure}

We detected 56 4FGL $\gamma$-ray sources showing a significant energy break in their spectrum between 50 MeV and 1 GeV confirmed by our studies of systematics. As can be seen in Figure~\ref{fig:lonlat}, the distribution of sources showing a significant break in their low-energy spectrum is more uniform in both latitude and longitude than the parent distribution even if there remains a peak at latitude 0 and in the Galactic Ridge.\\
The sources that we detect significantly with our analysis (${\rm TS} >25$) follow the same trend except for the region at $\sim 300^{\circ}$ longitude which contains more faint sources than the other regions of the plane. Figure~\ref{fig:signi} clearly shows that the sources that we do not detect with ${\rm TS} > 25$ in our pipeline have predominantly low significance in the 4FGL catalog in the 300 MeV -- 1 GeV energy band, which is reassuring. However, there is no correlation between the significance value in the 4FGL catalog and the detection of a break with our pipeline. It can be seen in this same Figure since the distribution for the sources presenting a significant break is uniform.\\

\begin{figure*}[ht]
\begin{center}
\begin{tabular}{ll}
\includegraphics[width=0.49\textwidth]{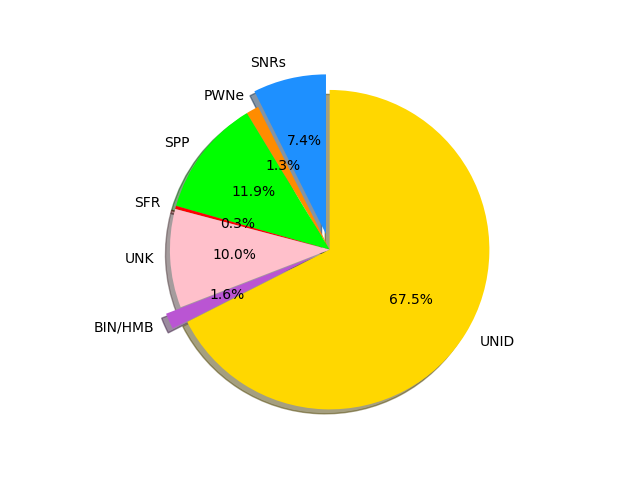}
\includegraphics[width=0.49\textwidth]{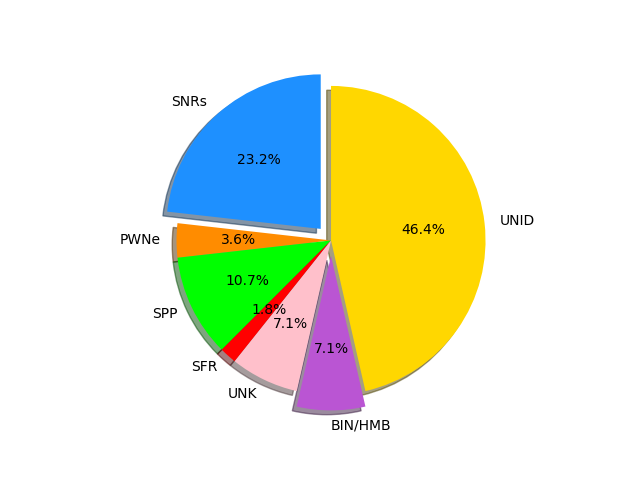}
\end{tabular}
\end{center}
\caption{
\label{fig:pop} Pie charts showing the classes of sources analyzed (Left) and those for which a significant break is detected (Right). The class names are those used in the 4FGL catalog: SNR stands for Supernova remnant, PWN for pulsar wind nebula, SFR for star-forming region, BIN for binary, HMB for high-mass binary. The designation SPP indicates potential association with SNR or PWN. The UNK class includes low-latitude blazar candidates of uncertain type associated solely via the Likelihood-Ratio (LR) method.
}
\end{figure*}

\begin{figure*}[ht]
\begin{center}
\includegraphics[width=0.98\textwidth]{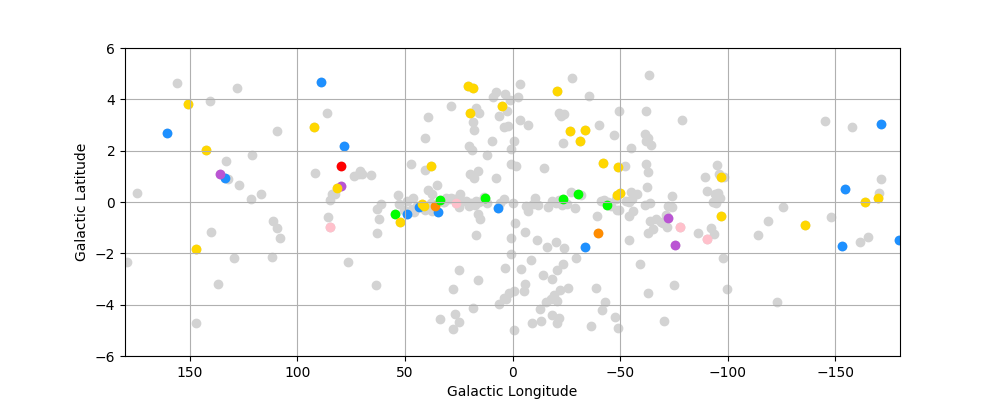}
\end{center}
\caption{
\label{fig:galplane} Distribution of sources in Galactic coordinates. Light gray markers indicate the 311 sources analyzed in this paper. Coloured markers indicate the position of the 56 sources for which a significant break is detected: yellow for UNIDs, blue for SNRs, orange for PWNe, green for SPPs, red for SFR, pink for UNKs and purple for BIN/HMB. The boundary of the latitude selection is $5^{\circ}$.
}
\end{figure*}

The association summary is given in Table~\ref{tab:class} and is illustrated by the pie charts in Figure~\ref{fig:pop}. Out of 311 candidates, 210 are unidentified, representing 67.5\% of the sources analyzed. It is striking to see that only 26 UNIDs show a spectral break confirmed with our systematic studies (which represents 46.4\% of the sources with significant breaks). The 30 remaining candidates out of 56 confirmed cases present an association reported by the 4FGL Catalog listed in Table~\ref{tab:assoc}.\\ 
On the other hand, the fraction of sources associated with supernova remnants (SNRs) increases from 7.4\% (23 out of 311 sources) to 23.2\% (13 out of 56 sources). This makes SNRs the dominant class of sources with significant low-energy spectral break. Similarly, the fraction of sources associated with binaries increases from 1.6\% (5 out of 311) to 7.1\% (4 out of 56), showing that almost all binaries except 4FGL J1826.2$-$1450 (also known as LS 5039), show a significant spectral break. Despite their small fractions, binaries could contribute significantly to our population of sources with low-energy spectral breaks; however it should be noted here that the spectral analysis is performed over 8 years and these sources often present variable $\gamma$-ray emission. A more thorough analysis of these sources would need to be done. Finally, only one star-forming region is analyzed (and confirmed) which prevents us from drawing a firm conclusion on this source class.\\ 
Figure~\ref{fig:galplane} illustrates the distribution over the sky of the 56 4FGL $\gamma$-ray sources showing a significant energy break. The lack of these sources at latitude smaller than $-2^{\circ}$ appears clearly. One can also note a large fraction of unidentified sources at longitude comprised between $-50^{\circ}$ and $50^{\circ}$. These sources are part of the large fraction of 4FGL unassociated sources located less than $10^{\circ}$ away from the Galactic plane with a wide latitude extension hard to reconcile with those of known classes of Galactic $\gamma$-ray sources.\\ 

\begin{figure}[ht]
\begin{center}
\begin{tabular}{ll}
\includegraphics[width=0.48\textwidth]{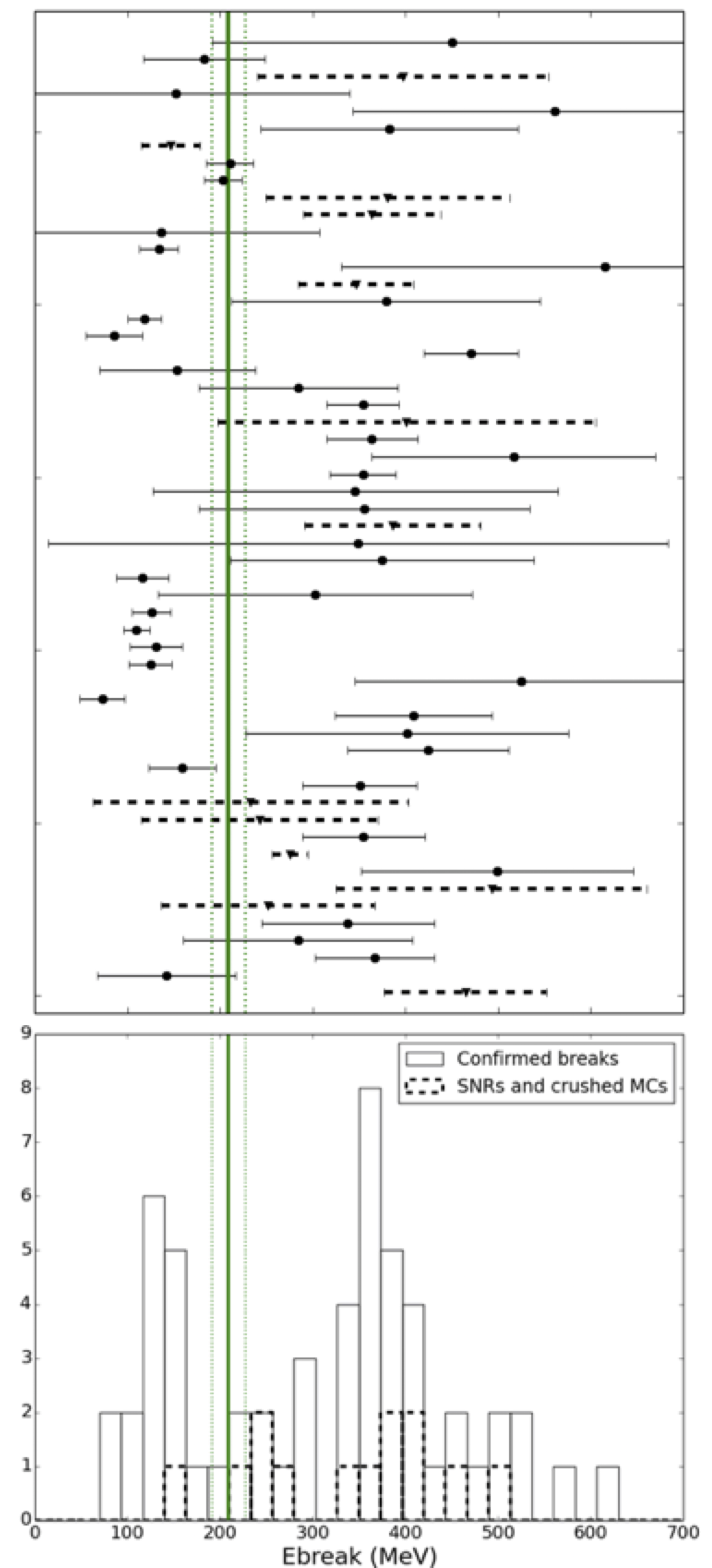}\\
\end{tabular}
\end{center}
\caption{
\label{fig:ebreak}Break energy for the 56 sources confirmed by our studies of systematics (black line) and for the identified SNRs and/or crushed molecular clouds (dotted line, see Section~\ref{subsection:snr}). The green line indicates the value of the break energy obtained using simulations based on the $naima$ package for a proton injection index of 2.0 and the two green dotted-dashed lines indicate the one sigma confidence interval derived (more details in the Appendix~\ref{appen:pion}). (Top) Individual values; (Bottom) Corresponding histograms.
}
\end{figure}

\begin{figure*}[ht]
\begin{center}
\begin{tabular}{ll}
\includegraphics[width=0.45\textwidth]{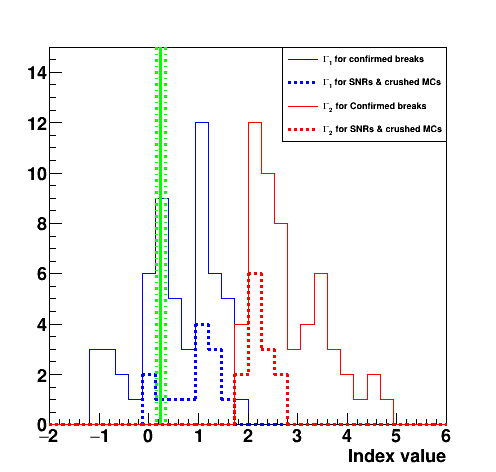}
\includegraphics[width=0.45\textwidth]{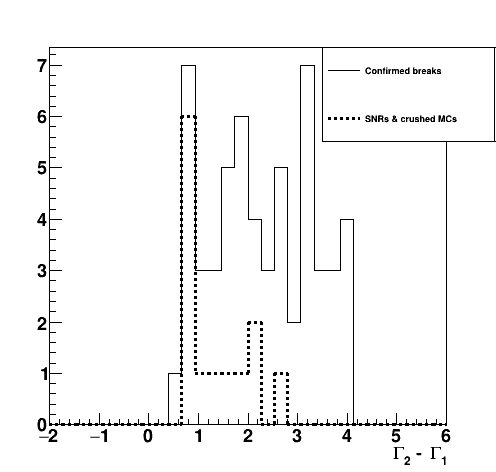}
\end{tabular}
\end{center}
\caption{
\label{fig:index}$\Gamma_1$ (blue line, left), $\Gamma_2$ (red line, left) and $\Gamma_2 - \Gamma_1$ (right) distributions for the 56 sources confirmed by our studies of systematics. In all cases, the dotted line corresponds to the same distribution presented with the solid line but restricted to SNRs (see Section~\ref{subsection:snr}). The green line indicates the value of $\Gamma_1$ obtained using simulations based on the $naima$ package for a proton injection index of 2.0 and the two green dotted-dashed lines indicate the one sigma confidence interval derived (more details in the Appendix~\ref{appen:pion}).
}
\end{figure*}

Looking now at the spectral parameters of the 56 confirmed sources, the distribution of the energy of the breaks detected by our analysis is relatively uniform between 70 MeV and 700 MeV, with no breaks detected below and above this energy interval (as a direct consequence of the energy interval analyzed here) and a higher proportion of breaks at $\sim$400 MeV as illustrated by Figure~\ref{fig:ebreak}. Interestingly, no low energy spectral breaks ($< 140$ MeV) are detected for the 13 sources associated with SNRs. As can be seen on the top panel of this Figure, the large error bars on this parameter prevent us from drawing any firm conclusion or even rejecting any candidate by a comparison with the standard value expected for proton-proton interaction indicated by the green line. On the other hand, there is a trend concerning the distributions of $\Gamma_1$ with a peak at $\sim 0.2$ and $\sim 1.0$. The peak at 0.2 is expected by proton-proton interaction (as indicated by the green line presenting the results of the simulations carried in Appendix~\ref{appen:pion}) but the peak at 1 is not predicted, though it is present for a large number of SNRs interacting with MCs. It might be due to some confusion by the Galactic and isotropic diffuse background. A double-peaked distribution is also visible in Figure~\ref{fig:index} for $\Gamma_2$ at $\sim 2.1$ and $\sim3.6$. For this parameter, the distribution restricted to SNRs contains a single peak at $\sim 2.1$. Looking now at the distribution of $\Gamma_2 - \Gamma_1$ in Figure~\ref{fig:index} (right), a peak at $\sim 0.9$ is highly pronounced for SNRs. This tends to show that the values obtained on $\Gamma_2$ and $\Gamma_2 - \Gamma_1$ could be used in the future to probe the type of particles radiating in a $\gamma$-ray source.  

% Table : Source association
\begin{deluxetable*}{ccc}
\tablecaption{Candidates with firm associations reported in the 4FGL Catalog\label{tab:assoc}}
\tablehead{
\colhead{4FGL Name} &
\colhead{Assoc1} &
\colhead{Assoc2} \\
}
\startdata
4FGL J0222.4+6156e & W 3 & HB 3 field \\ 
4FGL J0240.5+6113 & LS I +61 303 &  \\ 
4FGL J0500.3+4639e & HB 9 &  \\ 
4FGL J0540.3+2756e & Sim 147 &  \\ 
4FGL J0617.2+2234e & IC 443 &  \\ 
4FGL J0634.2+0436e & Rosette & Monoceros field \\ 
4FGL J0639.4+0655e & Monoceros &  \\ 
4FGL J0904.7$-$4908 & 1RXS J090505.3$-$490324 &  \\ 
4FGL J1008.1$-$5706 & 1RXS J100718.2$-$570335 &  \\ 
4FGL J1018.9$-$5856 & 1FGL J1018.6$-$5856 & FGES J1036.3$-$5833 field \\ 
4FGL J1045.1$-$5940 & Eta Carinae & FGES J1036.3$-$5833 field \\ 
4FGL J1442.2$-$6005 & SNR G316.3$-$00.0 &  \\ 
4FGL J1514.2$-$5909e & MSH 15$-$52 &  \\ 
4FGL J1552.9$-$5607e & MSH 15$-$56 &  \\ 
4FGL J1601.3$-$5224 & SNR G329.7+00.4 &  \\ 
4FGL J1633.0$-$4746e & HESS J1632$-$478 &  \\ 
4FGL J1801.3$-$2326e & W 28 &  \\ 
4FGL J1813.1$-$1737e & HESS J1813$-$178 &  \\ 
4FGL J1839.4$-$0553 & NVSS J183922$-$055321 & HESS J1841$-$055 field \\ 
4FGL J1852.4+0037e & Kes 79 &  \\ 
4FGL J1855.9+0121e & W 44 &  \\ 
4FGL J1857.7+0246e & HESS J1857+026 &  \\ 
4FGL J1911.0+0905 & W 49B &  \\ 
4FGL J1923.2+1408e & W 51C &  \\ 
4FGL J1934.3+1859 & SNR G054.4$-$00.3 &  \\ 
4FGL J2021.0+4031e & gamma Cygni & Cygnus Cocoon field \\ 
4FGL J2028.6+4110e & Cygnus X Cocoon &  \\ 
4FGL J2032.6+4053 & Cyg X$-$3 & Cygnus Cocoon field \\ 
4FGL J2045.2+5026e & HB 21 &  \\ 
4FGL J2056.4+4351 & 1RXS J205549.4+435216 &  \\
\enddata
\tablecomments{Columns 2 and 3 are derived from the Assoc1 and Assoc2 columns of the 4FGL Catalog. The latter provides an alternate designation or an indicator as to whether the source is inside an extended source. }
\end{deluxetable*}

\subsection{Supernova remnants and molecular clouds}\label{subsection:snr}
\begin{figure*}[ht]
\centering
\subfigure{
\begin{tabular}{ll}
\centering
\includegraphics[width=0.48\textwidth]{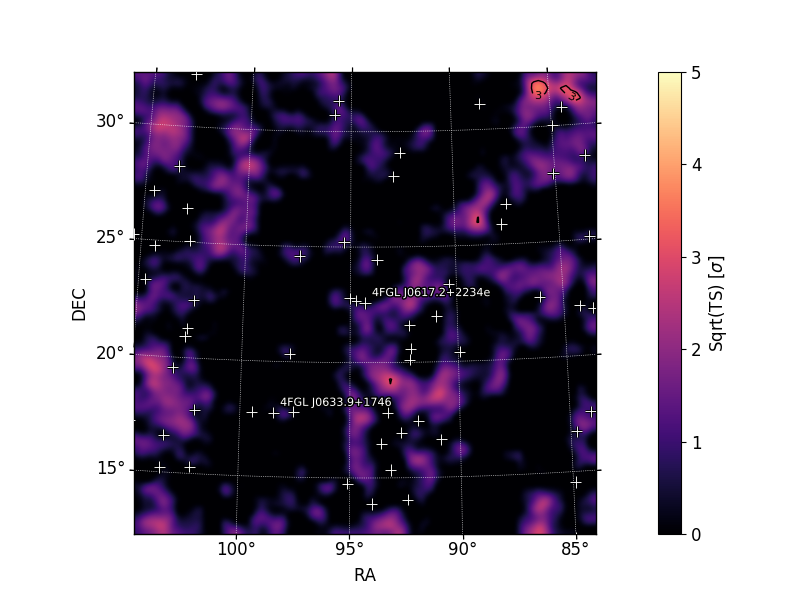}
\includegraphics[width=0.44\textwidth]{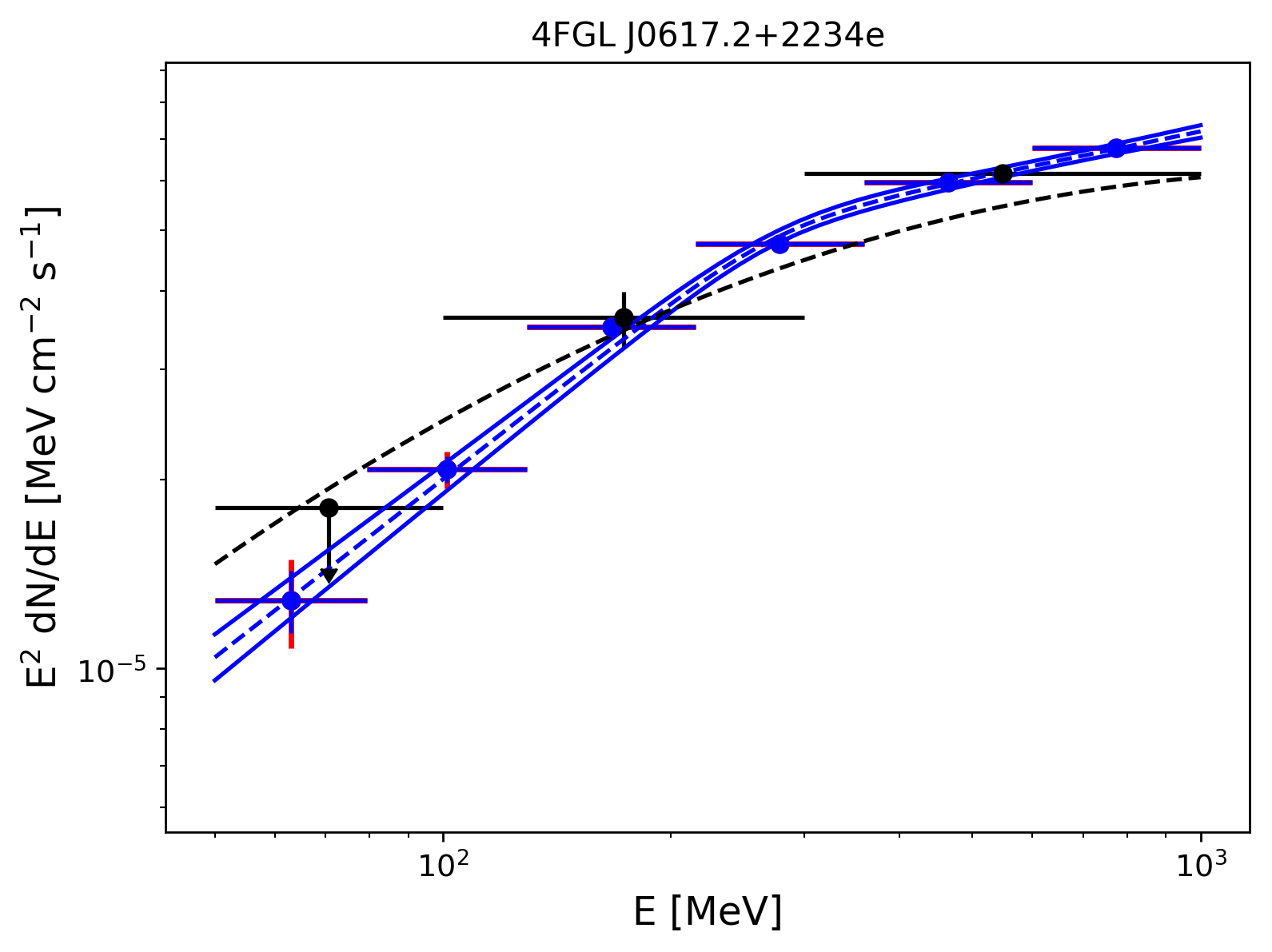}
\end{tabular}
}
\subfigure{
\begin{tabular}{ll}
\centering
\includegraphics[width=0.48\textwidth]{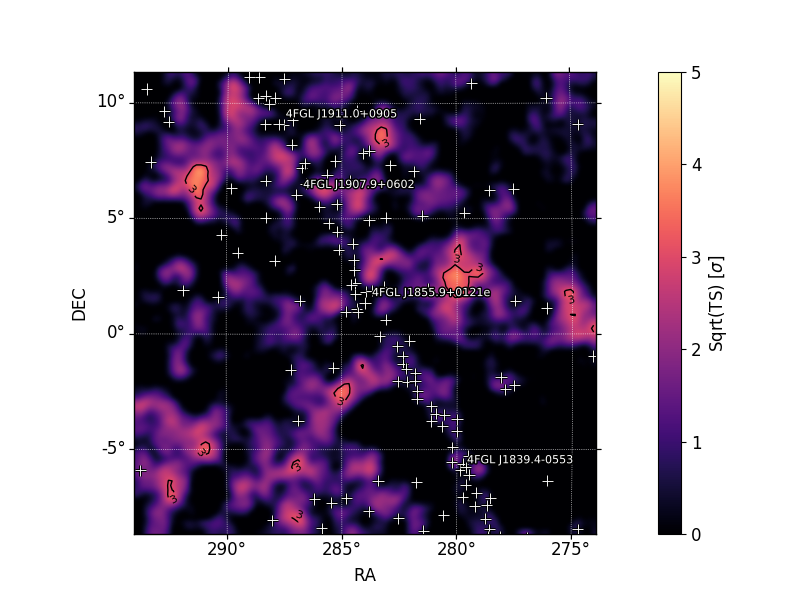}
\includegraphics[width=0.44\textwidth]{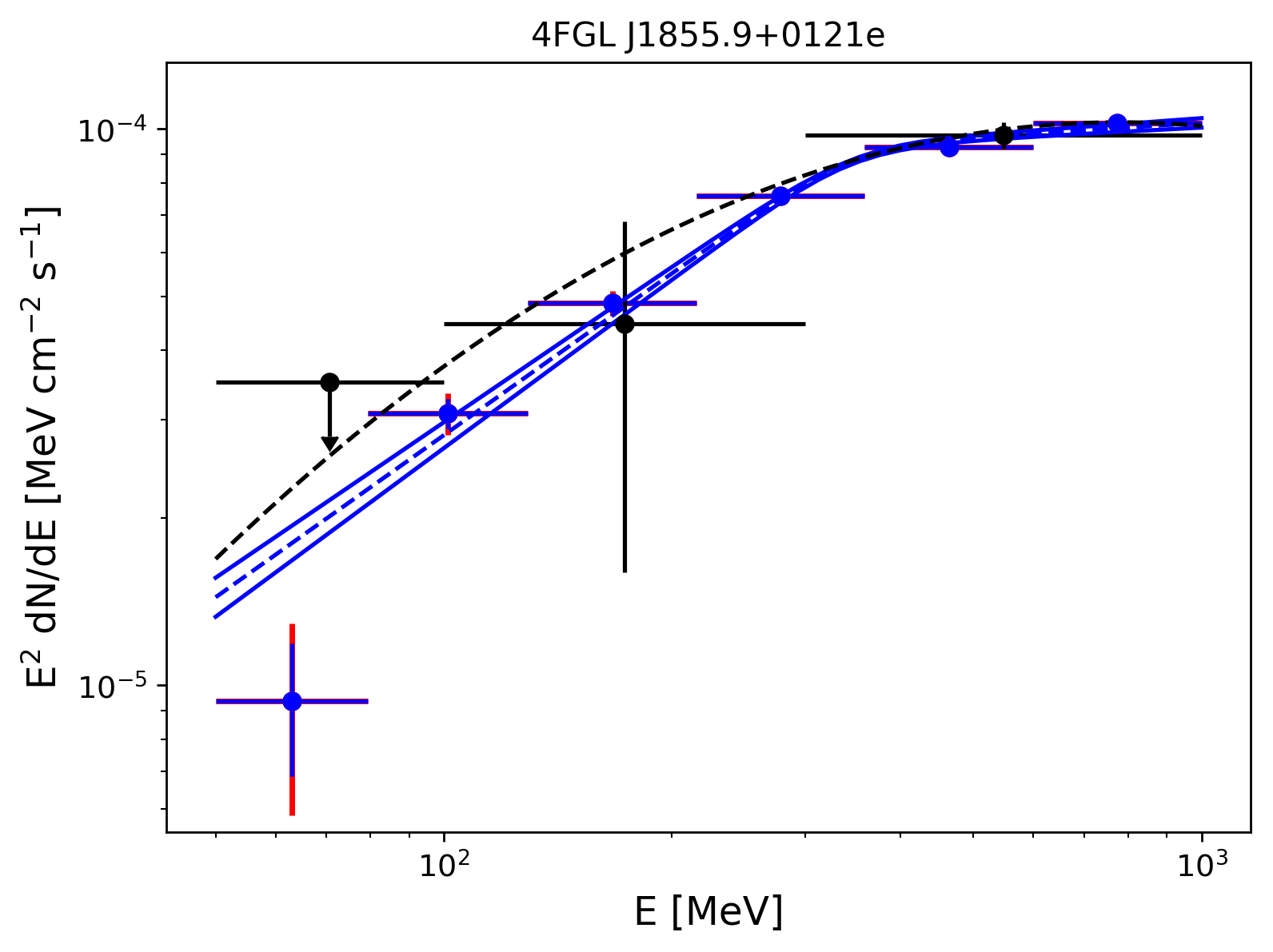}
\end{tabular}
}
\caption{\label{fig:}\label{fig:}\label{fig:snrsed}LAT residual TS maps in equatorial coordinates and significance units (left) and spectral energy distributions (right) of IC 443 (top) and W44 (bottom) between 50 MeV and 1 GeV. In the residual TS maps, all white crosses indicate the 4FGL sources included in the model of the region. For the SEDs, the blue points and butterflies are obtained in this analysis while the black points and dashed line are from the 4FGL catalog. The red lines take into account both the statistical and systematic errors added in quadrature. A 95\% C.L. upper limit is computed when the TS value is below 1.}
\end{figure*}

\begin{figure*}[ht]
\centering
\subfigure{
\begin{tabular}{ll}
\centering
\includegraphics[width=0.44\textwidth]{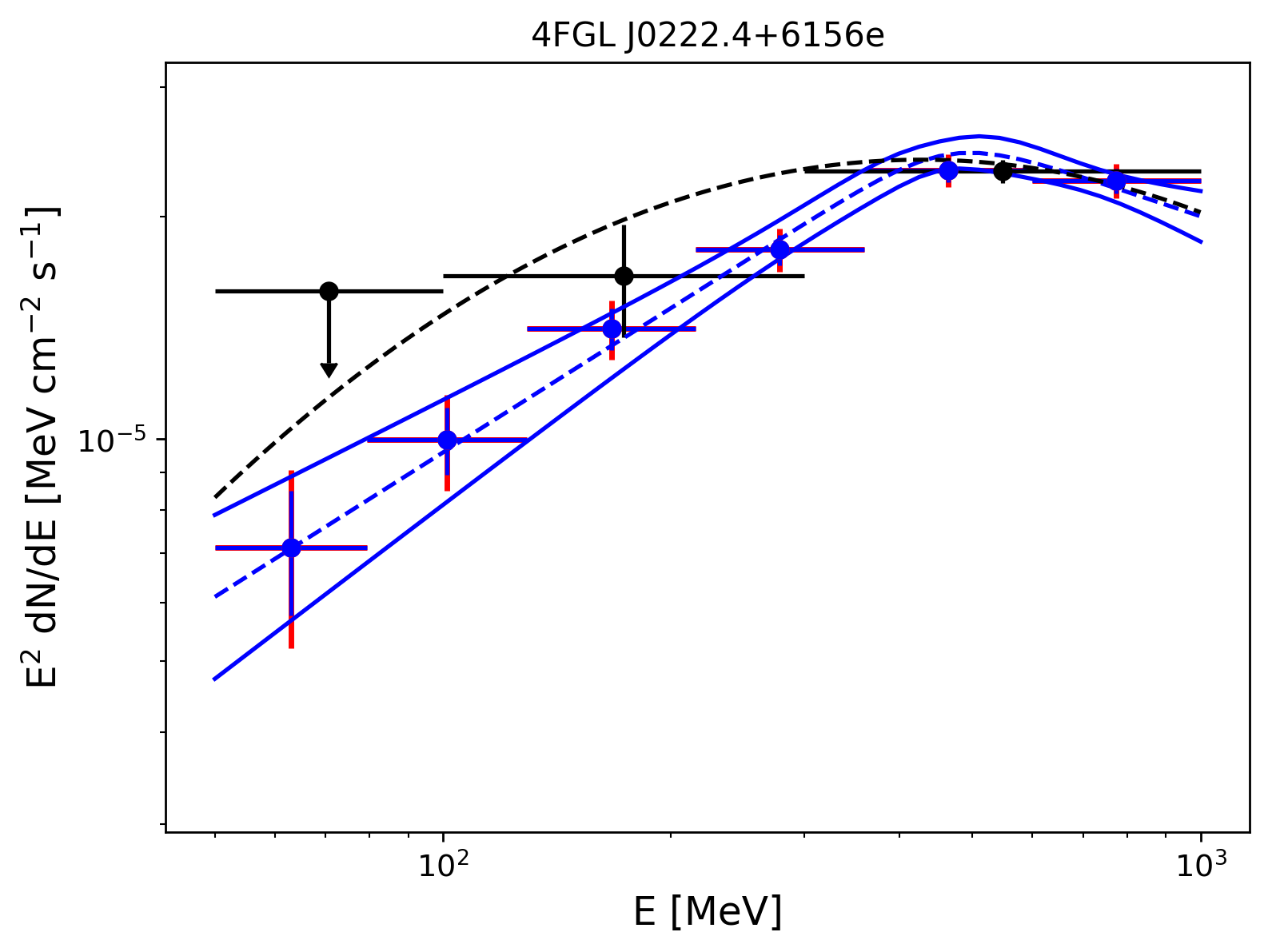}
\includegraphics[width=0.44\textwidth]{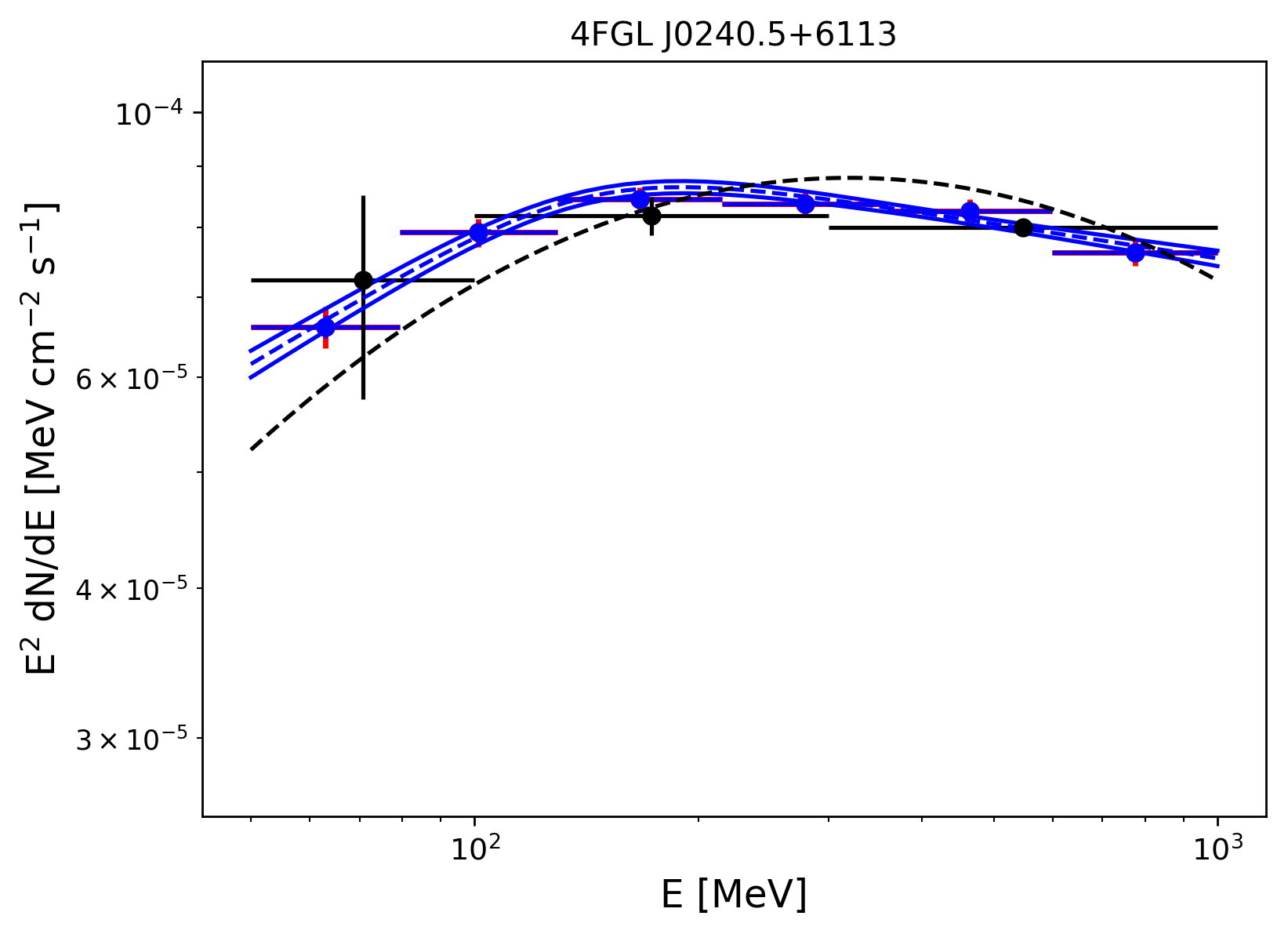}
\end{tabular}
}
\subfigure{
\begin{tabular}{ll}
\centering
\includegraphics[width=0.44\textwidth]{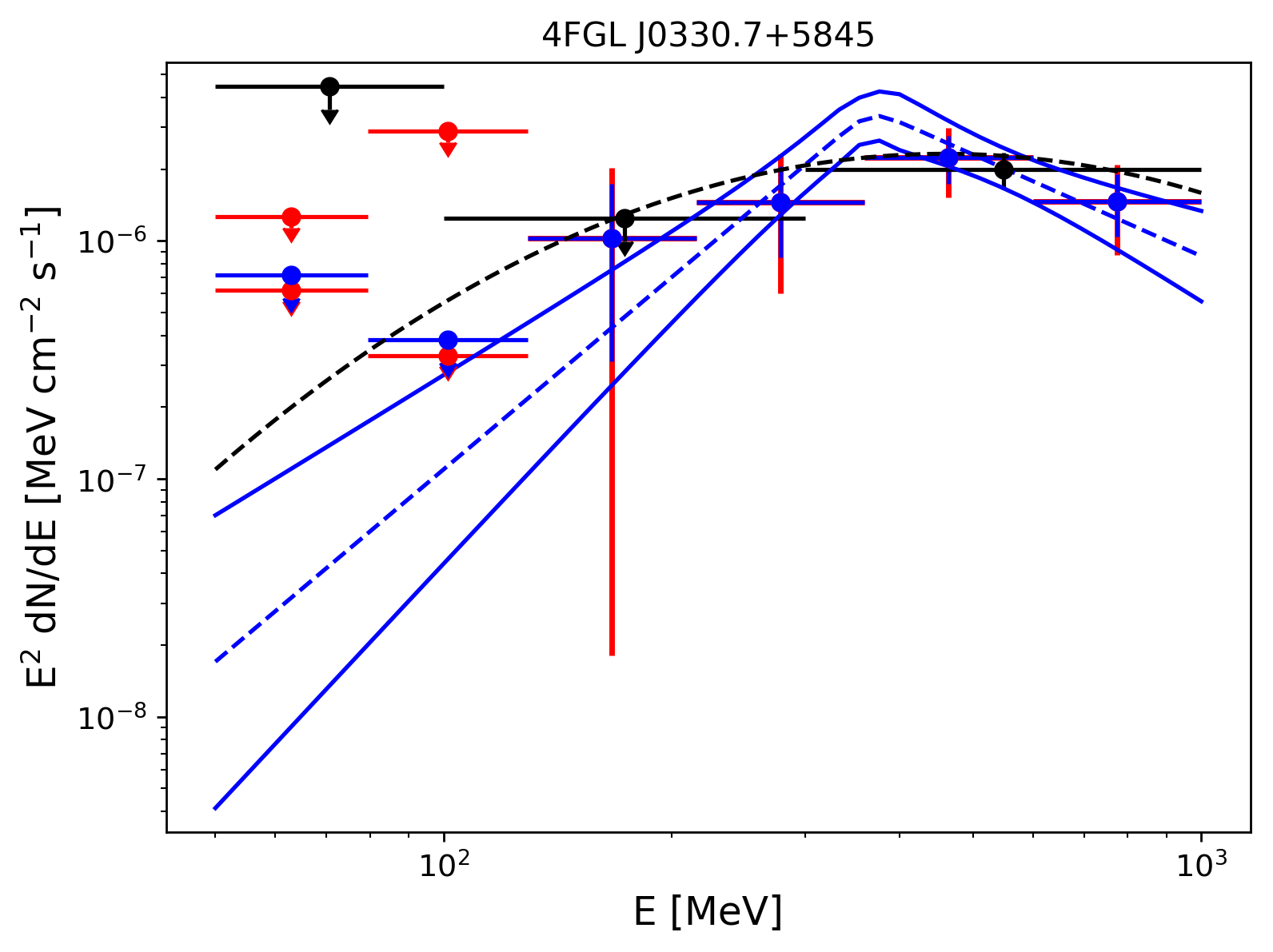}
\includegraphics[width=0.44\textwidth]{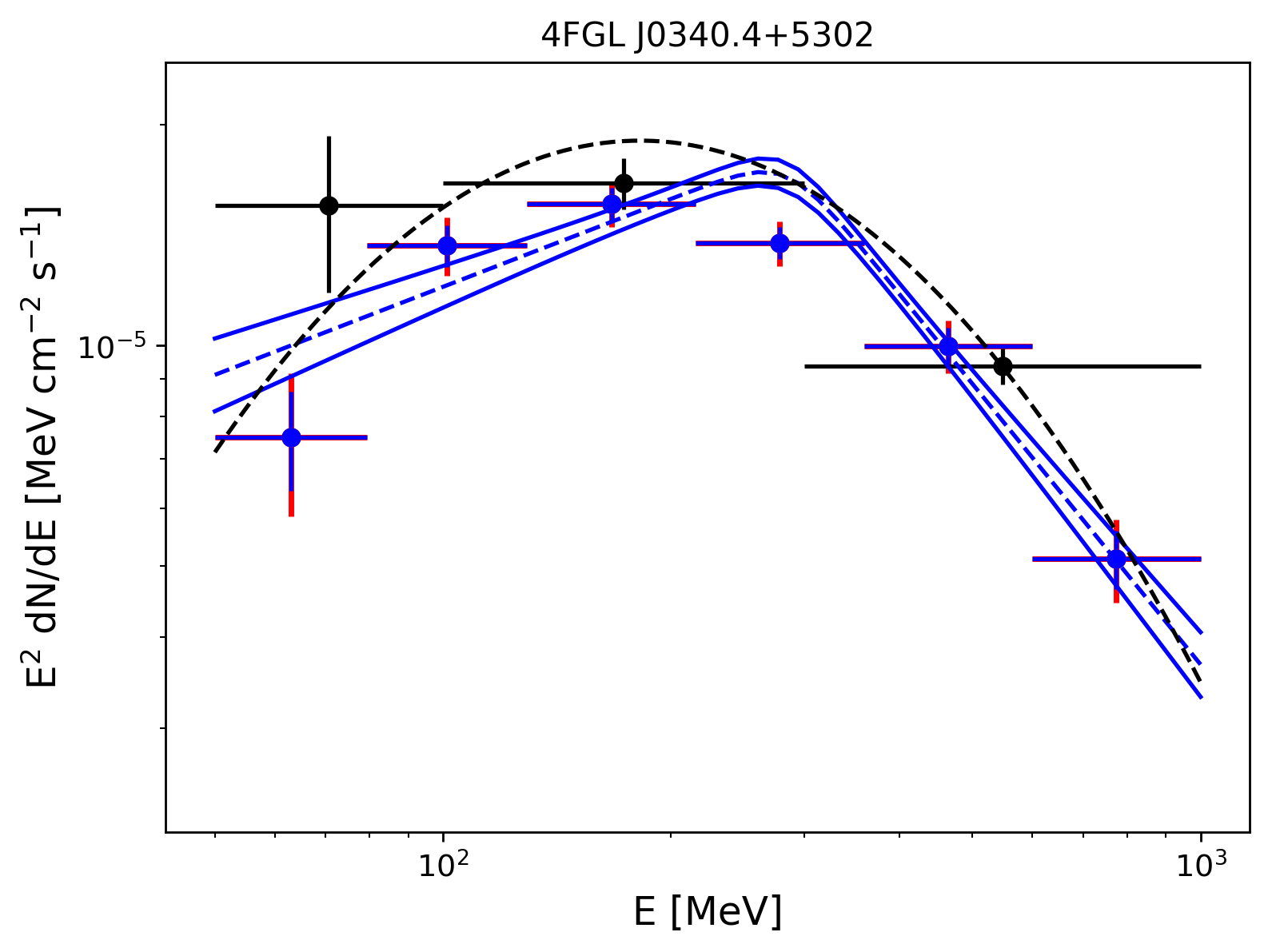}
\end{tabular}
}
\subfigure{
\begin{tabular}{ll}
\centering
\includegraphics[width=0.44\textwidth]{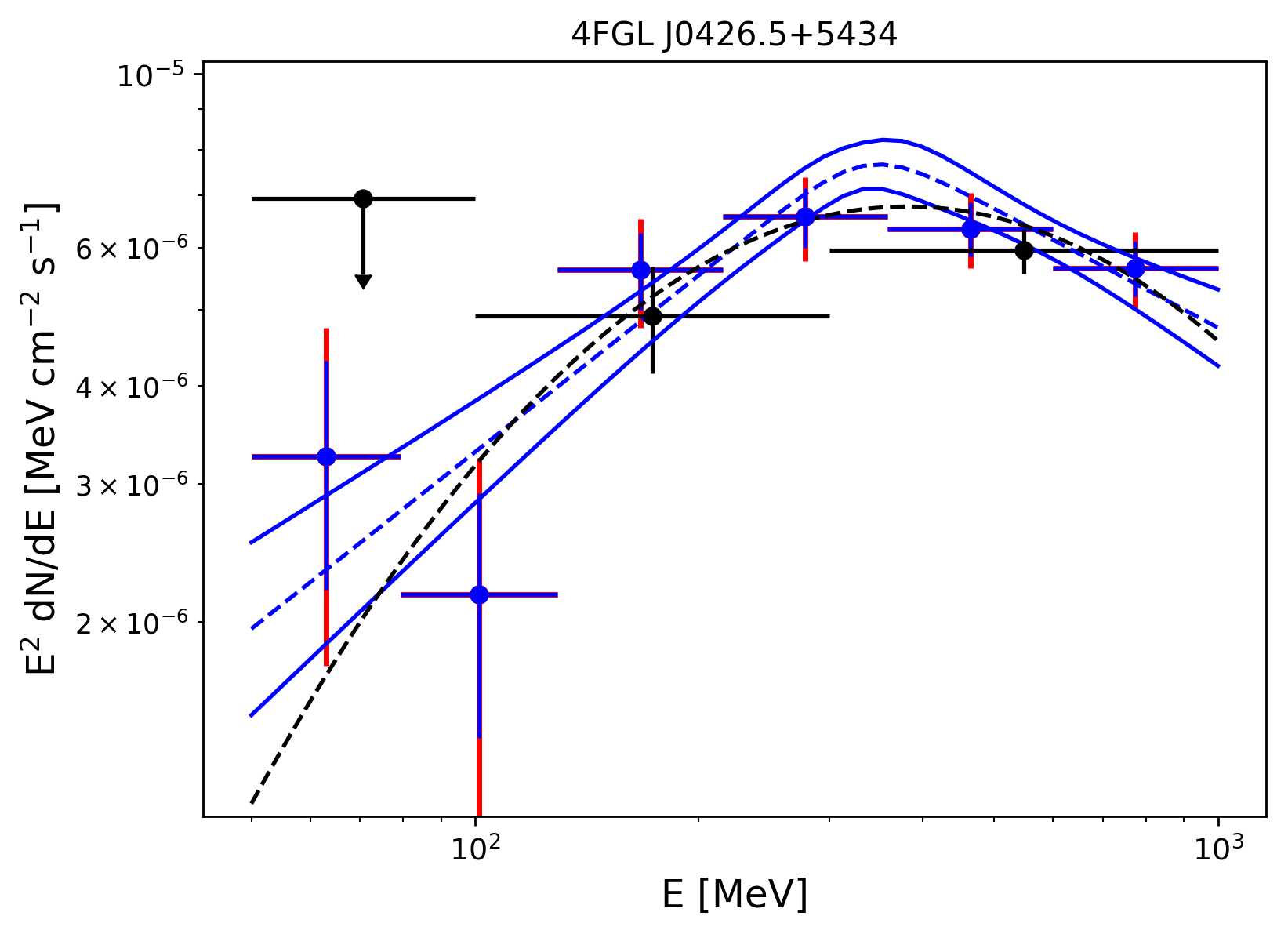}
\includegraphics[width=0.44\textwidth]{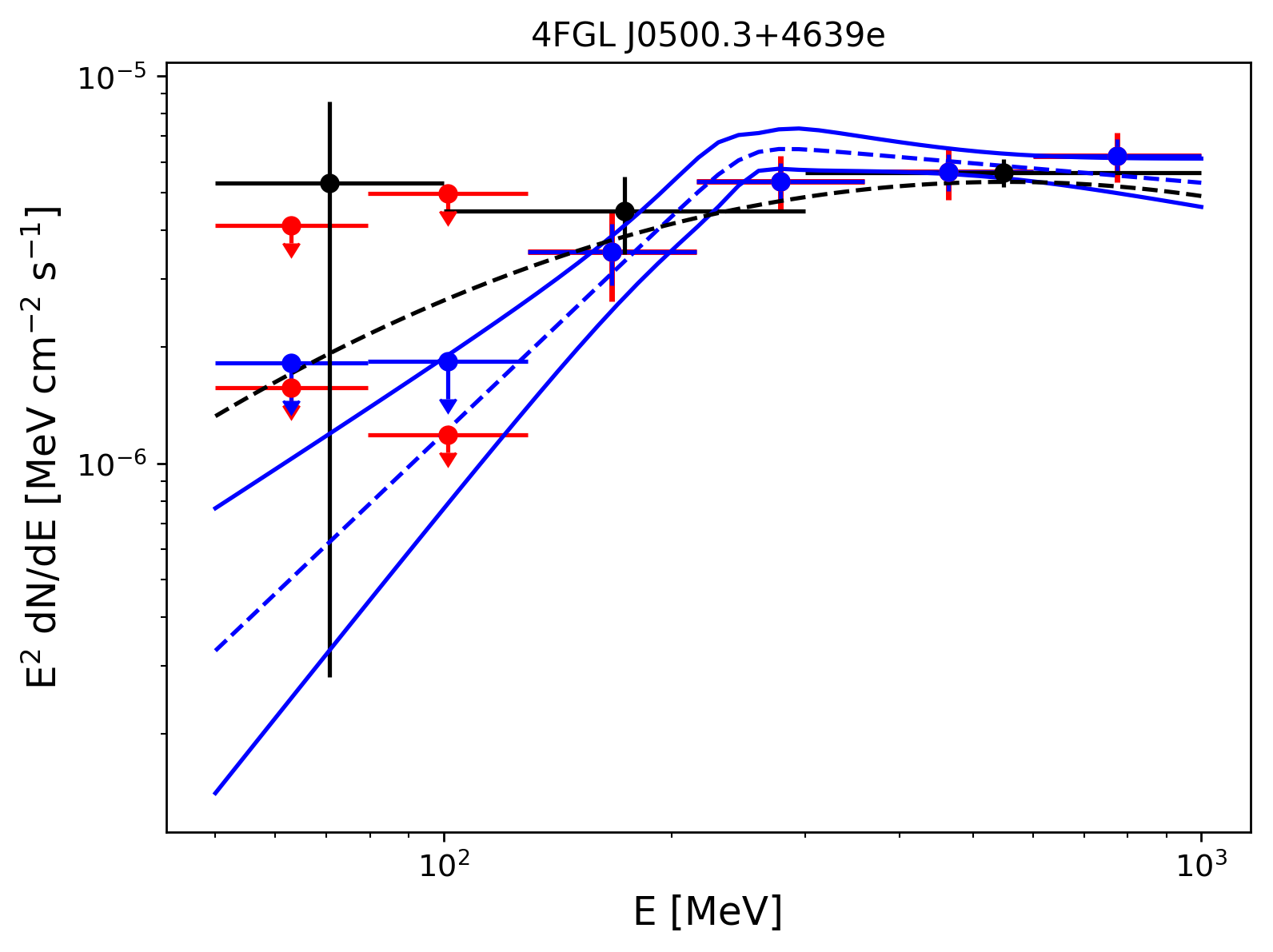}
\end{tabular}
}
\caption{\label{fig:sed1}LAT Spectral energy distributions of 4FGL J0222.4+6156e (top left), 4FGL J0240.5+6113 (top right), 4FGL J0330.7+5845 (middle left), 4FGL J0340.4+5302 (middle right), 4FGL J0426.5+5434 (bottom left), 4FGL J0500.3+4639e (bottom right) with the same conventions used in Figure~\ref{fig:snrsed}.}
\end{figure*}

\begin{figure*}[ht]
\centering
\subfigure{
\begin{tabular}{ll}
\centering
\includegraphics[width=0.44\textwidth]{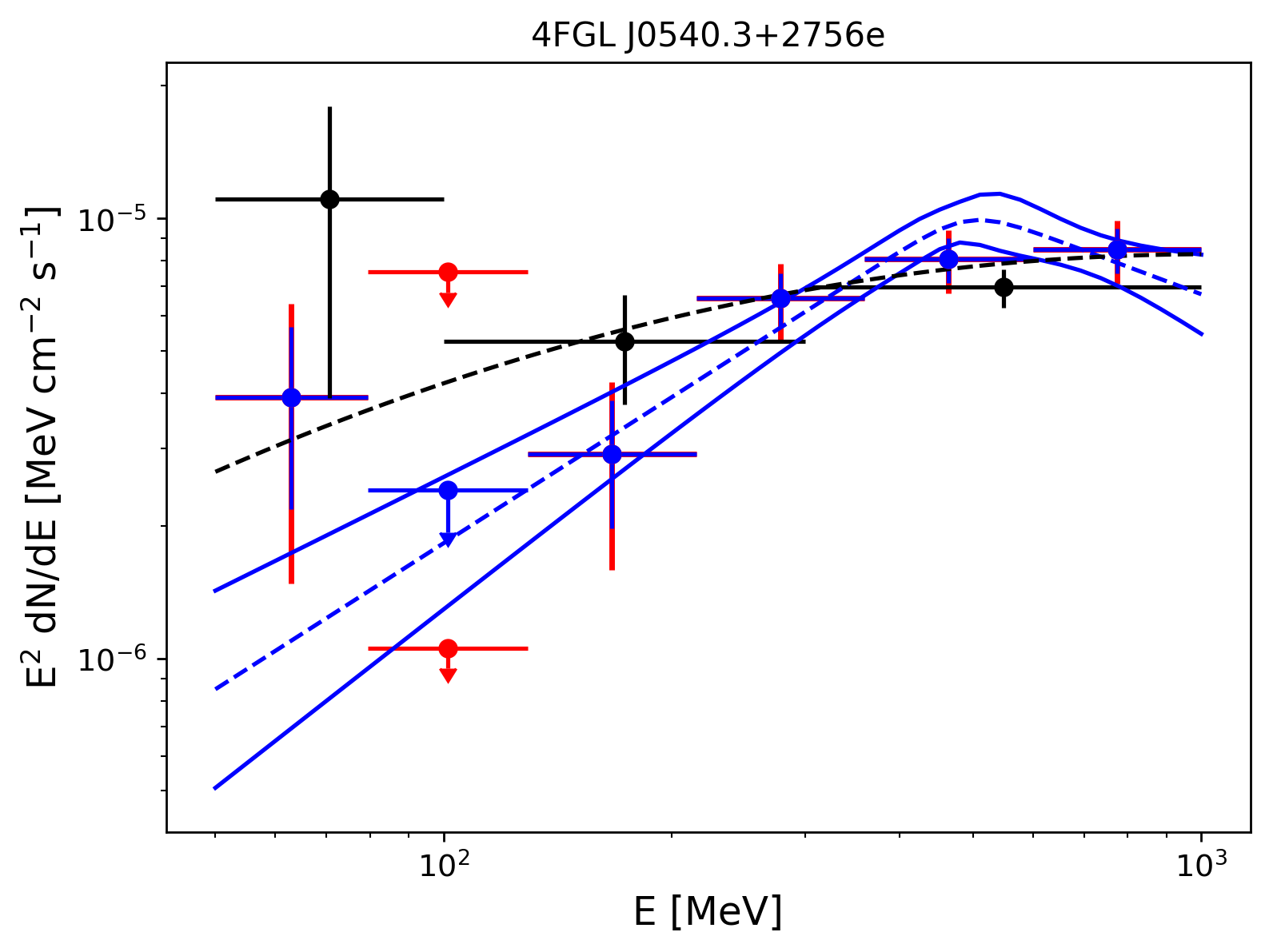}
\includegraphics[width=0.44\textwidth]{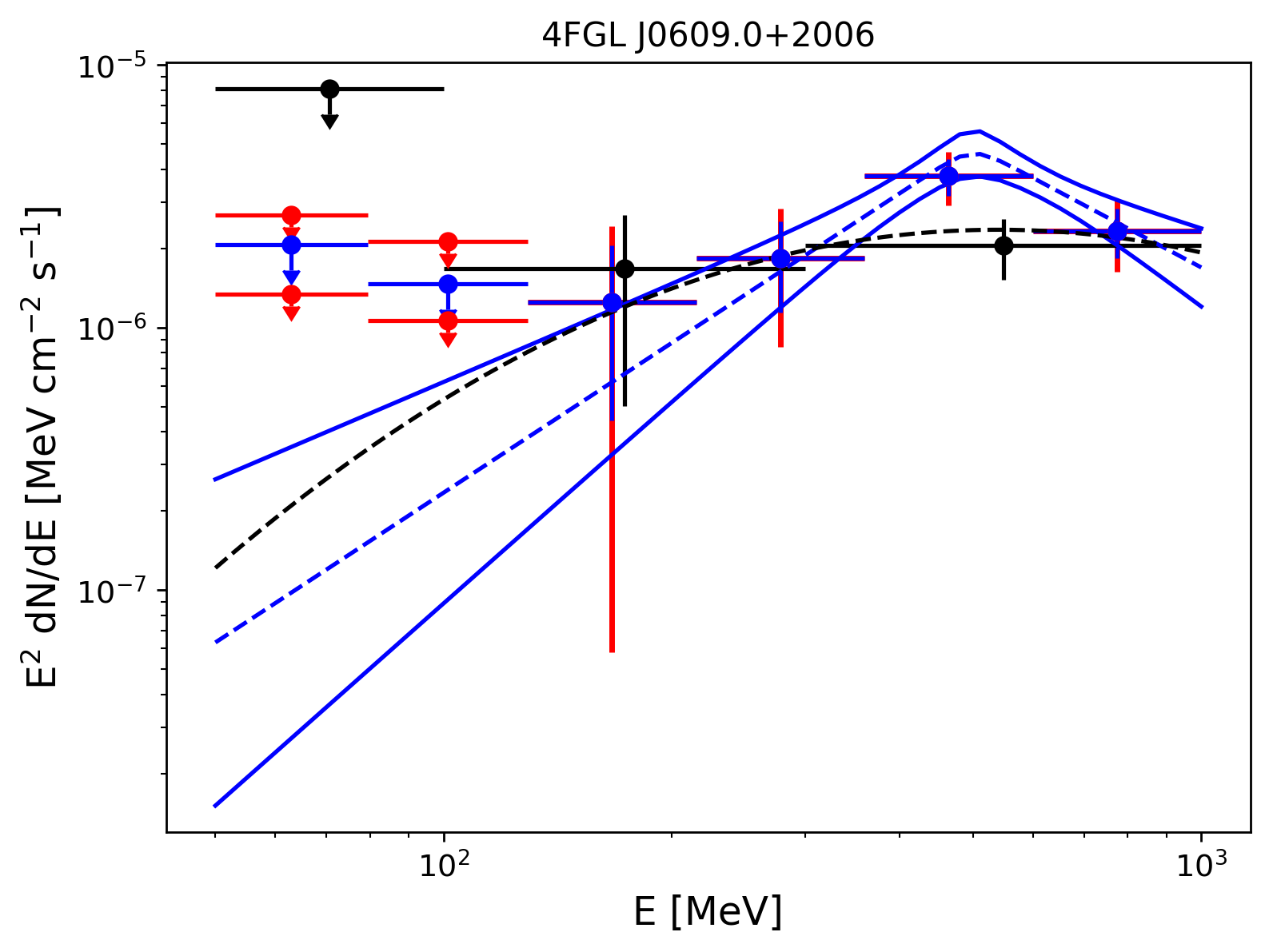}
\end{tabular}
}
\subfigure{
\begin{tabular}{ll}
\centering
\includegraphics[width=0.44\textwidth]{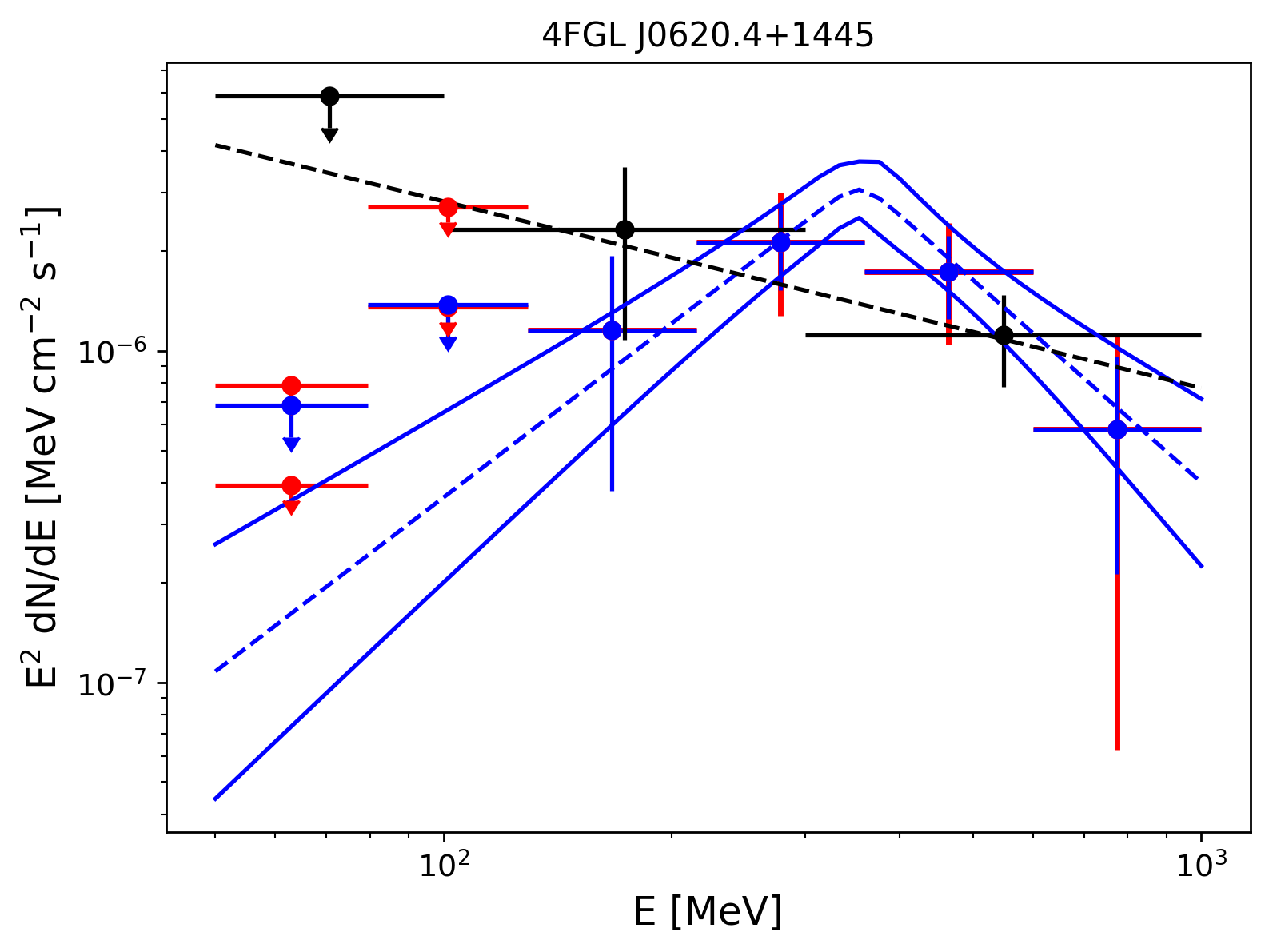}
\includegraphics[width=0.44\textwidth]{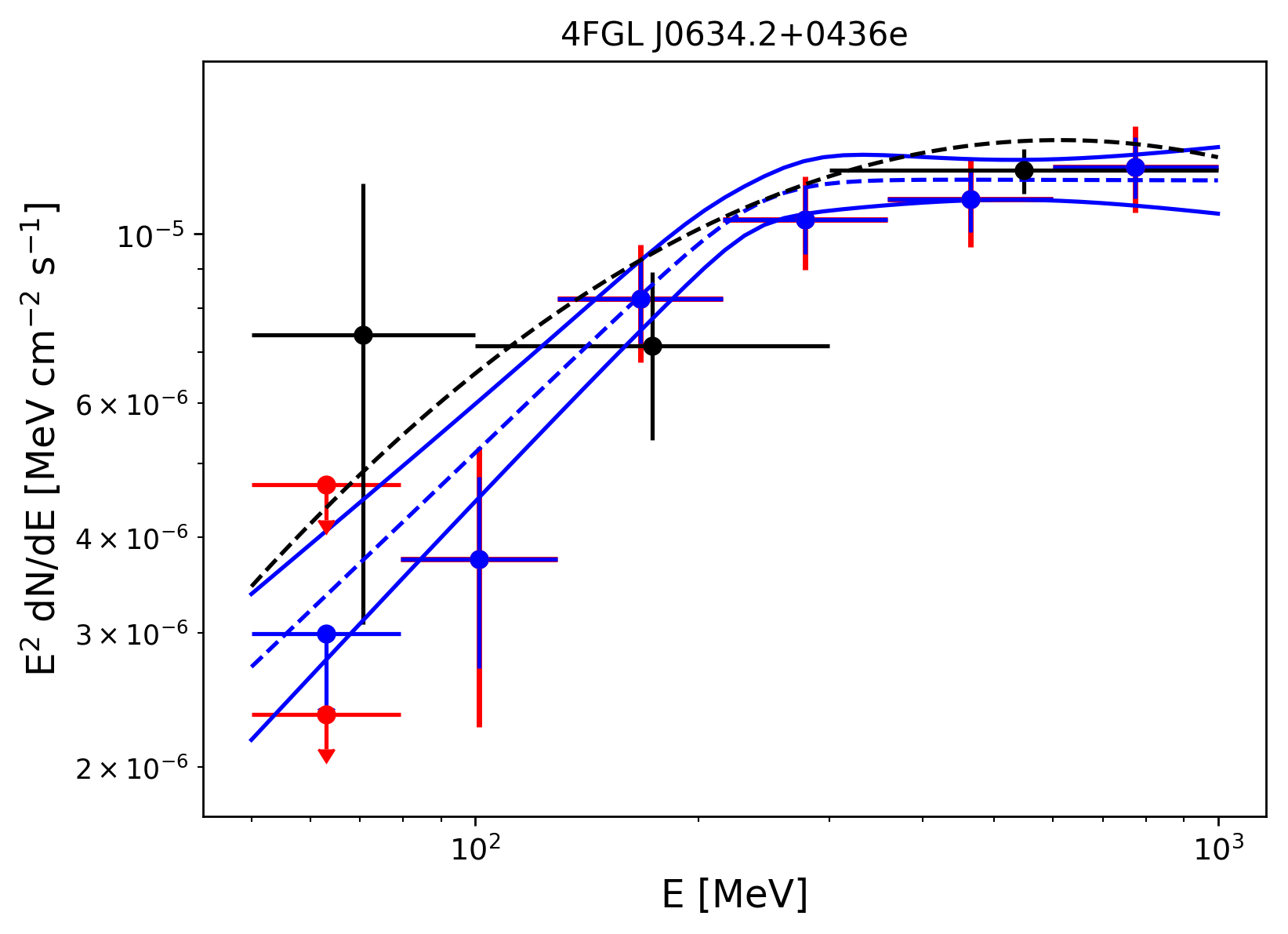}
\end{tabular}
}
\subfigure{
\begin{tabular}{ll}
\centering
\includegraphics[width=0.44\textwidth]{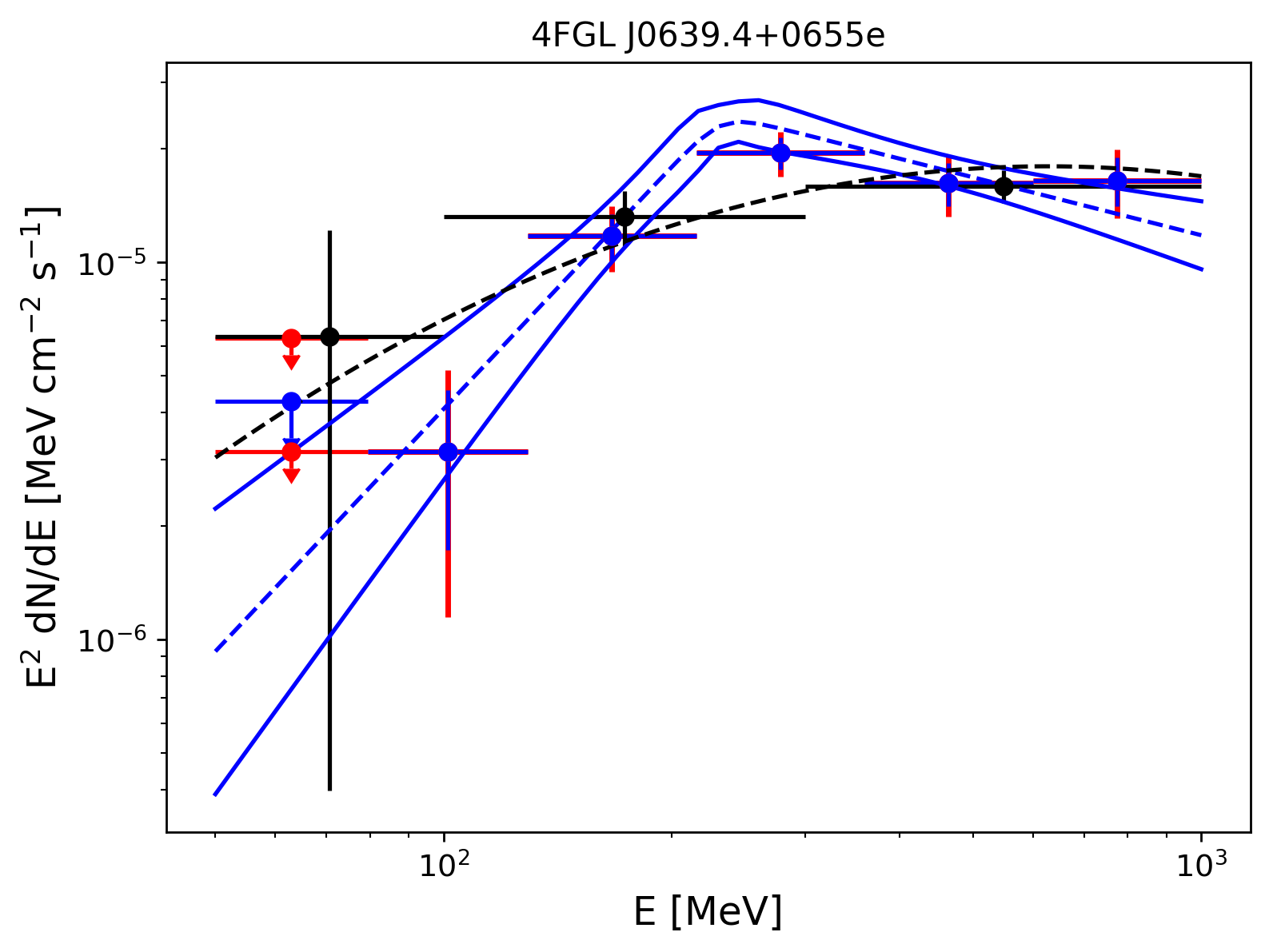}
\includegraphics[width=0.44\textwidth]{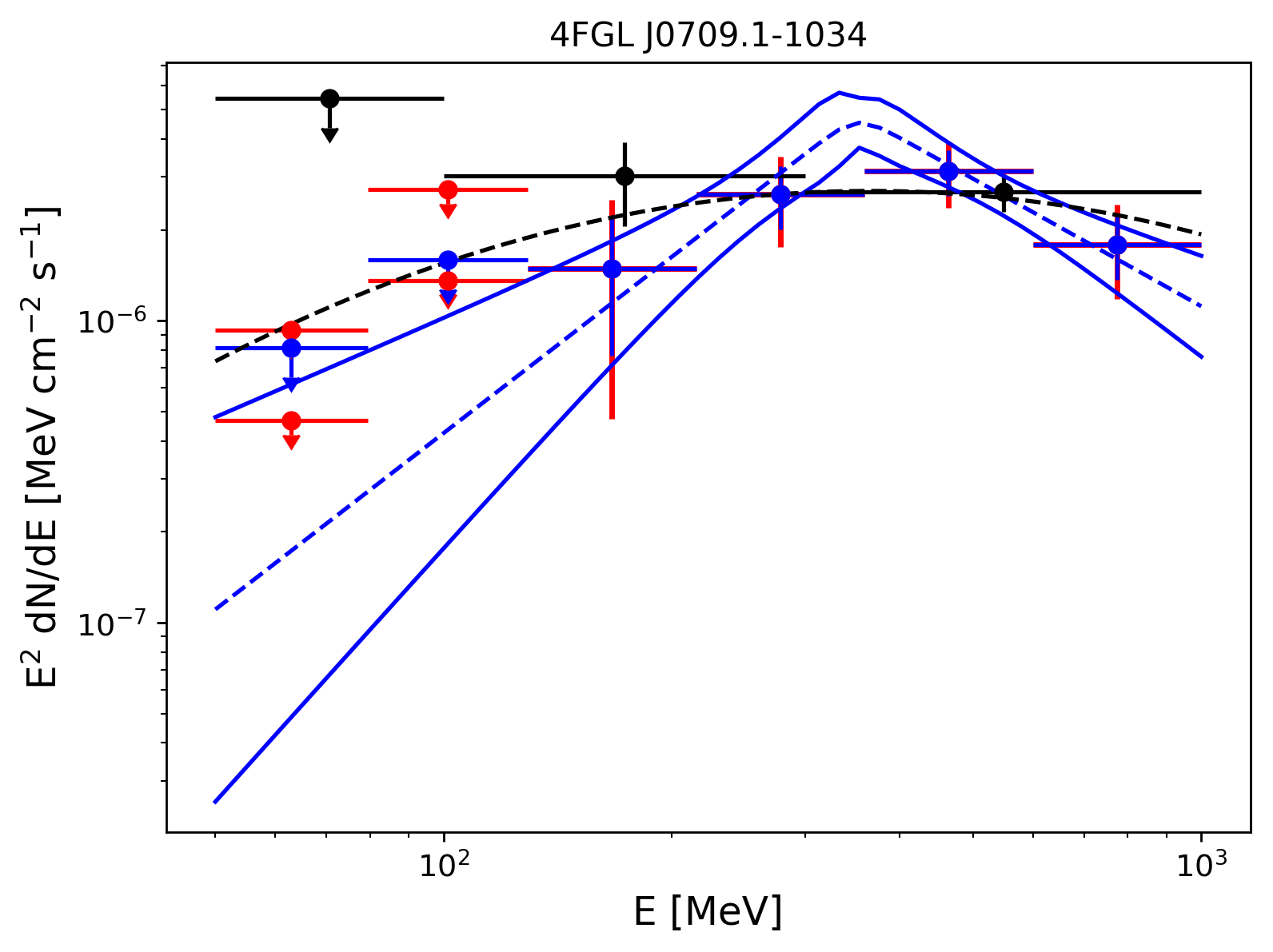}
\end{tabular}
}
\caption{\label{fig:sed2}LAT Spectral energy distributions of 4FGL J0540.3+2756e (top left), 4FGL J0609.0+2006 (top right), 4FGL J0620.4+1445 (middle left), 4FGL J0634.2+0436e (middle right), 4FGL J0639.4+0655e (bottom left), 4FGL J0709.1$-$1034 (bottom right) with the same conventions used in Figure~\ref{fig:snrsed}.}
\end{figure*}

\begin{figure*}[ht]
\centering
\subfigure{
\begin{tabular}{ll}
\centering
\includegraphics[width=0.44\textwidth]{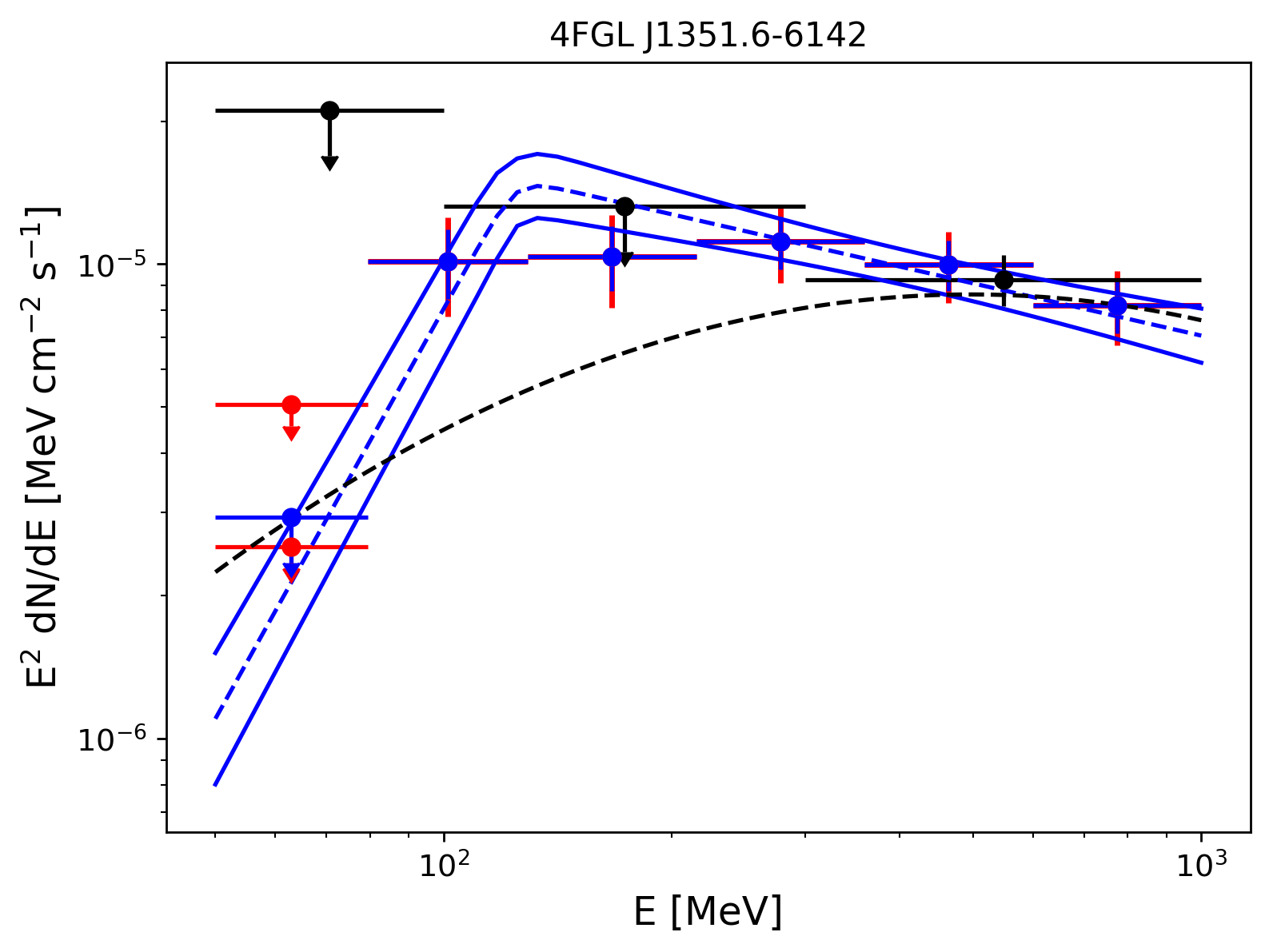}
\includegraphics[width=0.44\textwidth]{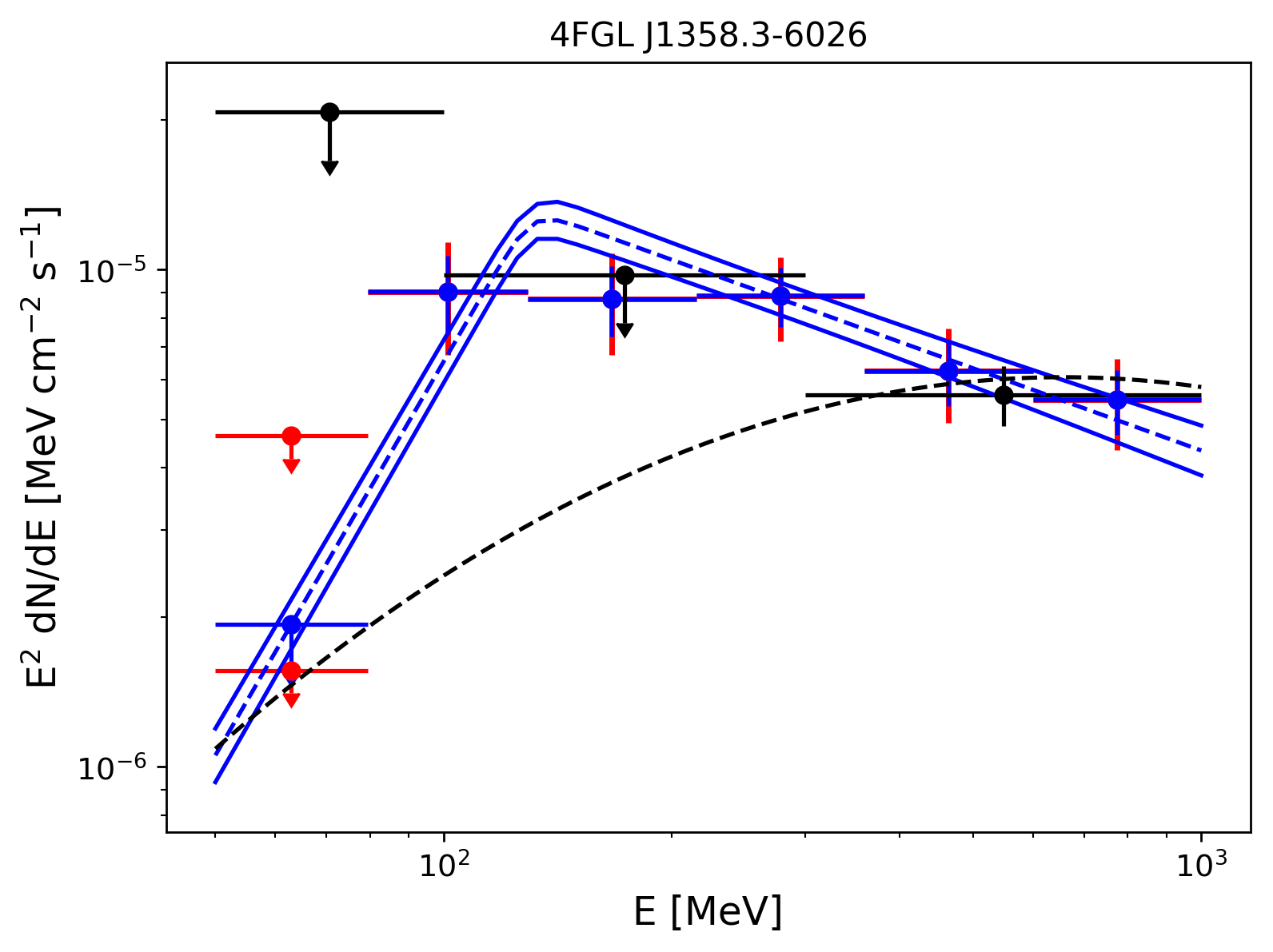}
\end{tabular}
}
\subfigure{
\begin{tabular}{ll}
\centering
\includegraphics[width=0.44\textwidth]{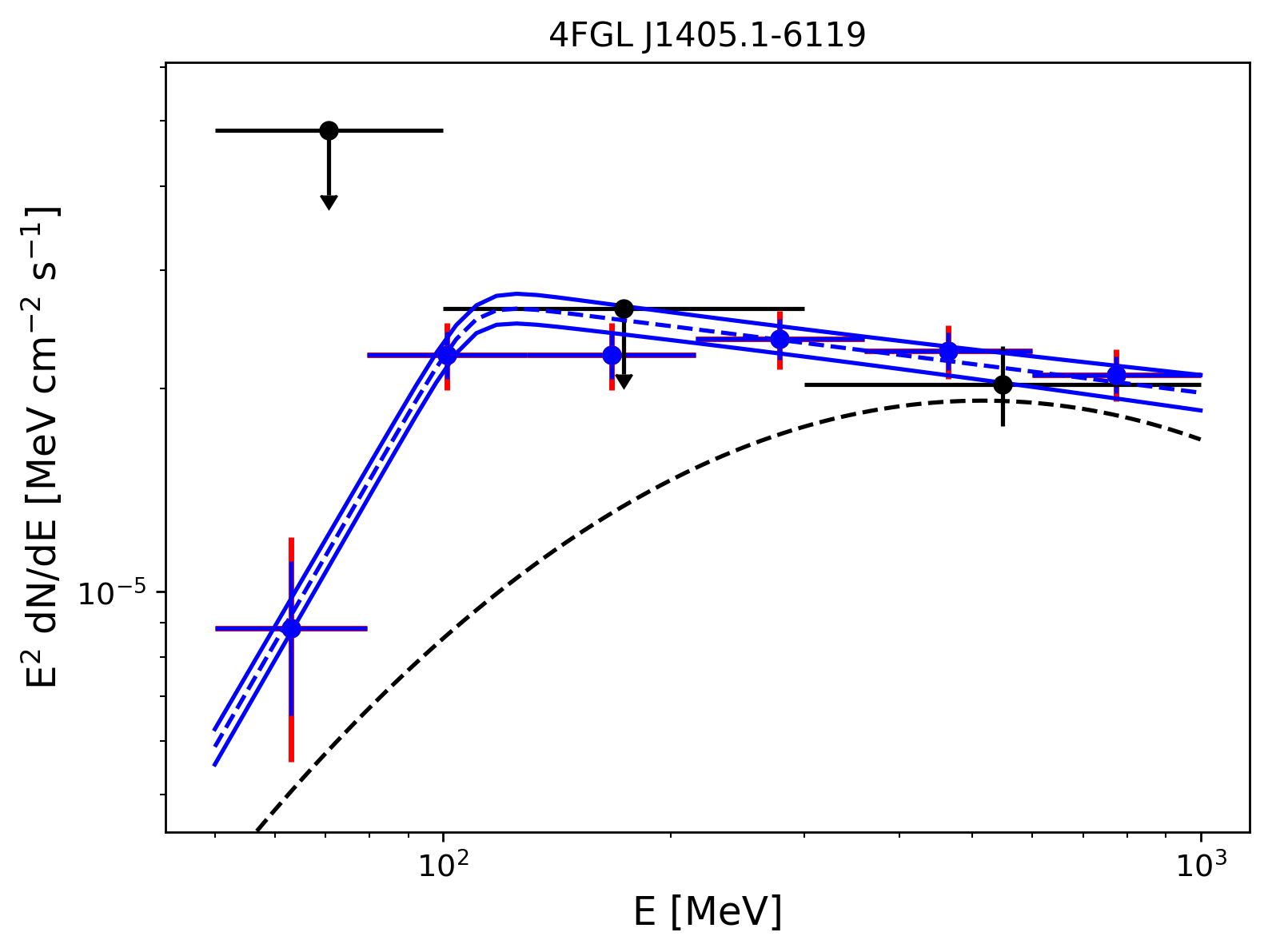}
\includegraphics[width=0.44\textwidth]{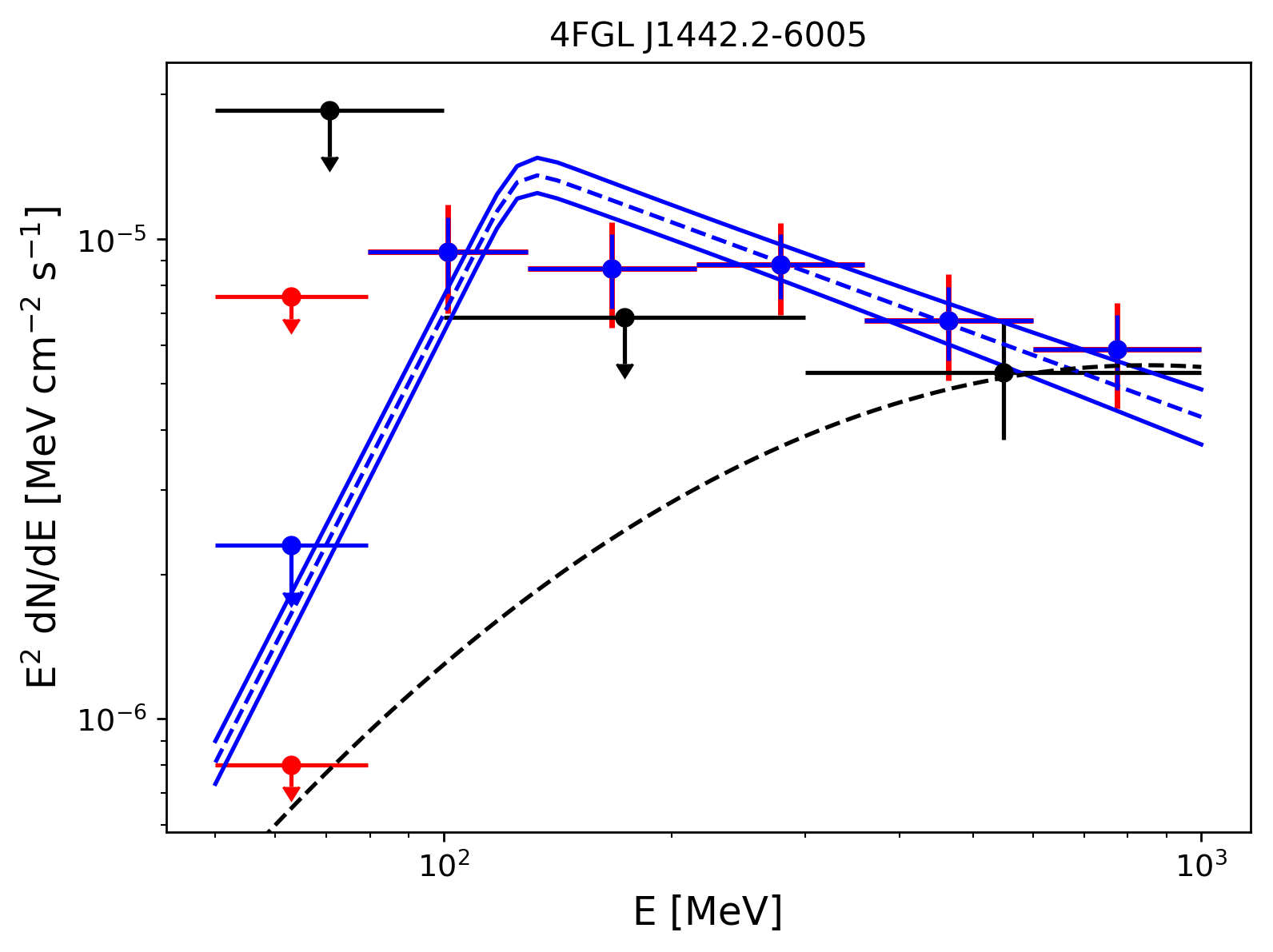}
\end{tabular}
}
\subfigure{
\begin{tabular}{ll}
\centering
\includegraphics[width=0.44\textwidth]{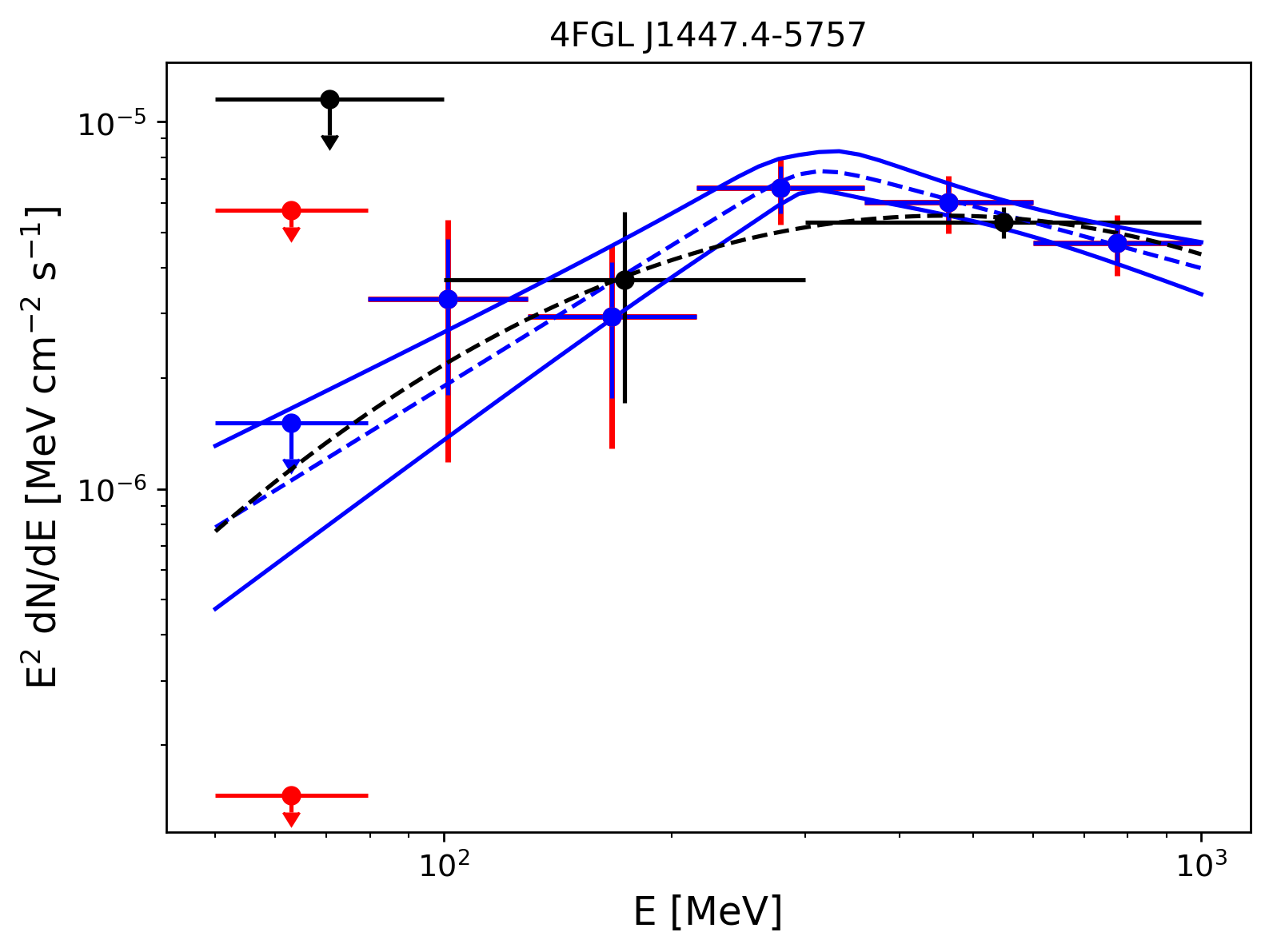}
\includegraphics[width=0.44\textwidth]{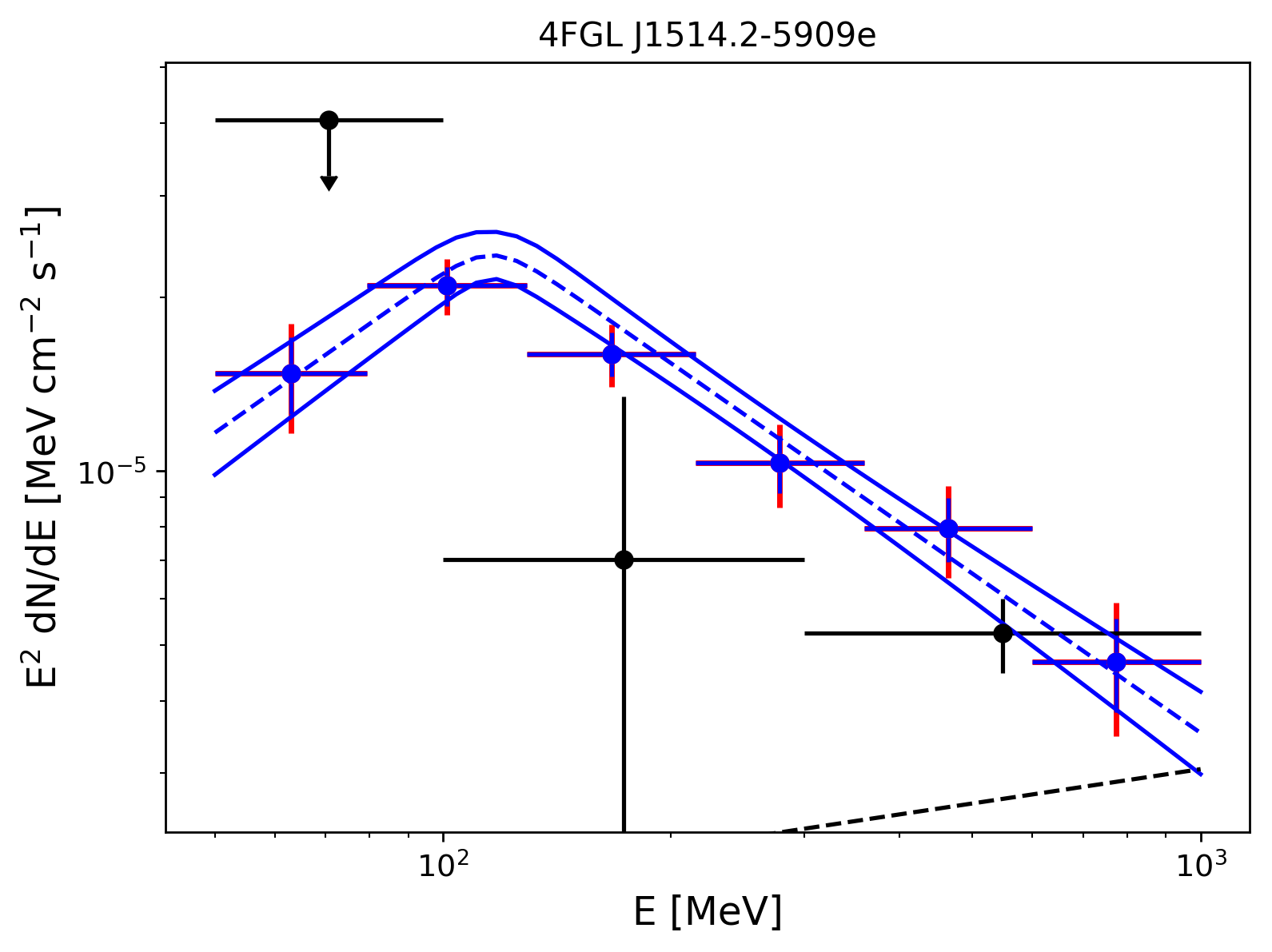}
\end{tabular}
}
\caption{\label{fig:sed4}LAT Spectral energy distributions of 4FGL J1351.6$-$6142 (top left), 4FGL J1358.3$-$6026 (top right), 4FGL J1405.1$-$6119 (middle left), 4FGL J1442.2$-$6005 (middle right), 4FGL J1447.4$-$5757 (bottom left), 4FGL J1514.2$-$5909e (bottom right) with the same conventions used in Figure~\ref{fig:snrsed}.}
\end{figure*}

The most famous sources with ``pion bump" signature are the middle-aged remnants IC 443 and W44. Figure~\ref{fig:snrsed} presents the residual TS maps of the region of IC 443 (4FGL J0617.2+2234e) and W44 (4FGL J1855.9+0121e) as well as their spectral energy distributions, showing the overall agreement with the 4FGL SED points superimposed. This Figure also illustrates the advantages of using a restricted energy range and different spectral shape than the 4FGL to better reproduce the significant energy break at low energy since we are not dominated here by photons at high energies. The spectral parameters reported in Table~\ref{tab:spectra} for these two sources are in reasonable agreement with those published by \cite{2013Sci...339..807A}, knowing that this first analysis did not take into account the effect of energy dispersion and no systematic uncertainties were evaluated at that time.\\ 
Among the 56 sources with significant breaks, one can see from the 4FGL Classification column listed in Table~\ref{tab:candidates} that ten sources are firm SNR identifications and three are associated with SNRs. Among the three SNR associations, 4FGL J1911.0+0905 (Figure~\ref{fig:sed8} top right) is associated to W49B and thus can be safely identified as a SNR since it is one of the few other sources for which a ``pion decay bump" signature was published with W51C (4FGL J1923.2+1408e, Figure~\ref{fig:sed8} middle left) and HB 21 (4FGL J2045.2+5026e, Figure~\ref{fig:sed9} middle right). The only missing source for which a low-energy break has been published is Cassiopeia A (4FGL~J2323.4+5849) but the break energy reported by \cite{2013ApJ...779..117Y} is at $1.72^{+1.35}_{-0.89}$~GeV which seems consistent with our non-detection in the 50 MeV -- 1 GeV energy interval. The five sources confirmed by our analysis are all supernova remnants interacting with molecular clouds (MCs). These molecular clouds are excellent targets for cosmic-ray interactions and subsequent pion-decay.\\ The hadronic scenario was also preferred for other LAT-detected SNRs interacting with MCs, though their $\gamma$-ray analysis starting above a few hundred MeV did not allow rejection of a leptonic scenario: the SNR HB3 and the W3 HII complex~\citep{2016ApJ...818..114K}, S147 \citep{2012ApJ...752..135K}, HB9~\citep{2014MNRAS.444..860A}, the SNR G326.3$-$1.8~\citep{2018A&A...617A...5D} and the SNR W28~\citep{2014ApJ...786..145H}. Our low-energy analysis presents a rapid turn-over of the spectrum at low energy which confirms the conclusions of the previous publications for 4FGL~J0222.4+6156e (W3, see Figure~\ref{fig:sed1} top left), 4FGL~J0500.3+4639e (HB9, see Figure~\ref{fig:sed1} bottom right), 4FGL~J0540.3+2756e (S147, Figure~\ref{fig:sed2} top left) 4FGL~J1552.9$-$5607e (G326.3$-$1.8, Figure~\ref{fig:sed5} middle left) and 4FGL~J1801.3-2326e (W28, see Figure~\ref{fig:sed6} middle left). No significant curvature is detected for the SNR HB3 but it should be noted that its $\gamma$-ray emission is much fainter than the adjacent molecular cloud W3 (TS value of 75.9 with respect to 1307.1 for W3) and more data would be needed to constrain the low-energy spectrum of the SNR. A hadronic scenario was also invoked for the SNR Monoceros Loop \citep{2016ApJ...831..106K}.
%but in this case the particles accelerated by the SNR interact with the interstellar gas to produce the $\gamma$-ray emission.
In this case, the brightest gamma-ray peak is spatially correlated with the Rosette Nebula, a young stellar cluster and molecular cloud complex located at the edge of the southern shell of the SNR which has a role similar to W3 for the HB3/W3 complex. The interaction between the SNR and the molecular cloud provides the target to naturally produce gamma rays via proton-proton interaction and it is not a surprise that we confirm a spectral break at low energy for the Monoceros SNR (4FGL~J0639.4+0655e, see Figure~\ref{fig:sed2} bottom left) and for the Rosette complex (4FGL~J0634.2+0436e, see Figure~\ref{fig:sed2} middle right). More recently, modeling of the non-thermal emission of the gamma Cygni SNR \citep{2016ApJ...826...31F, 2019ICRC...36..675F}, associated with the source 4FGL J2021.0+4031e, also suggested that the $\gamma$-ray emission (analyzed above 100 MeV) might be of hadronic nature with enhanced GeV emission spatially coincident with the TeV source VER J2019+407. Here again, our low energy analysis detects a low-energy break in the spectrum of this SNR (see Figure~\ref{fig:sed8} bottom right) but it should be noted that the bright $\gamma$-ray emission from the pulsar PSR J2021+4026, lying near the center of the remnant, is very difficult to disentangle from the signal of the SNR at these low energies which could lead to some contamination in the SNR spectrum. A follow-up study in the off-pulse of the pulsar would therefore be needed to confirm the results obtained with our pipeline. This applies not only to supernova remnants but also to all sources coincident with (or very close to) a bright gamma-ray pulsar. It is even more clear for 4FGL~J1514.2$-$5909e associated with the pulsar wind nebula MSH 15$-$52 and coincident with the soft gamma-ray pulsar PSR B1509$-$58. The very high low-energy flux visible in Figure~\ref{fig:sed4} (bottom right) is most likely to the associated pulsar PSR B1509$-$58 which is not included in the 4FGL Catalog and would be hard to disentangle from the PWN at these energies.

\subsection{Constraints on other identified sources}\label{subsection:id}
\begin{figure*}[ht]
\centering
\subfigure{
\begin{tabular}{ll}
\centering
\includegraphics[width=0.44\textwidth]{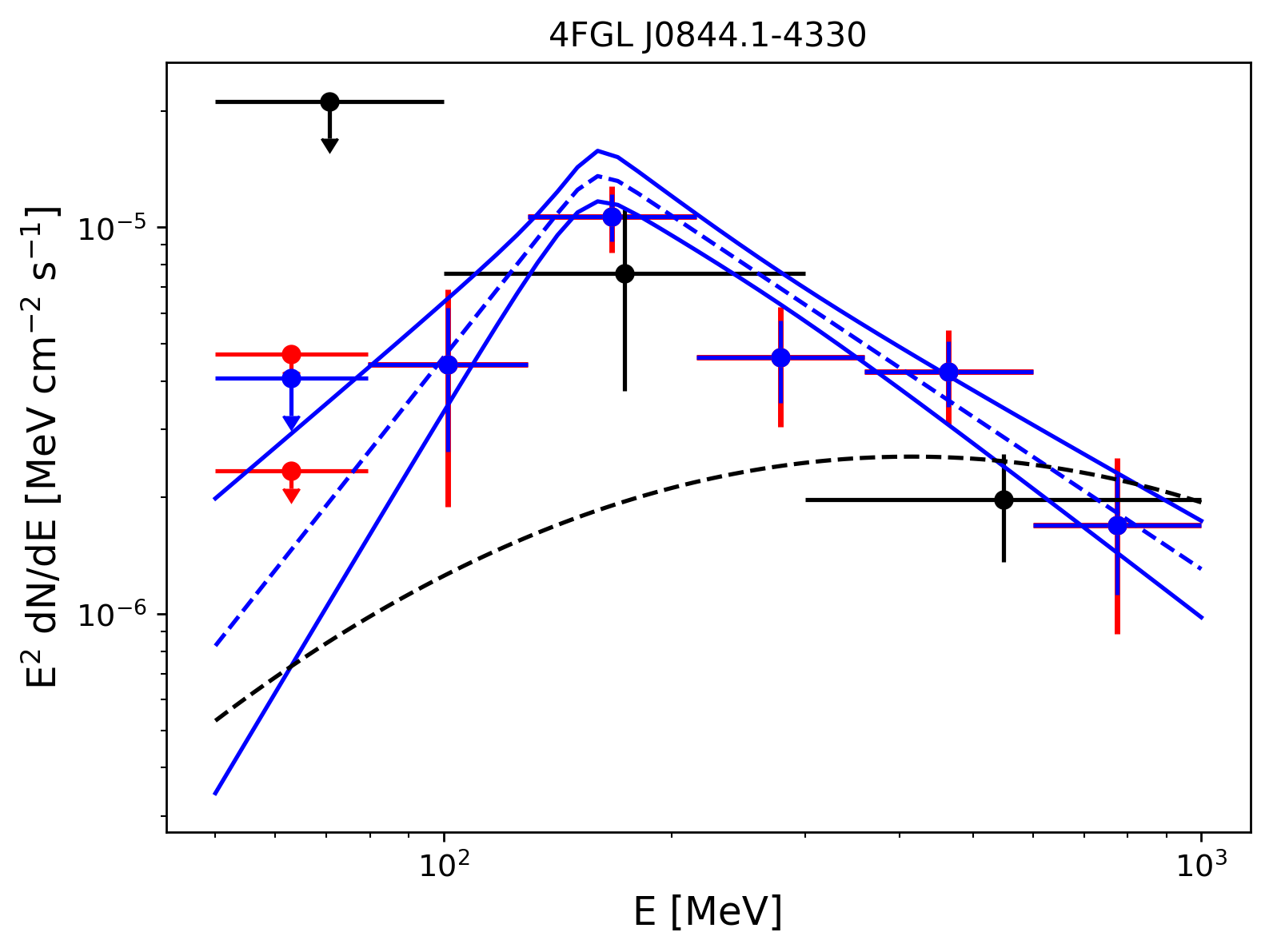}
\includegraphics[width=0.44\textwidth]{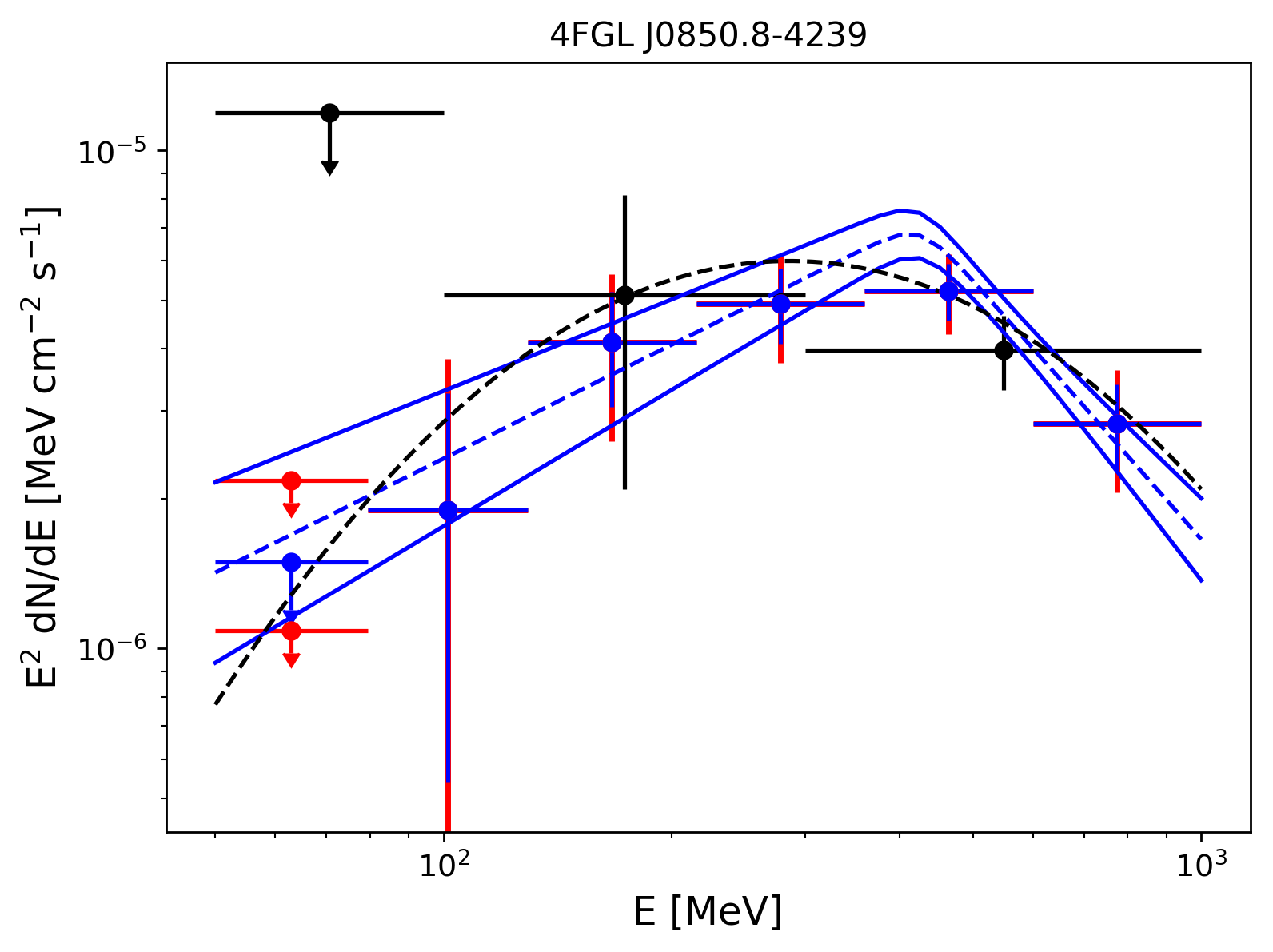}
\end{tabular}
}
\subfigure{
\begin{tabular}{ll}
\centering
\includegraphics[width=0.44\textwidth]{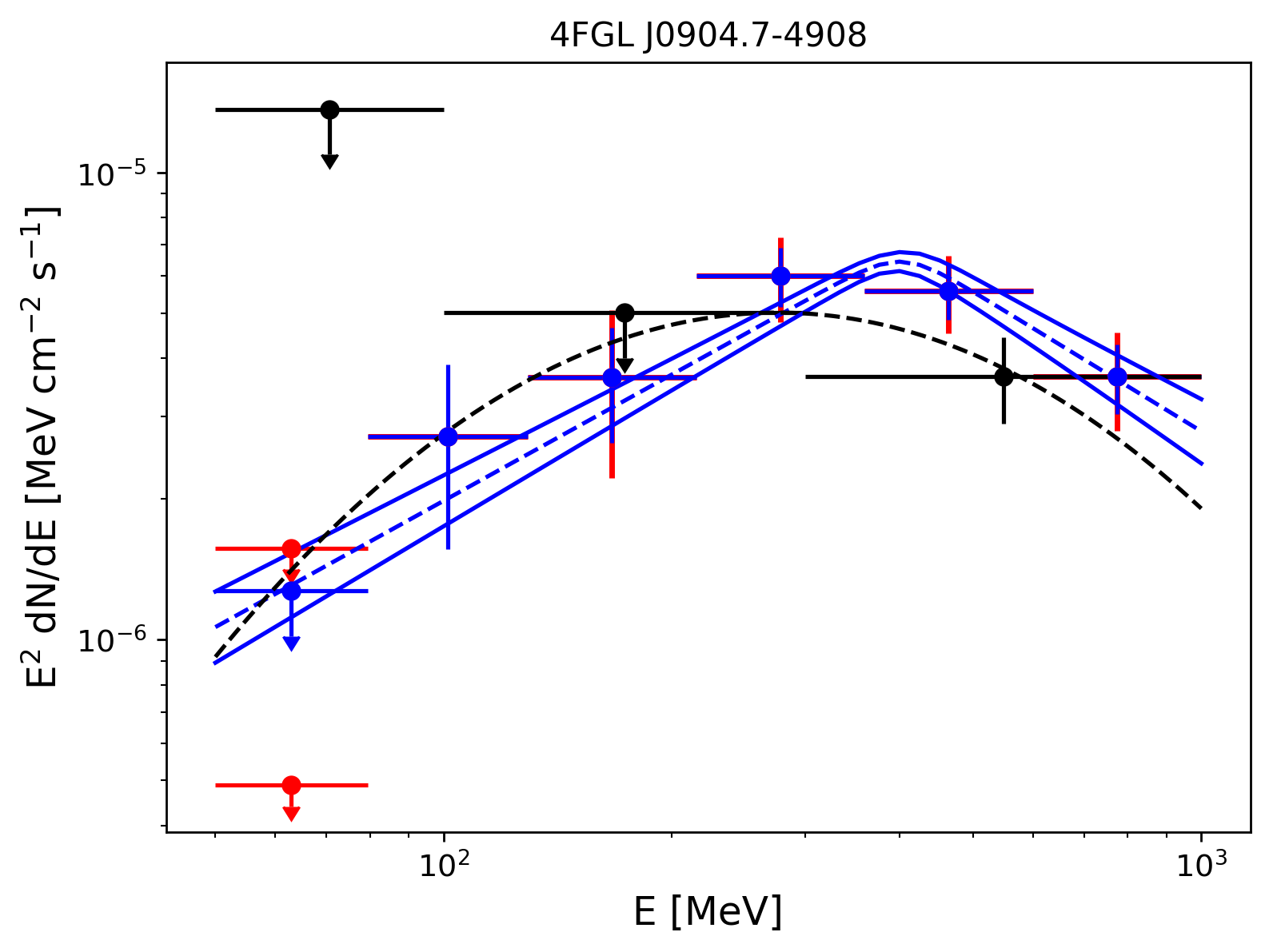}
\includegraphics[width=0.44\textwidth]{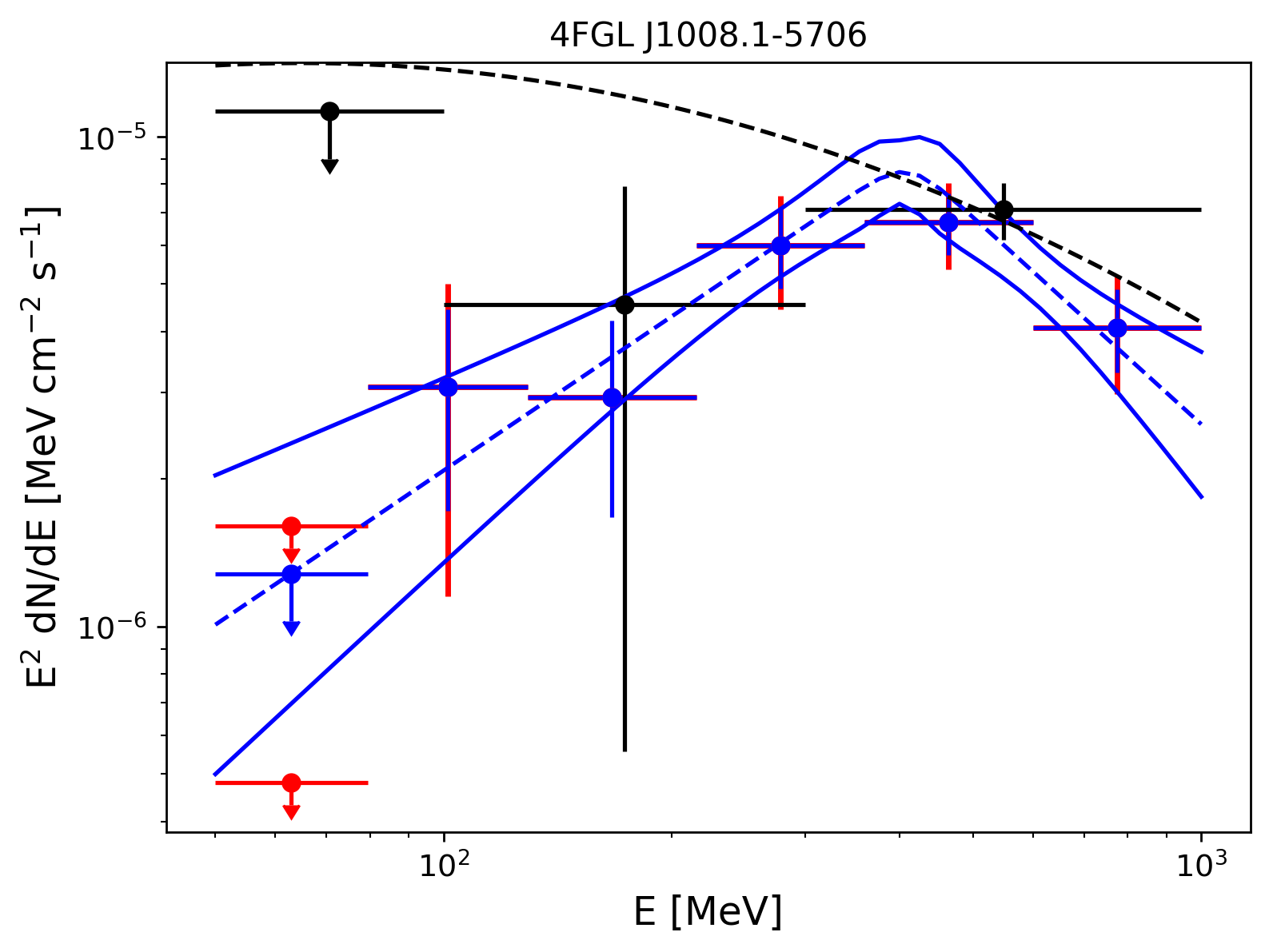}
\end{tabular}
}
\subfigure{
\begin{tabular}{ll}
\centering
\includegraphics[width=0.44\textwidth]{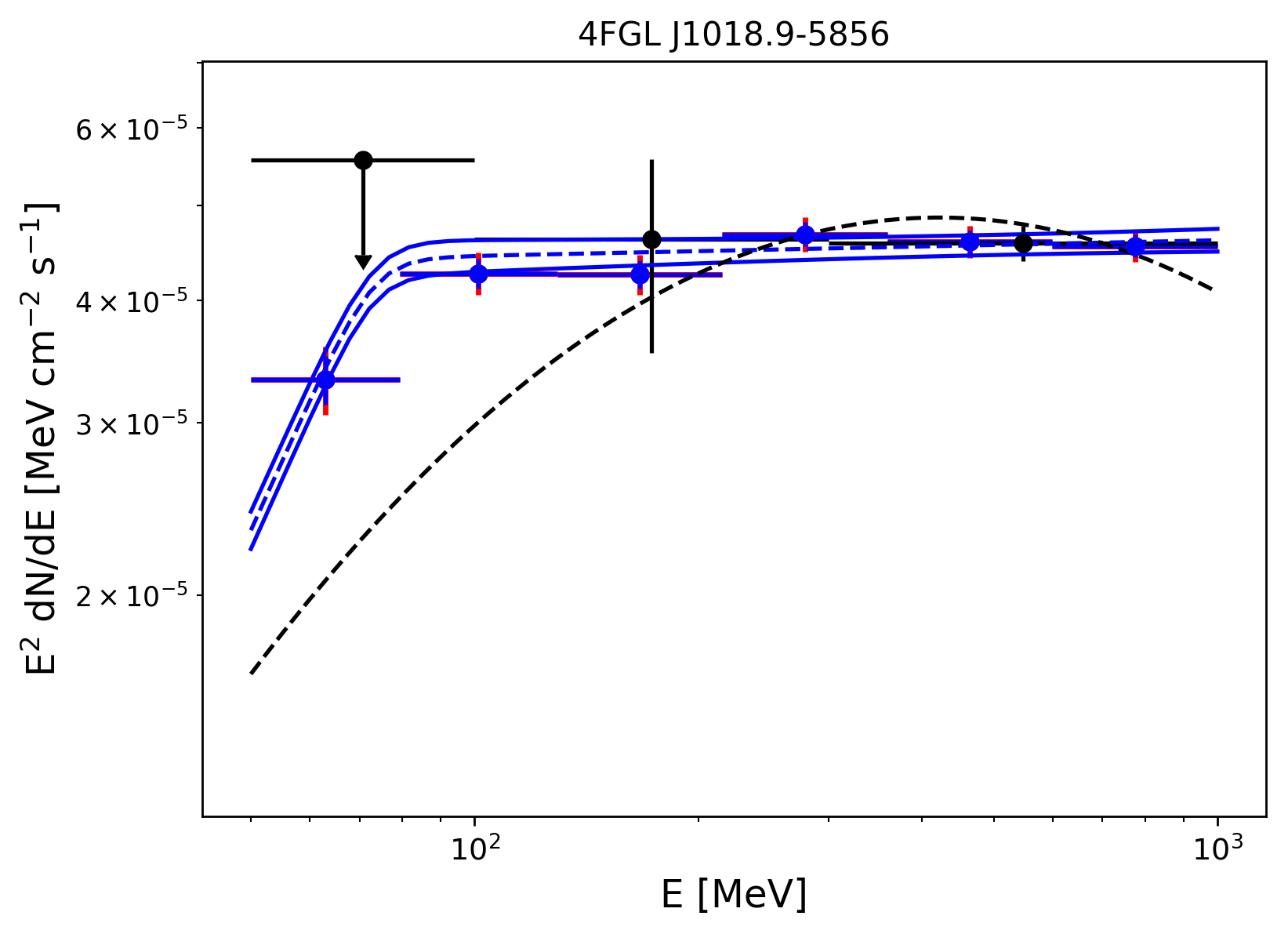}
\includegraphics[width=0.44\textwidth]{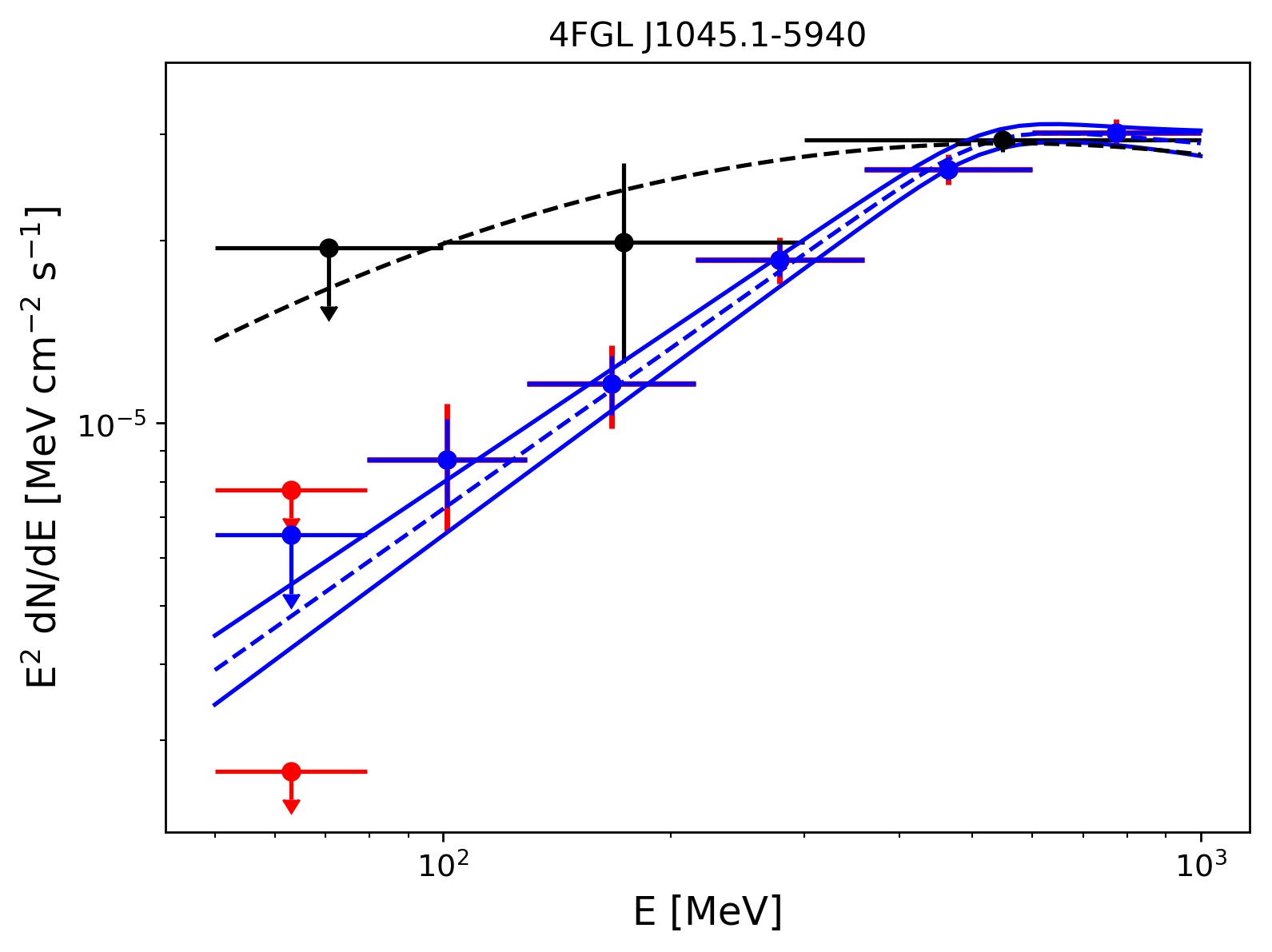}
\end{tabular}
}
\caption{\label{fig:sed3}LAT Spectral energy distributions of 4FGL J0844.1$-$4330 (top left), 4FGL J0850.8$-$4239 (top right), 4FGL J0904.7$-$4908 (middle left), 4FGL J1008.1$-$5706 (middle right), 4FGL J1018.9$-$5856 (bottom left), 4FGL J1045.1$-$5940 (bottom right) with the same conventions used in Figure~\ref{fig:snrsed}.}
\end{figure*}

\begin{figure*}[ht]
\centering
\subfigure{
\begin{tabular}{ll}
\centering
\includegraphics[width=0.44\textwidth]{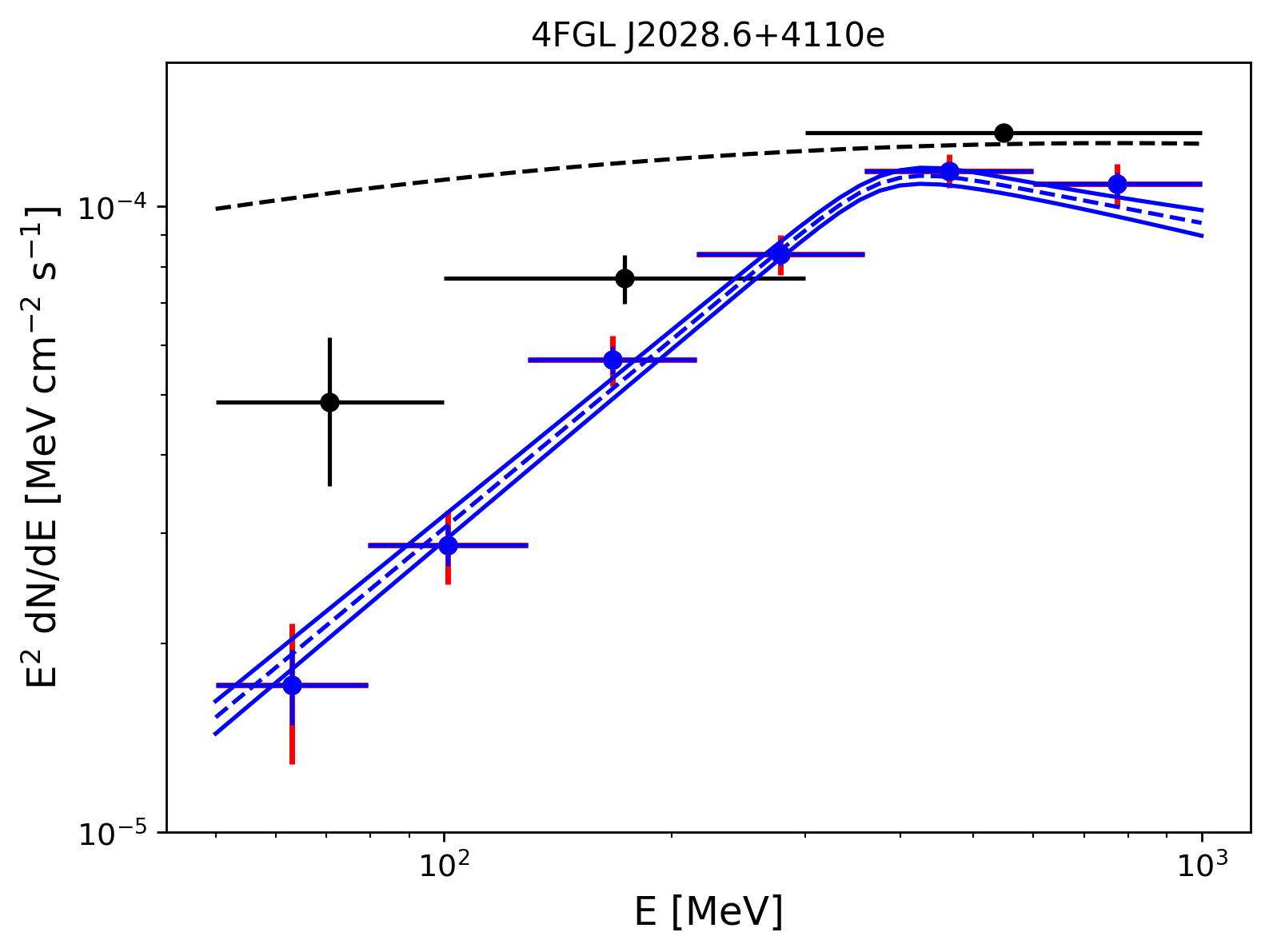}
\includegraphics[width=0.44\textwidth]{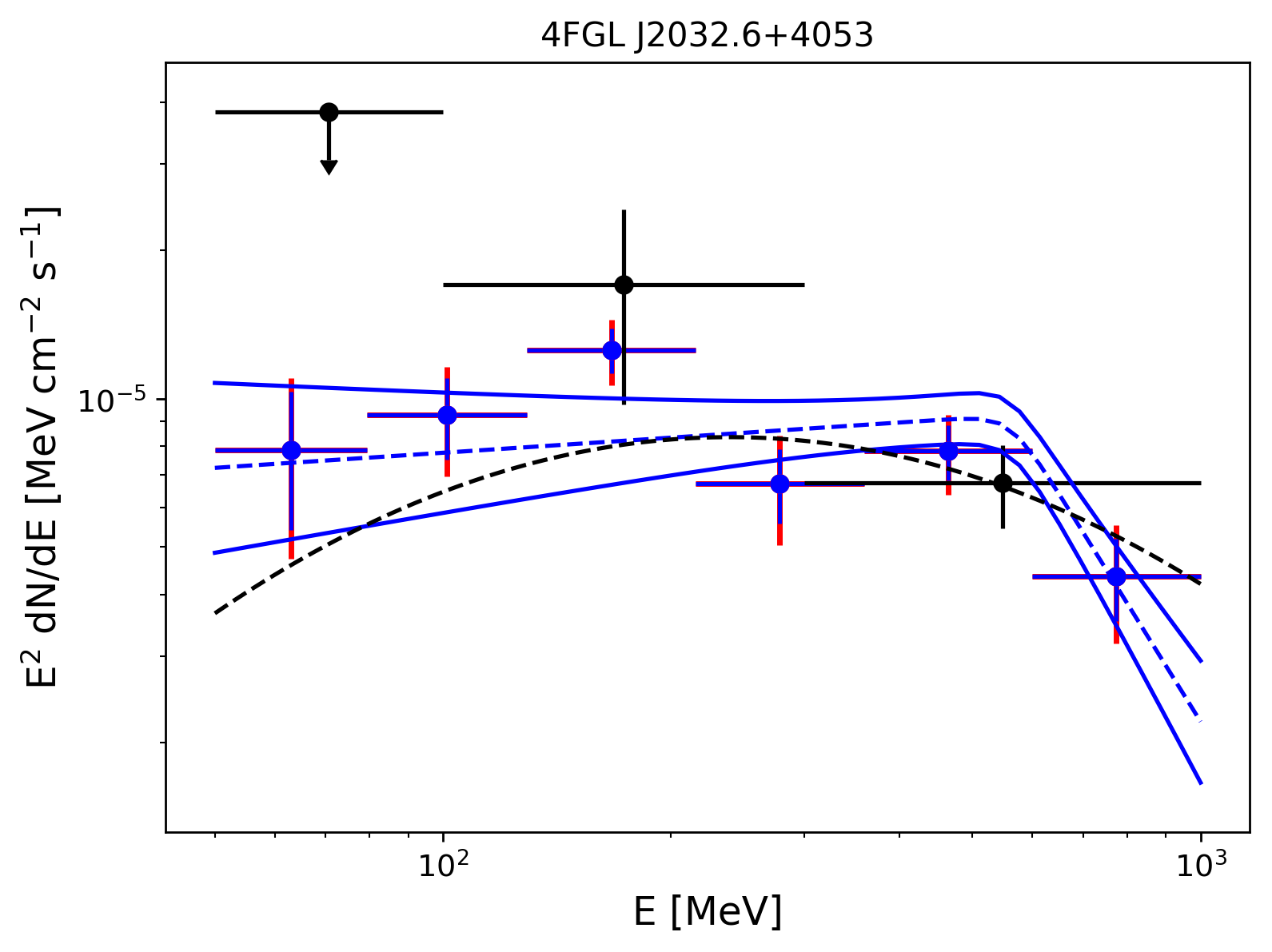}
\end{tabular}
}
\subfigure{
\begin{tabular}{ll}
\centering
\includegraphics[width=0.44\textwidth]{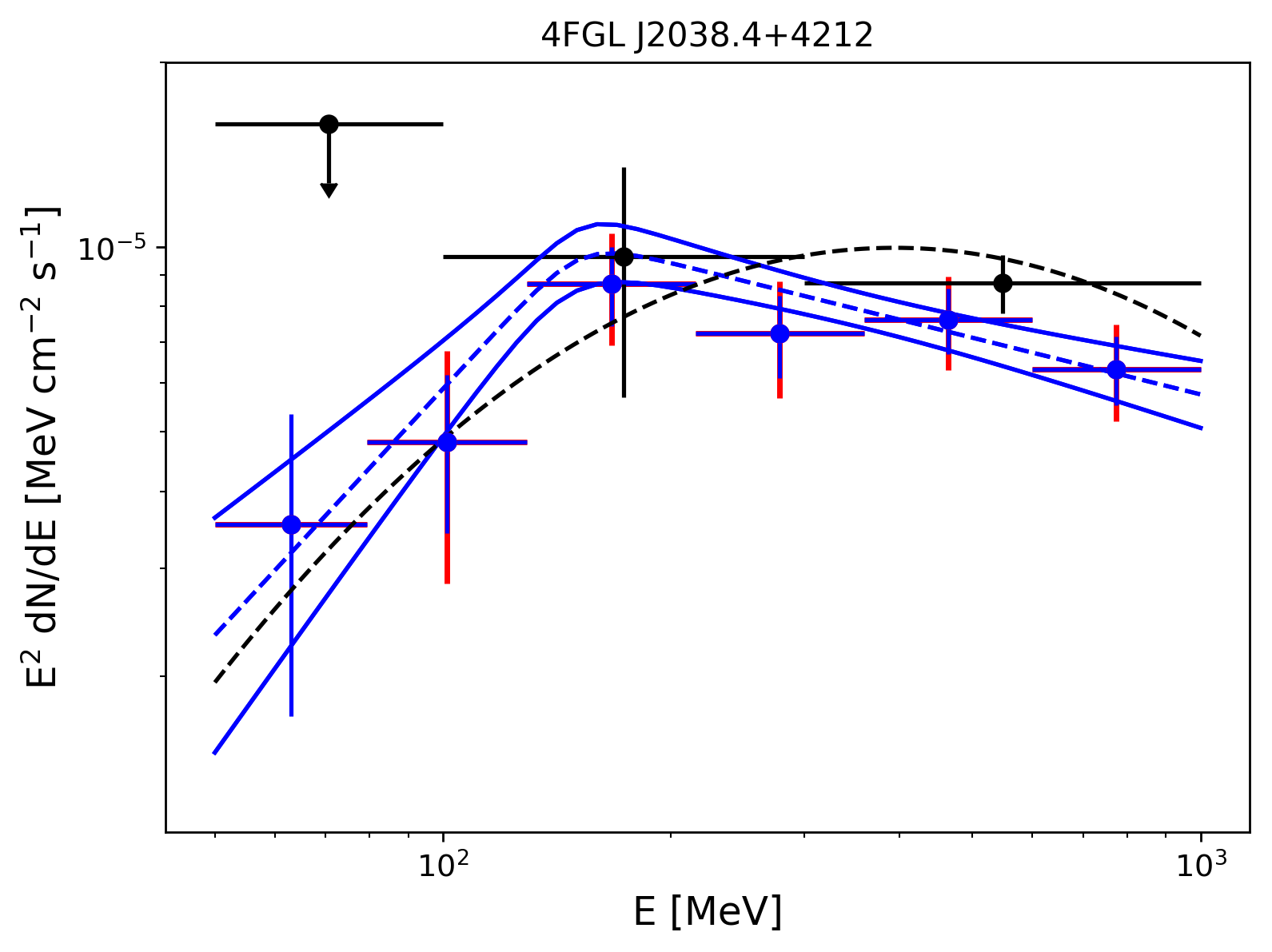}
\includegraphics[width=0.44\textwidth]{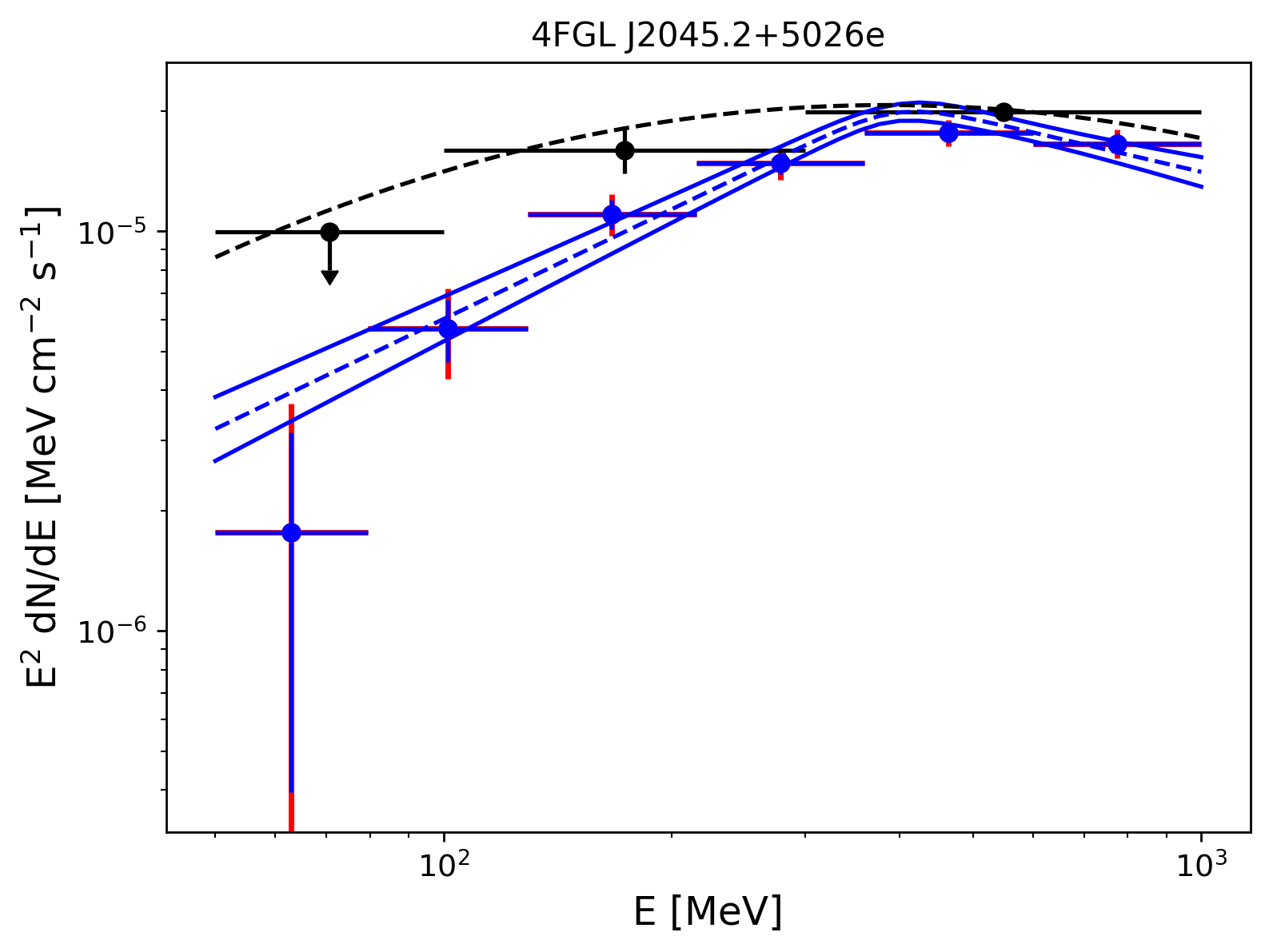}
\end{tabular}
}
\subfigure{
\begin{tabular}{ll}
\centering
\includegraphics[width=0.44\textwidth]{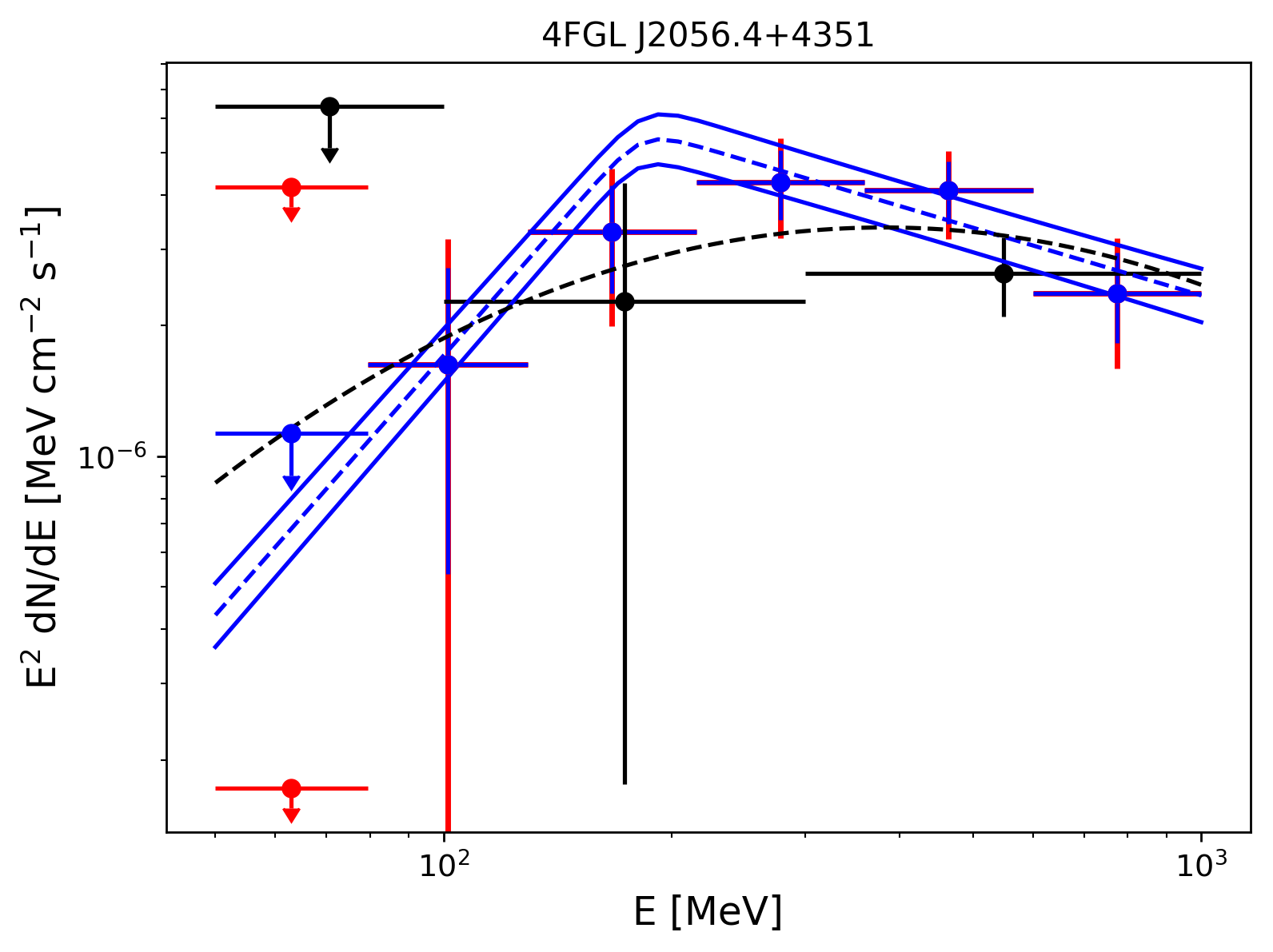}
\includegraphics[width=0.44\textwidth]{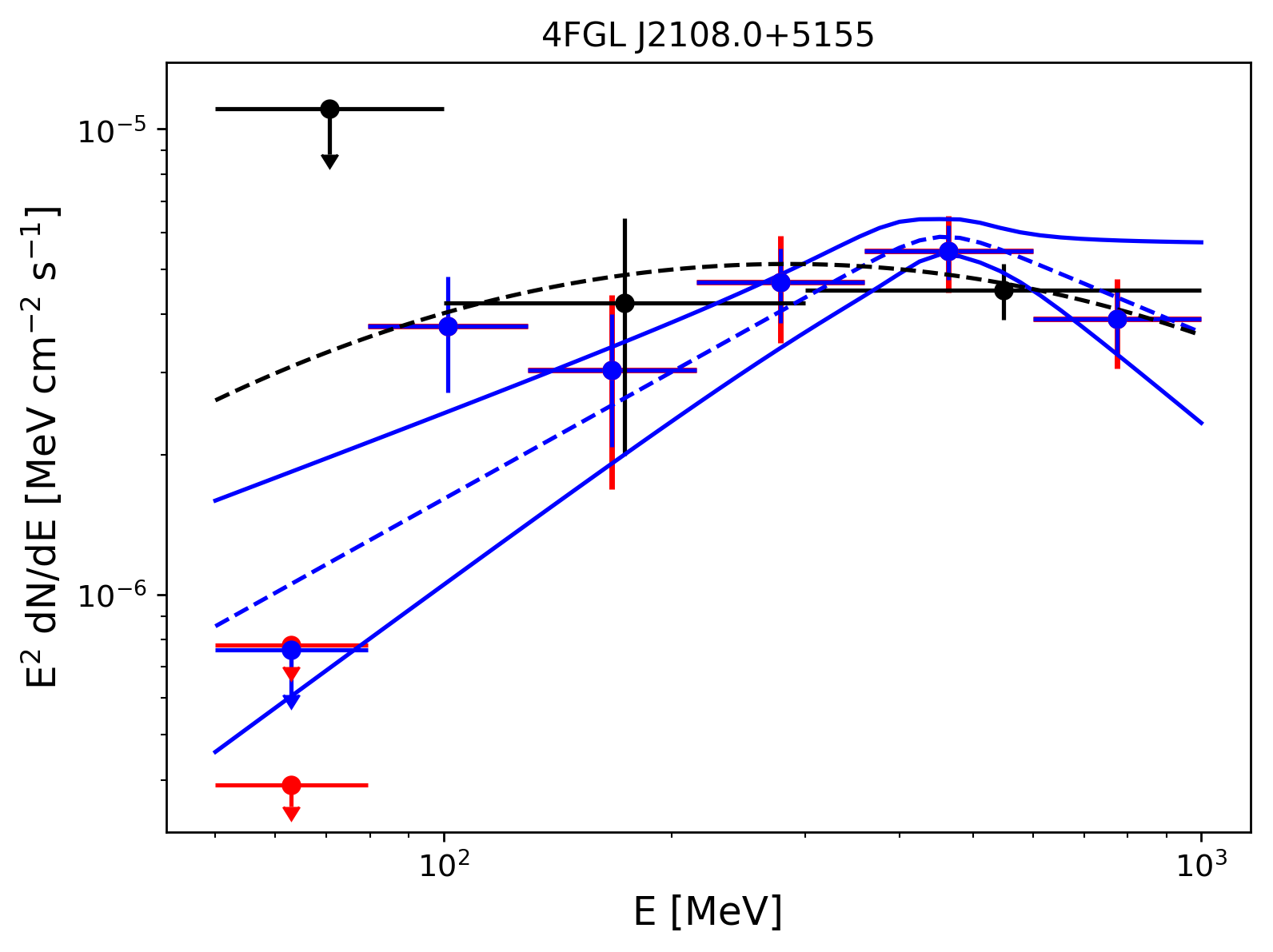}
\end{tabular}
}
\caption{\label{fig:sed9}LAT Spectral energy distributions of 4FGL J2028.6+4110e (top left), 4FGL J2032.6+4053 (top right), 4FGL J2038.4+4212 (middle left), 4FGL J2045.2+5026e (middle right), 4FGL J2056.4+4351 (bottom left), 4FGL J2108.0+5155 (bottom right) with the same conventions used in Figure~\ref{fig:snrsed}.}
\end{figure*}

As discussed in Section~\ref{subsection:snr}, gamma-ray observations are suggestive of hadron acceleration in a number of SNRs: the young SNRs Tycho and Cassiopeia A, and the middle-aged remnants with ``pion decay signature" cited above. However, definite proof of proton acceleration, especially at PeV energies, is still missing and alternative Galactic sources of cosmic rays could play a significant role.\\
 The shocks generated by the stellar winds of massive stars or star-forming regions are among these cosmic-ray accelerators. In this respect, the detection in gamma rays of the Cygnus region by the LAT \citep{2011Sci...334.1103A} opened new perspectives by revealing the presence of a cocoon of freshly-accelerated CRs over a scale of $\sim50$ pc. Our analysis revealed a spectral break for the star-forming region analyzed, 4FGL~J2028.6+4110e (see Figure~\ref{fig:sed9} top left), which is associated with the cocoon. A very hard index $\Gamma_1 = 1.00 \pm 0.02_{\rm stat} \pm 0.37_{\rm syst}$ is detected up to a break energy at $383 \pm 13_{\rm stat} \pm 138_{\rm syst}$ MeV, followed by a spectral index $\Gamma_2 = 2.23 \pm 0.06_{\rm stat} \pm 0.24_{\rm syst}$, similar to those observed for the population of identified SNRs as can be seen in Figure~\ref{fig:index}. A complete modeling of the source at gamma-ray energies is beyond the scope of this paper but our results tend to favour the hadronic scenario, thus reinforcing the long-standing hypothesis that massive-star-forming regions house particle accelerators.\\ 
Gamma-ray binaries, microquasars and colliding wind binaries could also contribute to the sea of Galactic cosmic rays and, at least contribute significantly to the population of sources with significant breaks as reported in Section~\ref{subsection:pop}. Spectral breaks have been detected for these three types of sources with 4FGL~J0240.5+6113 associated with the high-mass $\gamma$-ray binary (HMB) LS~I +61 303 (Figure~\ref{fig:sed1} top right), the HMB 4FGL~J1018.9$-$5856 (Figure~\ref{fig:sed3} bottom left), 4FGL~J1045.1-5940 associated with the colliding wind binary $\eta$ Carinae (Figure~\ref{fig:sed3} bottom right) and 4FGL~J2032.6+4053 associated with the microquasar Cyg X-3 (Figure~\ref{fig:sed9} top right). However, this last source presents the highest value of spectral index $\Gamma_1$ ($1.90 \pm 0.16_{\rm stat} \pm 0.07_{\rm syst}$) among the 56 candidates, which does not really look like the standard ``pion bump" signature observed for interacting SNRs. Finally, the source 4FGL J1405.1$-$6119 was recently identified as a high-mass gamma-ray binary using \emph{Fermi}-LAT observations \citep{corbet}, and should therefore be added to the small set of gamma-ray binaries detected in our analysis. Since significant variability was detected by the LAT for these five $\gamma$-ray sources, an individual analysis taking into account their orbital period would be needed to see if the spectral break detected is a signature of proton-proton acceleration.

\subsection{Interesting new cases: potential proton accelerators ?}\label{subsection:new}
\begin{figure*}[ht]
\centering
\subfigure{
\begin{tabular}{ll}
\centering
\includegraphics[width=0.44\textwidth]{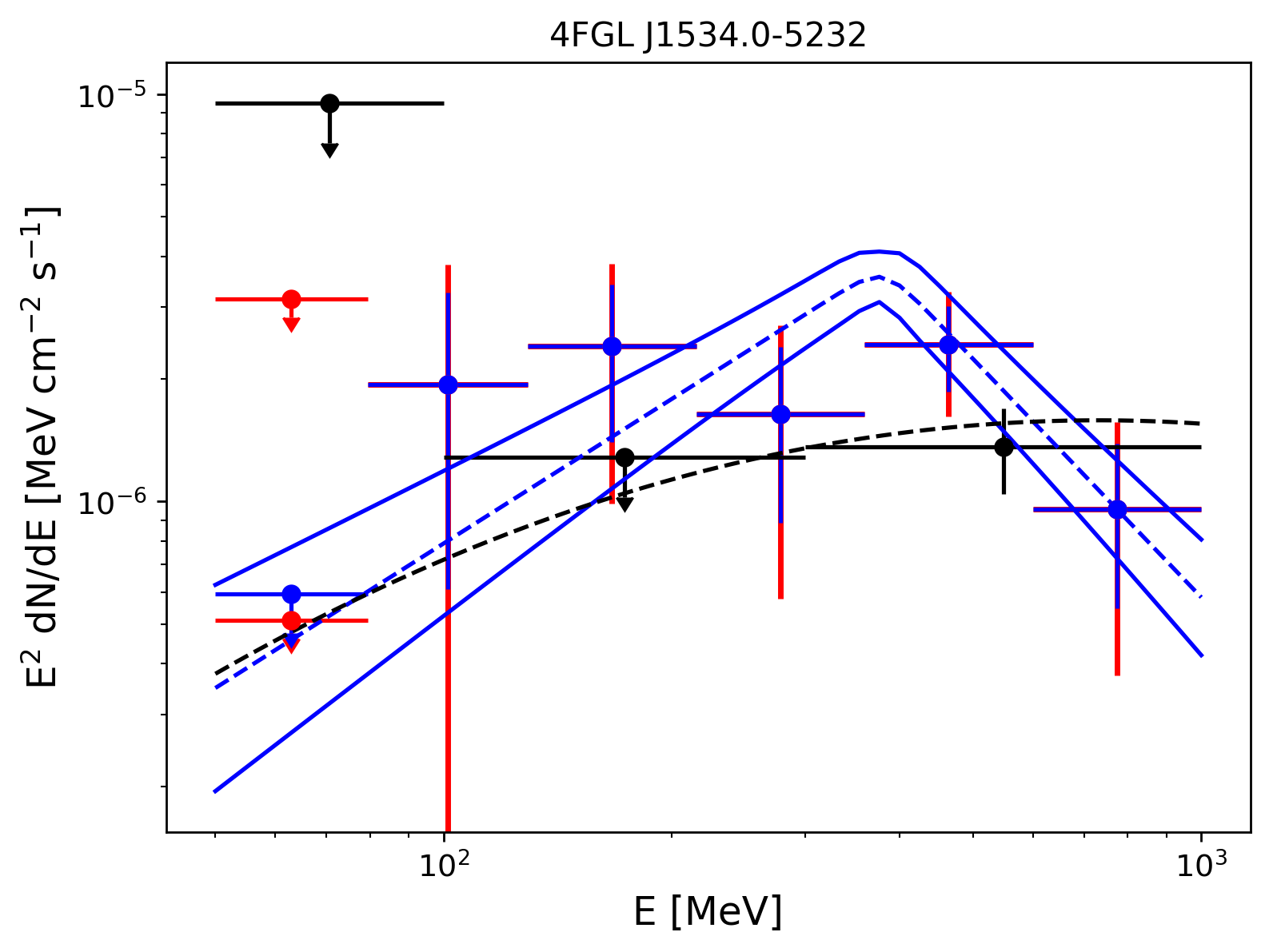}
\includegraphics[width=0.44\textwidth]{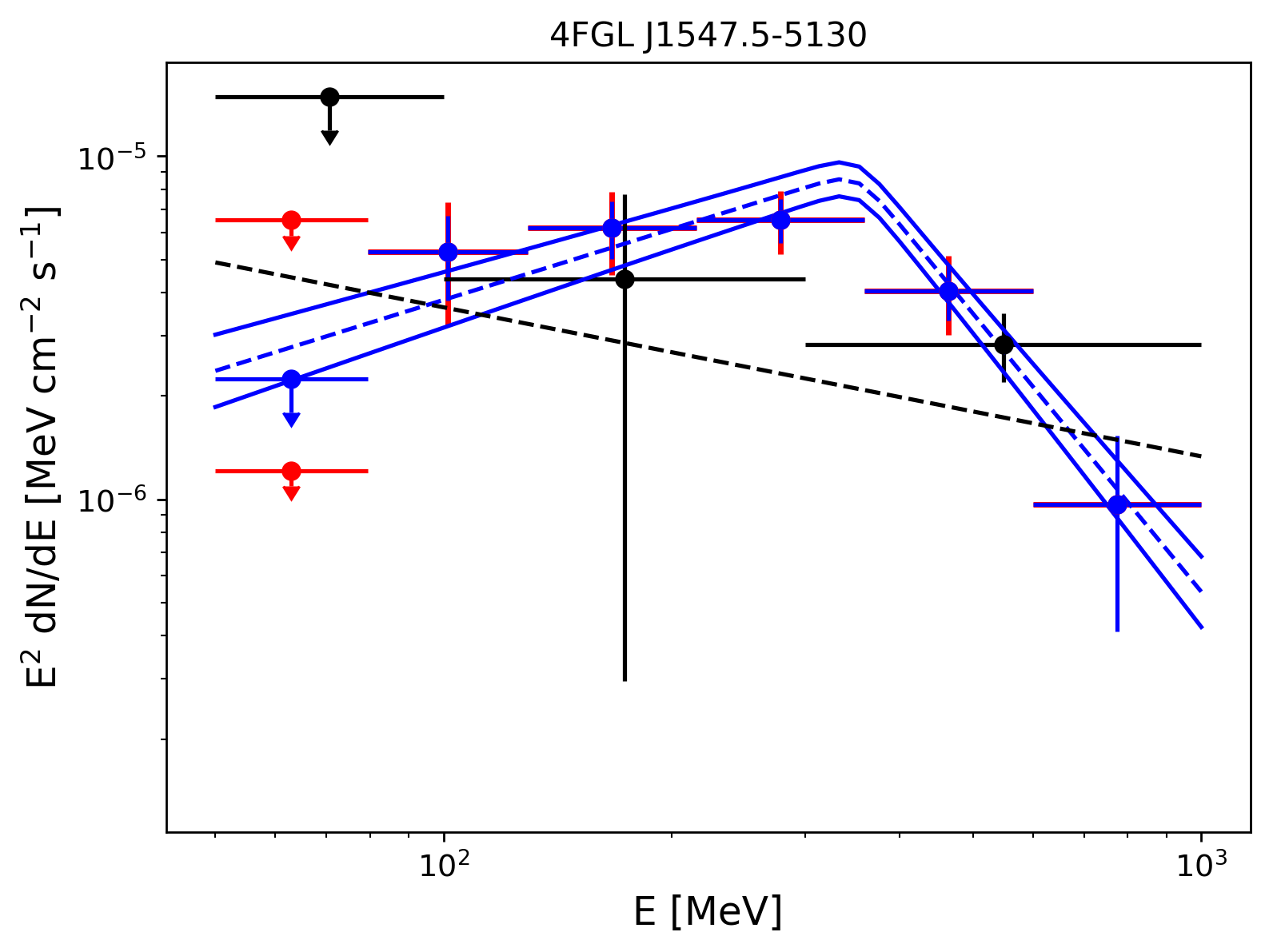}
\end{tabular}
}
\subfigure{
\begin{tabular}{ll}
\centering
\includegraphics[width=0.44\textwidth]{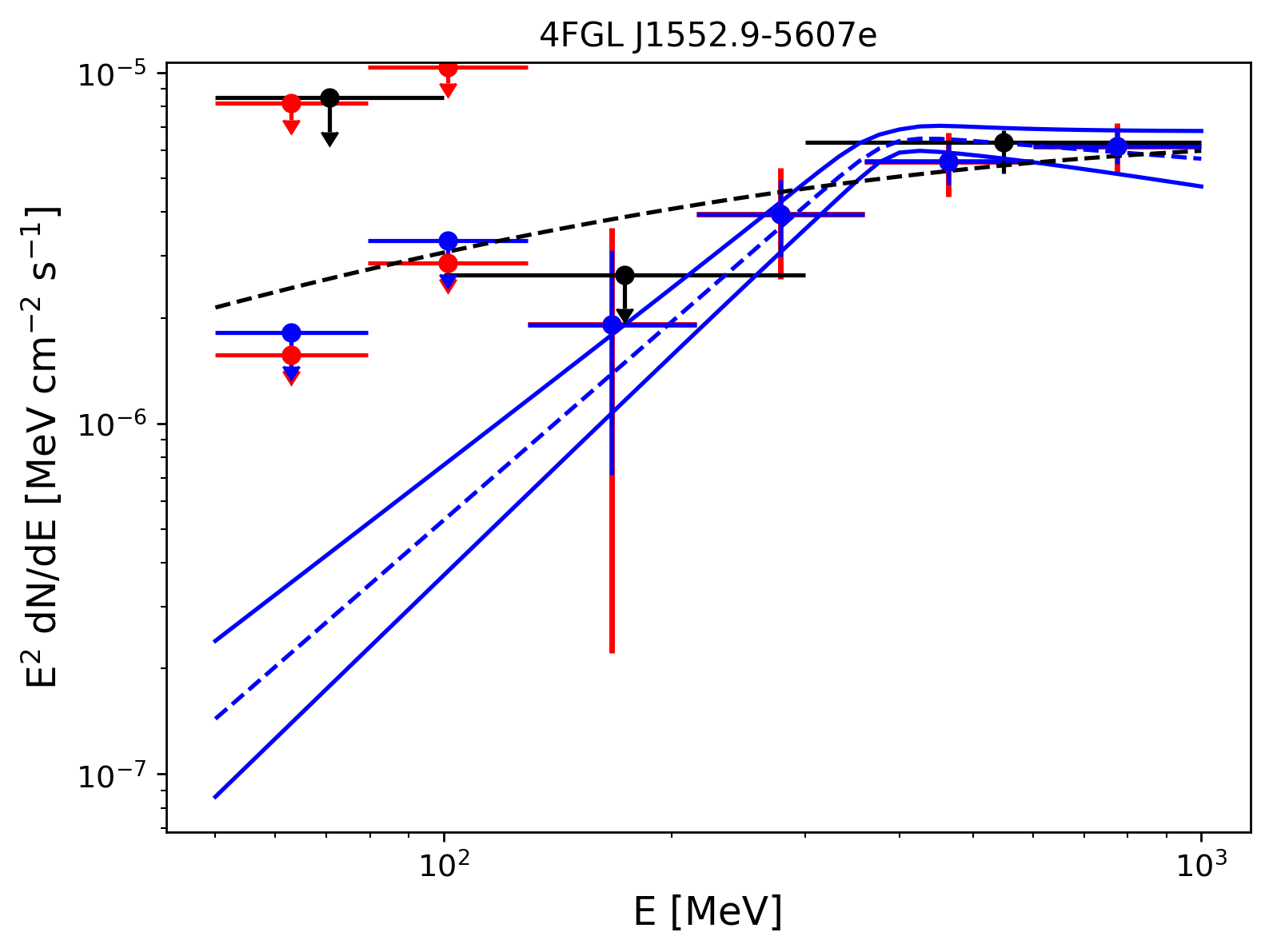}
\includegraphics[width=0.44\textwidth]{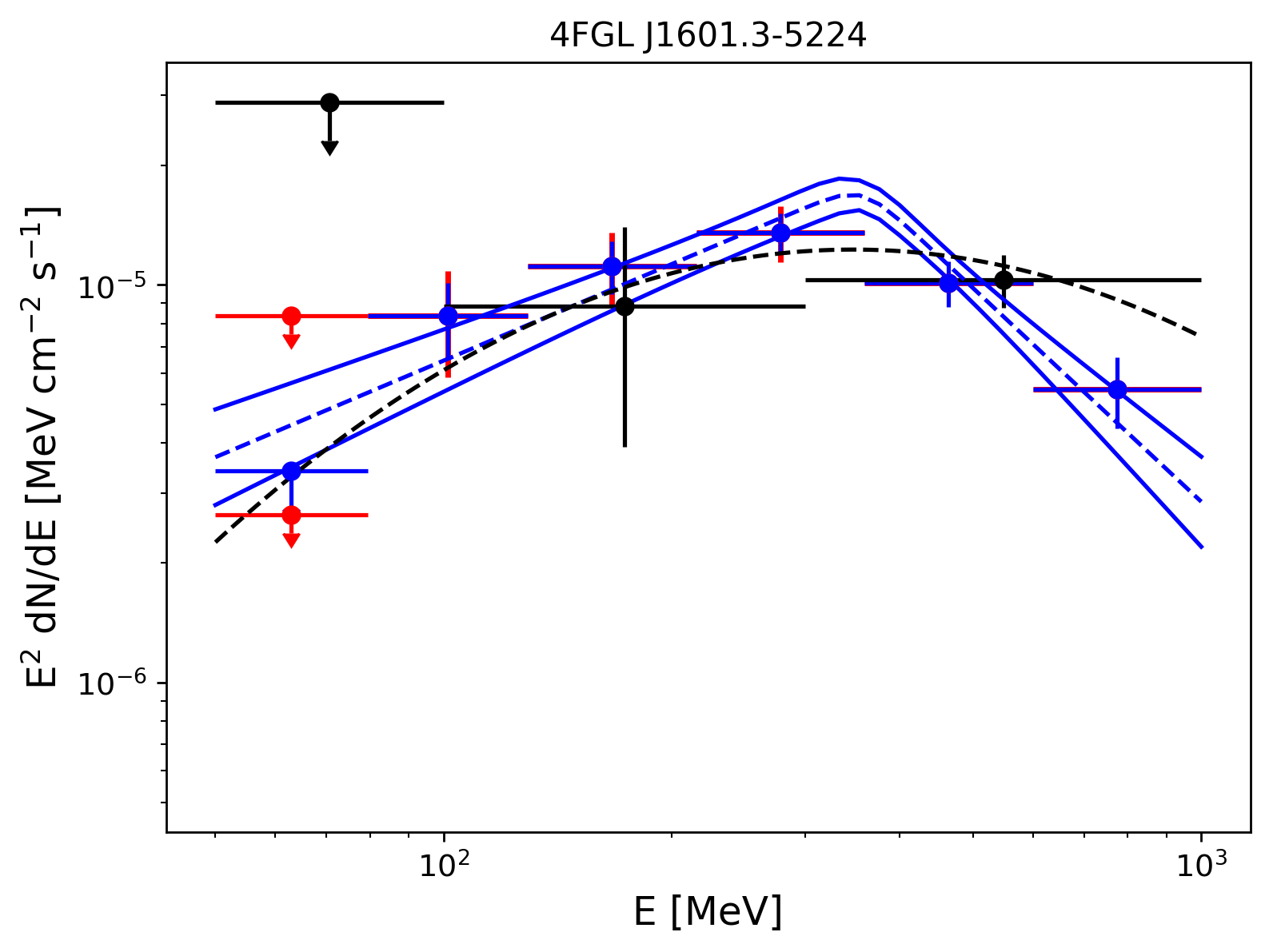}
\end{tabular}
}
\subfigure{
\begin{tabular}{ll}
\centering
\includegraphics[width=0.44\textwidth]{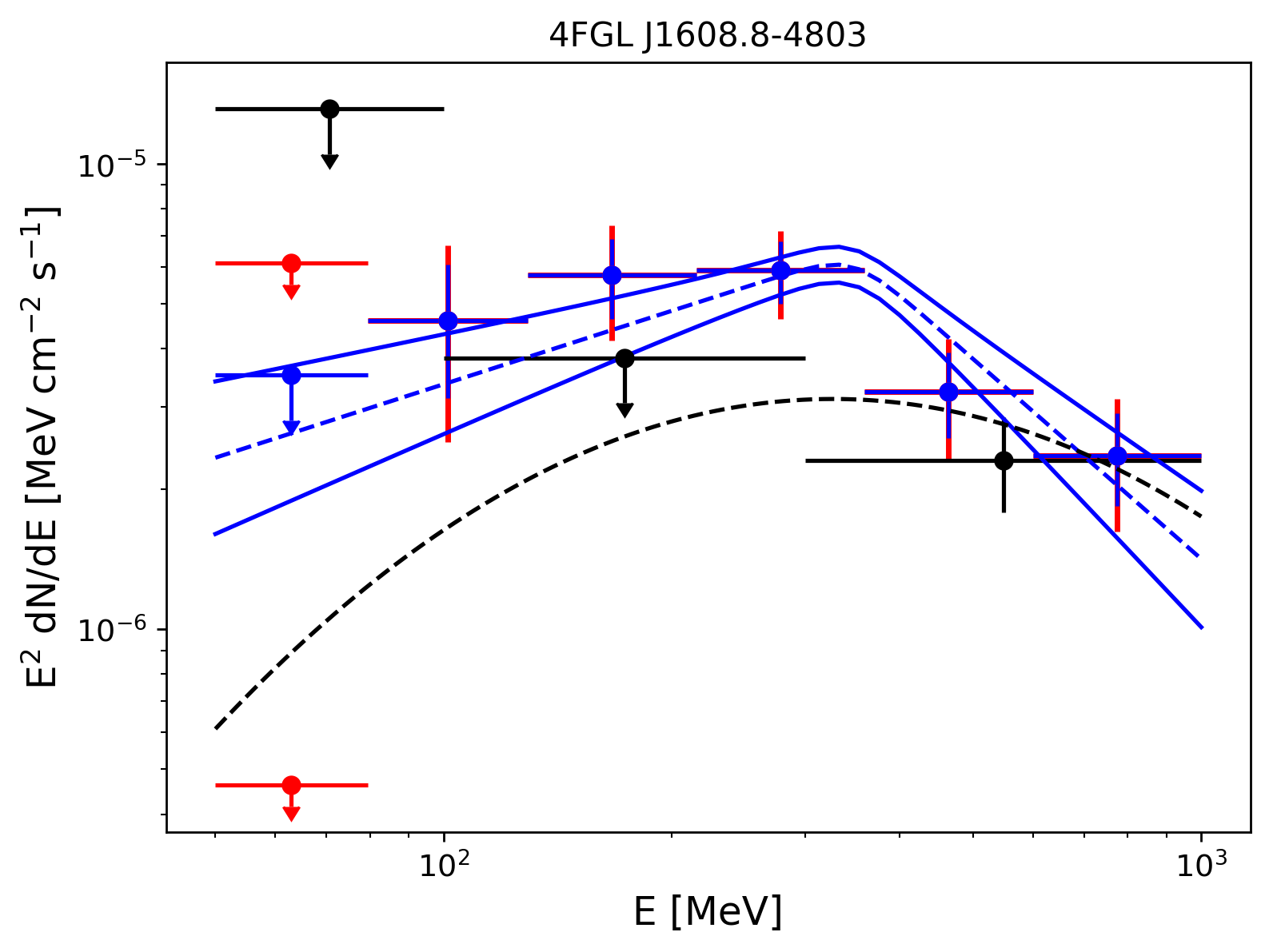}
\includegraphics[width=0.44\textwidth]{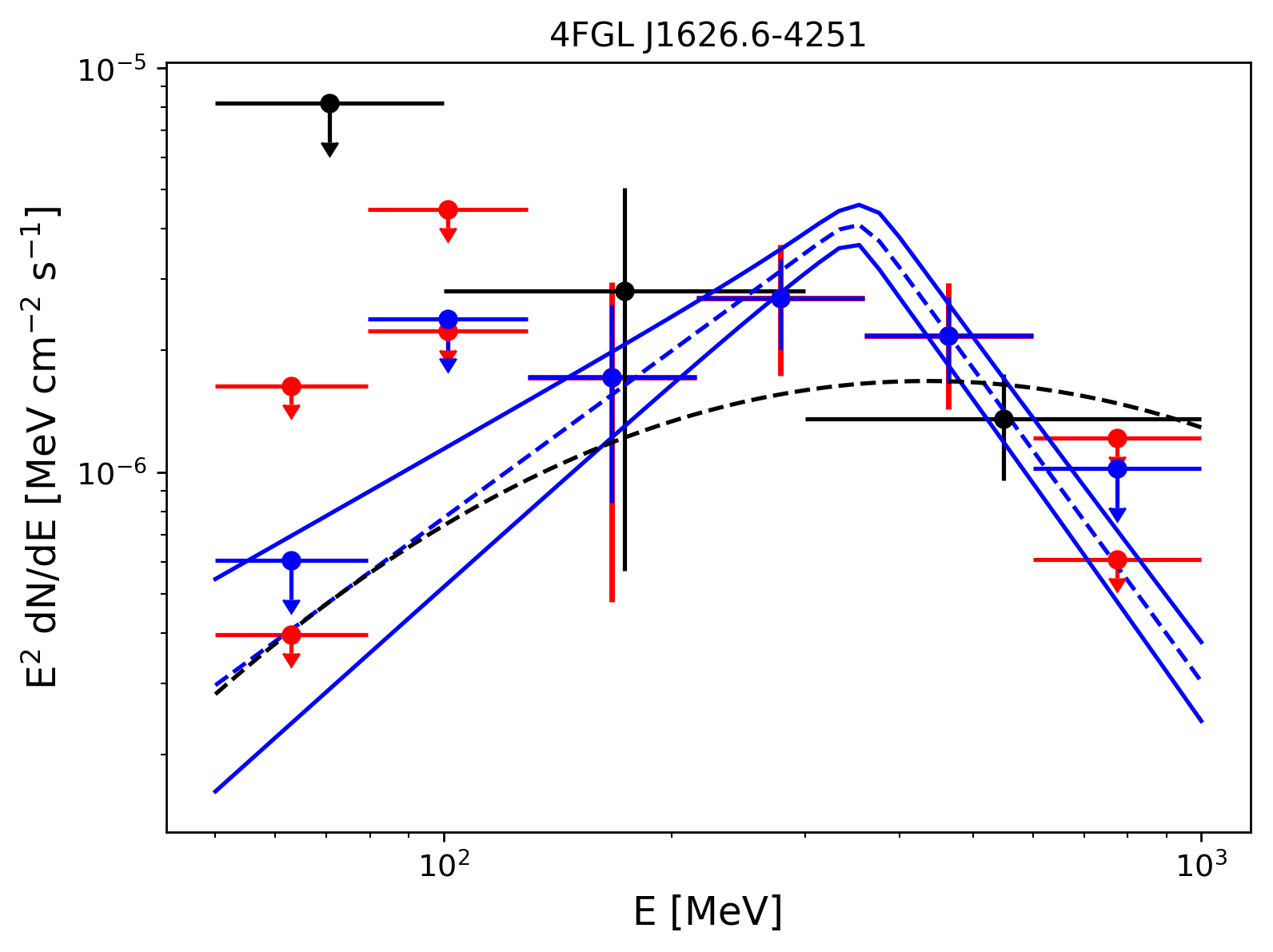}
\end{tabular}
}
\caption{\label{fig:sed5}LAT Spectral energy distributions of 4FGL J1534.0$-$5232 (top left), 4FGL J1547.5$-$5130 (top right), 4FGL J1552.9$-$5607e (middle left), 4FGL J1601.3$-$5224 (middle right), 4FGL J1608.8$-$4803 (bottom left), 4FGL J1626.6$-$4251 (bottom right) with the same conventions used in Figure~\ref{fig:snrsed}.}
\end{figure*}

\begin{figure*}[ht]
\centering
\subfigure{
\begin{tabular}{ll}
\centering
\includegraphics[width=0.44\textwidth]{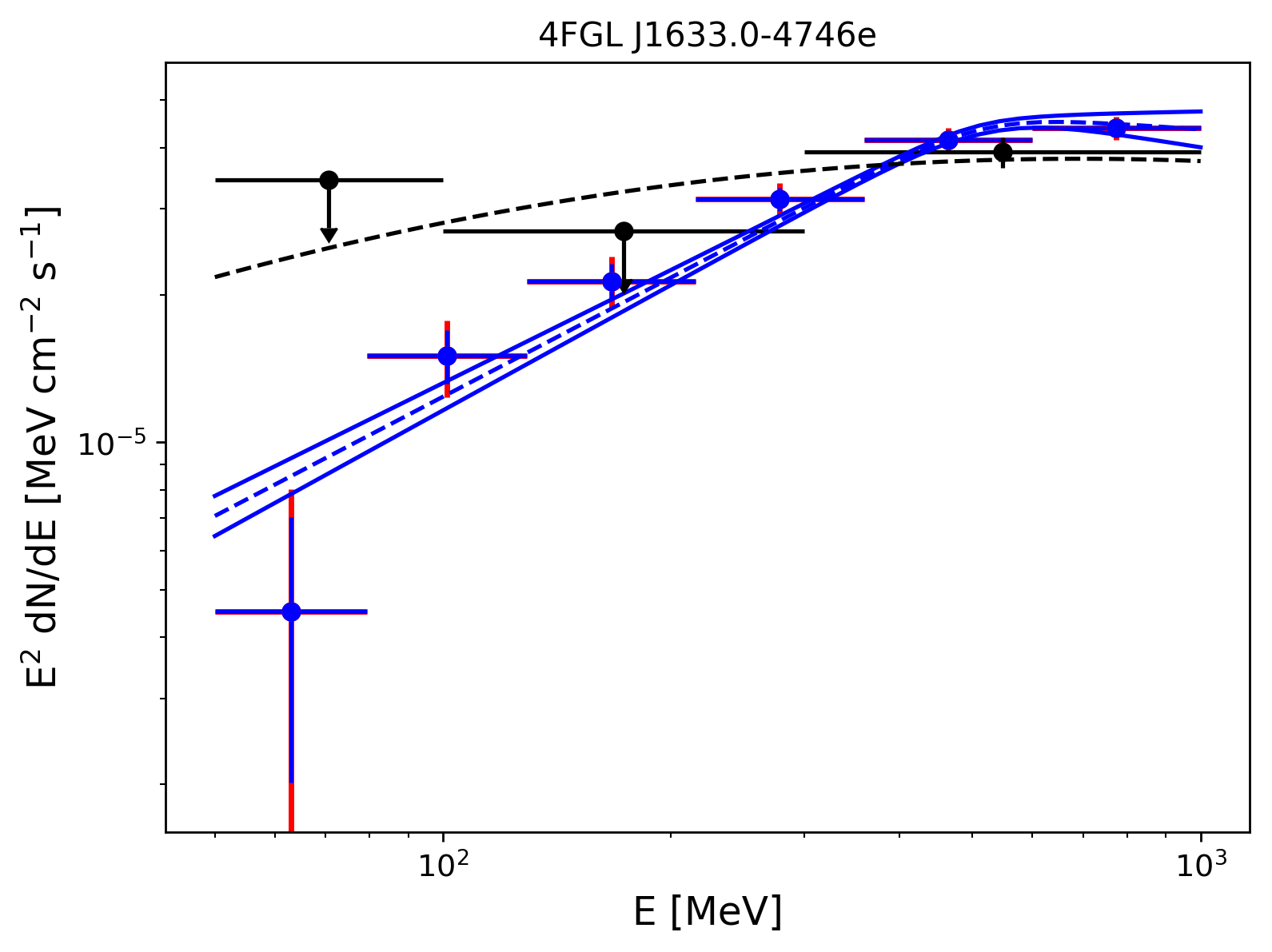}
\includegraphics[width=0.44\textwidth]{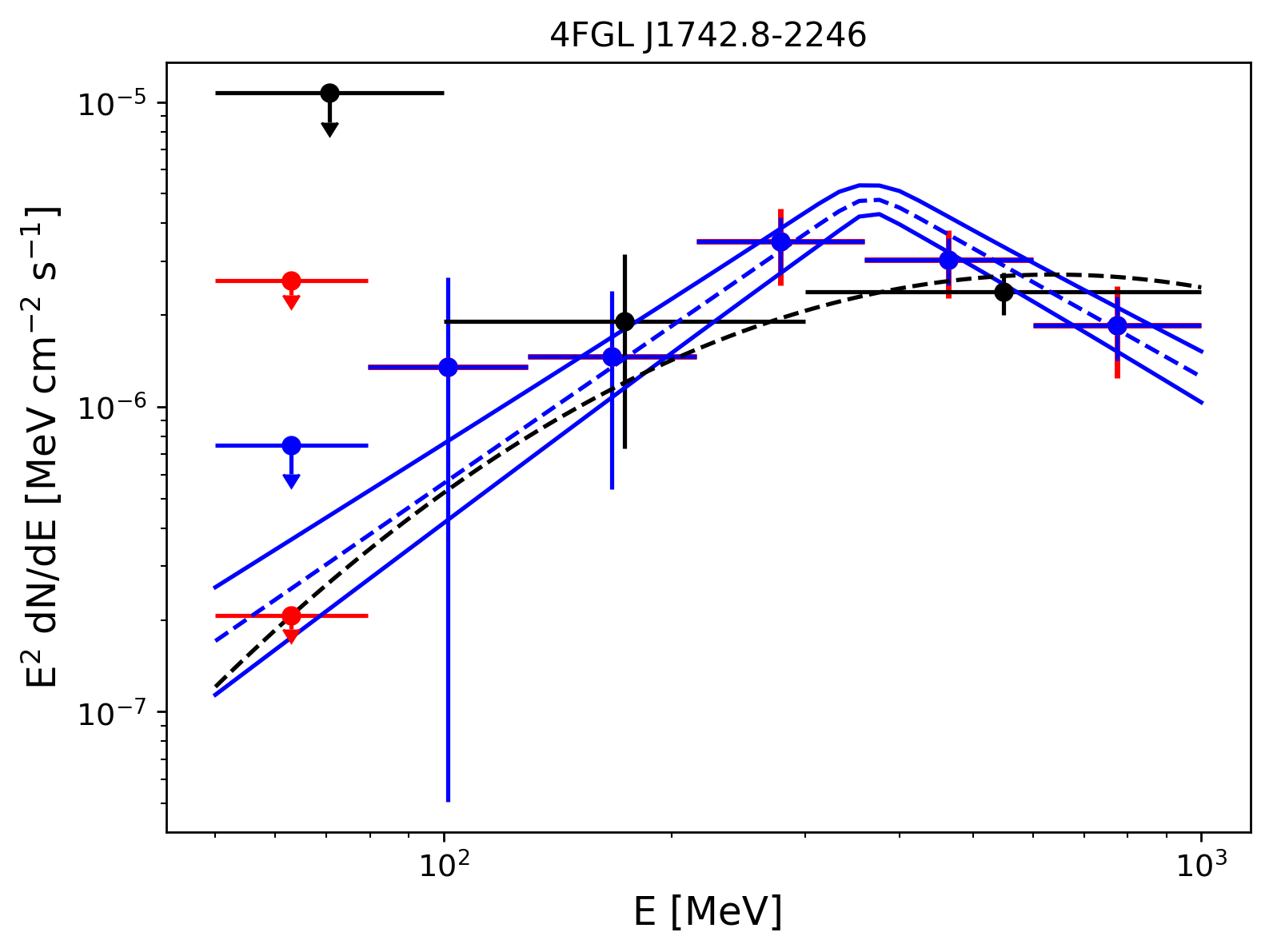}
\end{tabular}
}
\subfigure{
\begin{tabular}{ll}
\centering
\includegraphics[width=0.44\textwidth]{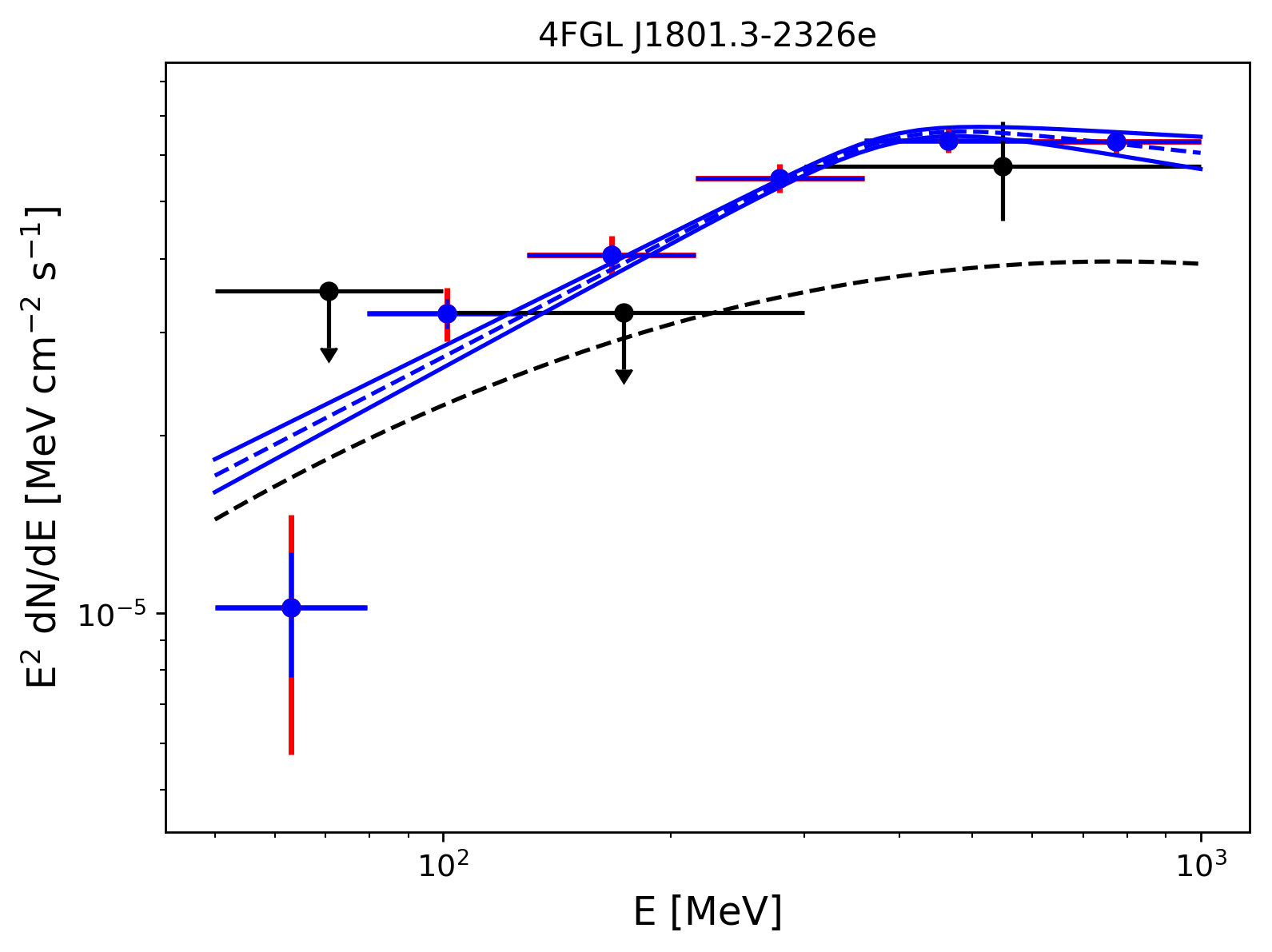}
\includegraphics[width=0.44\textwidth]{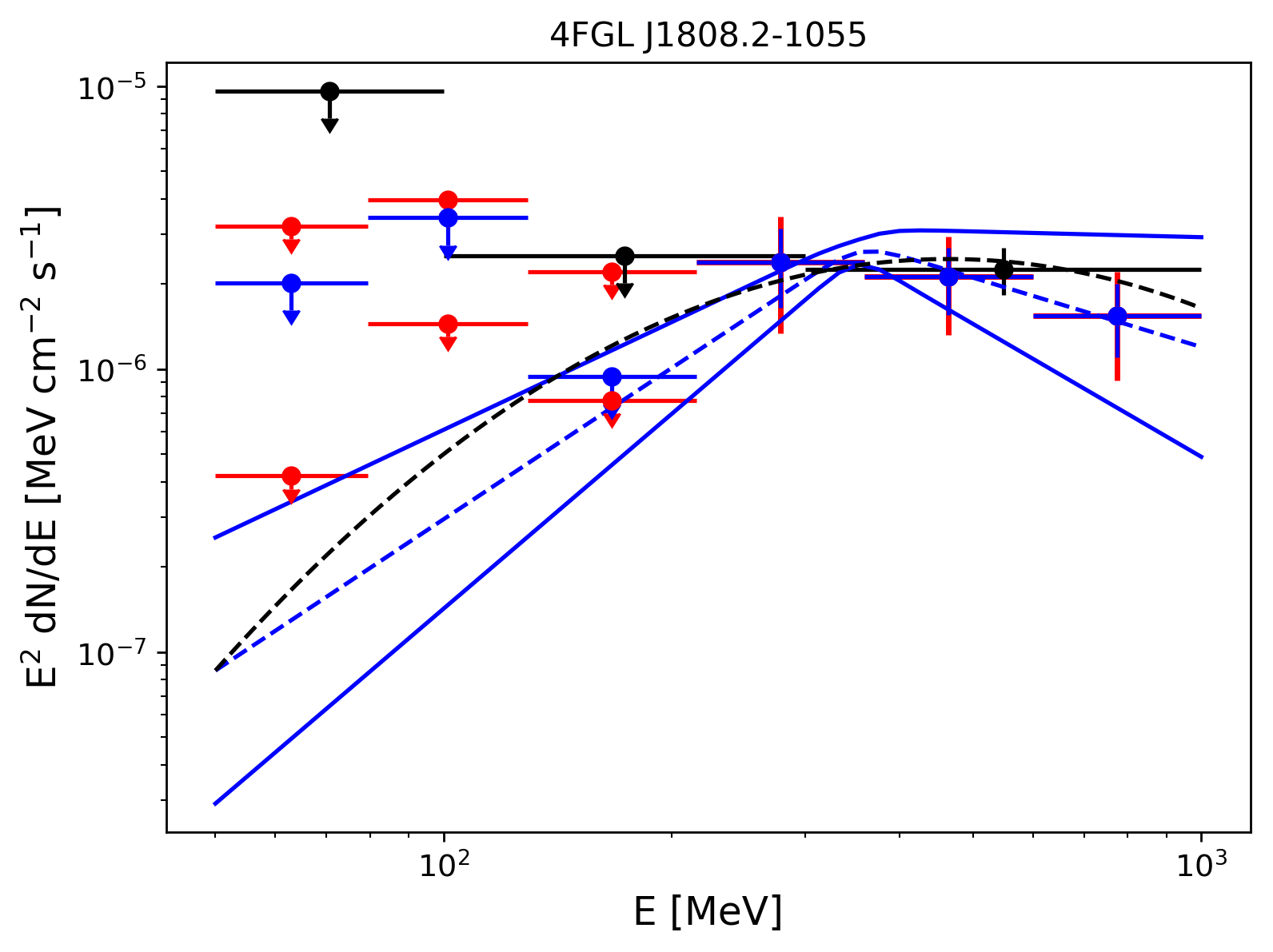}
\end{tabular}
}
\subfigure{
\begin{tabular}{ll}
\centering
\includegraphics[width=0.44\textwidth]{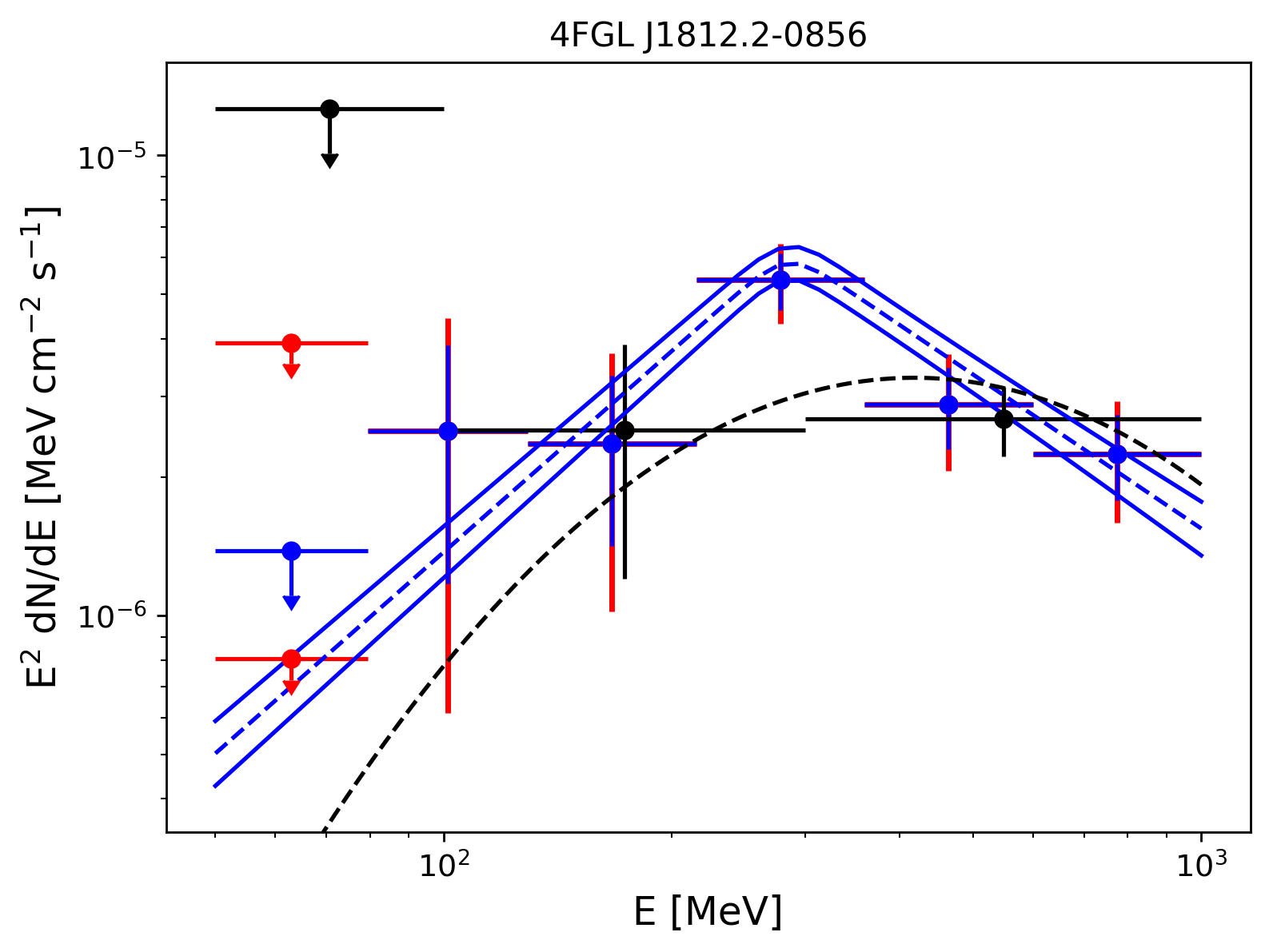}
\includegraphics[width=0.44\textwidth]{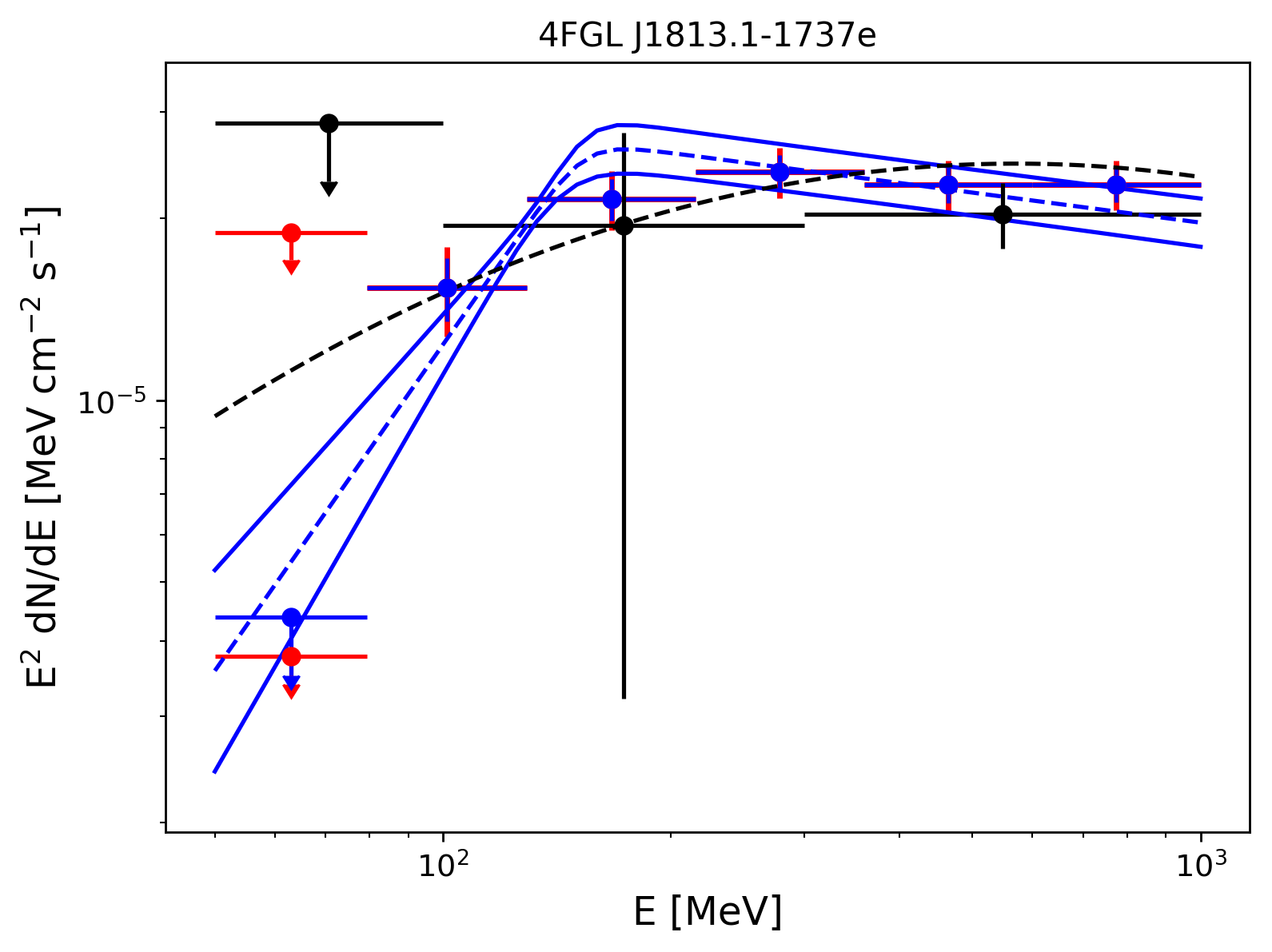}
\end{tabular}
}
\caption{\label{fig:sed6}LAT Spectral energy distributions of 4FGL J1633.0$-$4746 (top left), 4FGL J1742.8$-$2246 (top right), 4FGL J1801.3$-$2326e (middle left), 4FGL J1808.2$-$1055 (middle right), 4FGL J1812.2$-$0856 (bottom left), 4FGL J1813.1$-$1737e (bottom right) with the same conventions used in Figure~\ref{fig:snrsed}.}
\end{figure*}

\begin{figure*}[ht]
\centering
\subfigure{
\begin{tabular}{ll}
\centering
\includegraphics[width=0.44\textwidth]{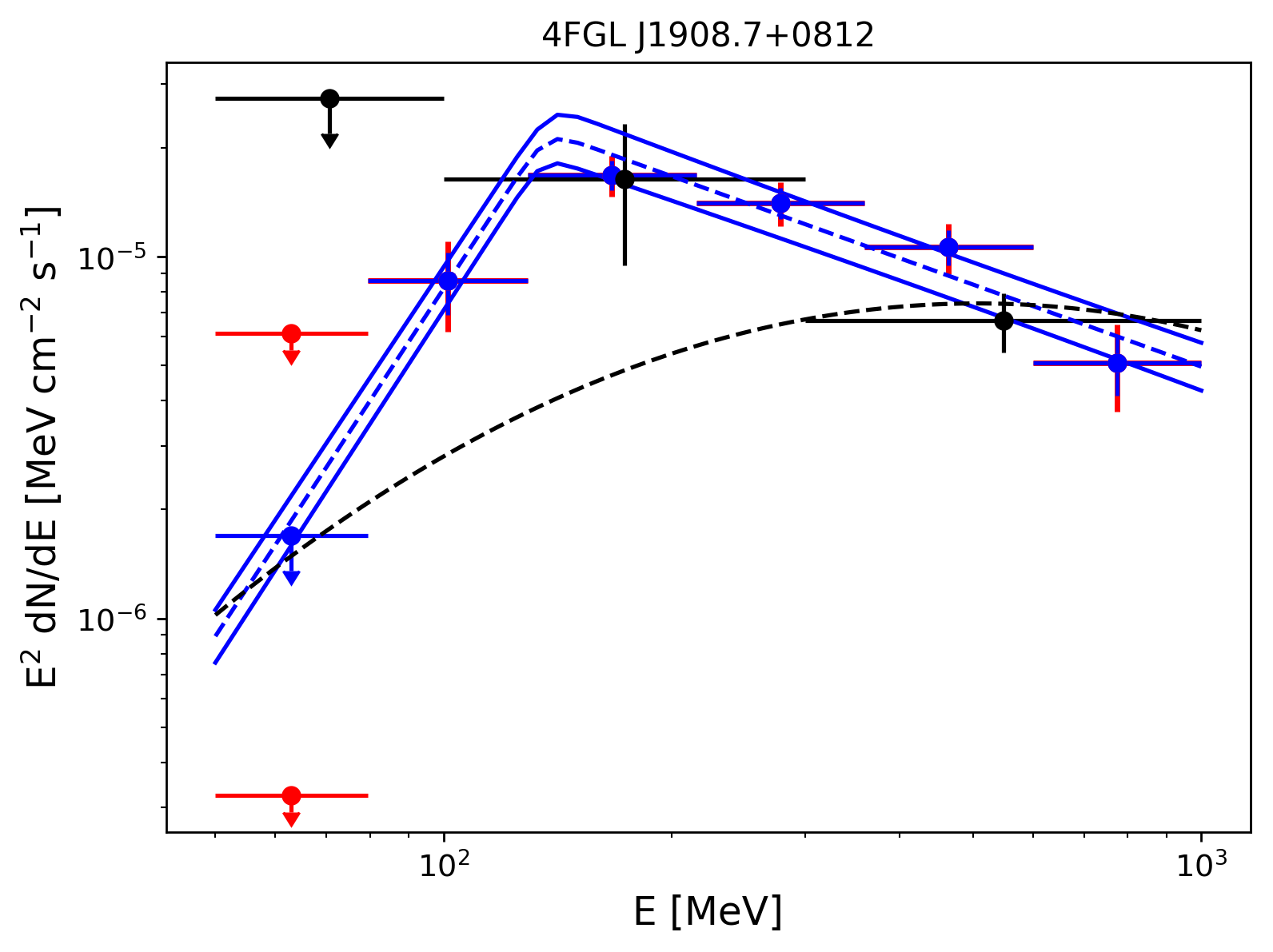}
\includegraphics[width=0.44\textwidth]{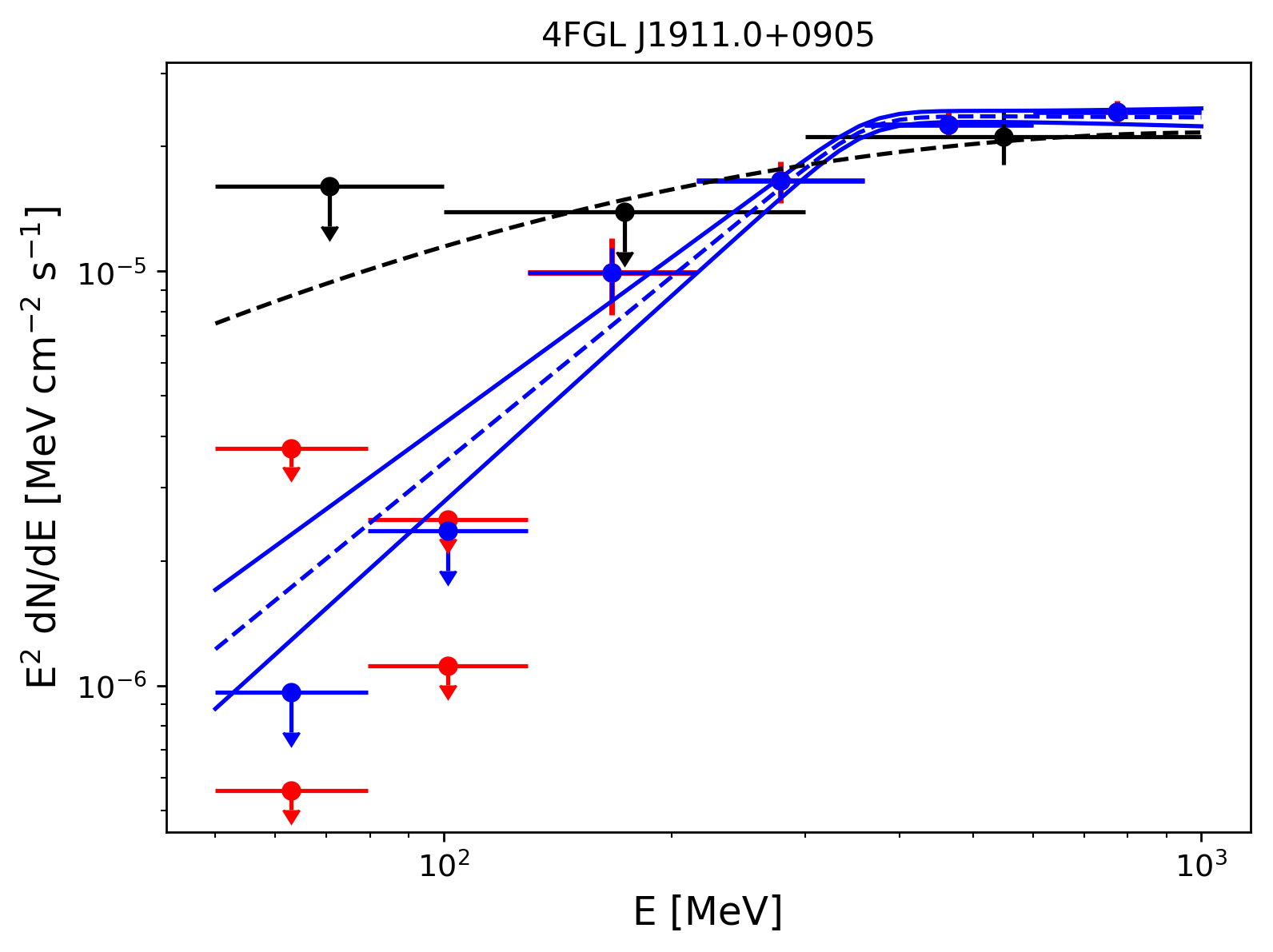}
\end{tabular}
}
\subfigure{
\begin{tabular}{ll}
\centering
\includegraphics[width=0.44\textwidth]{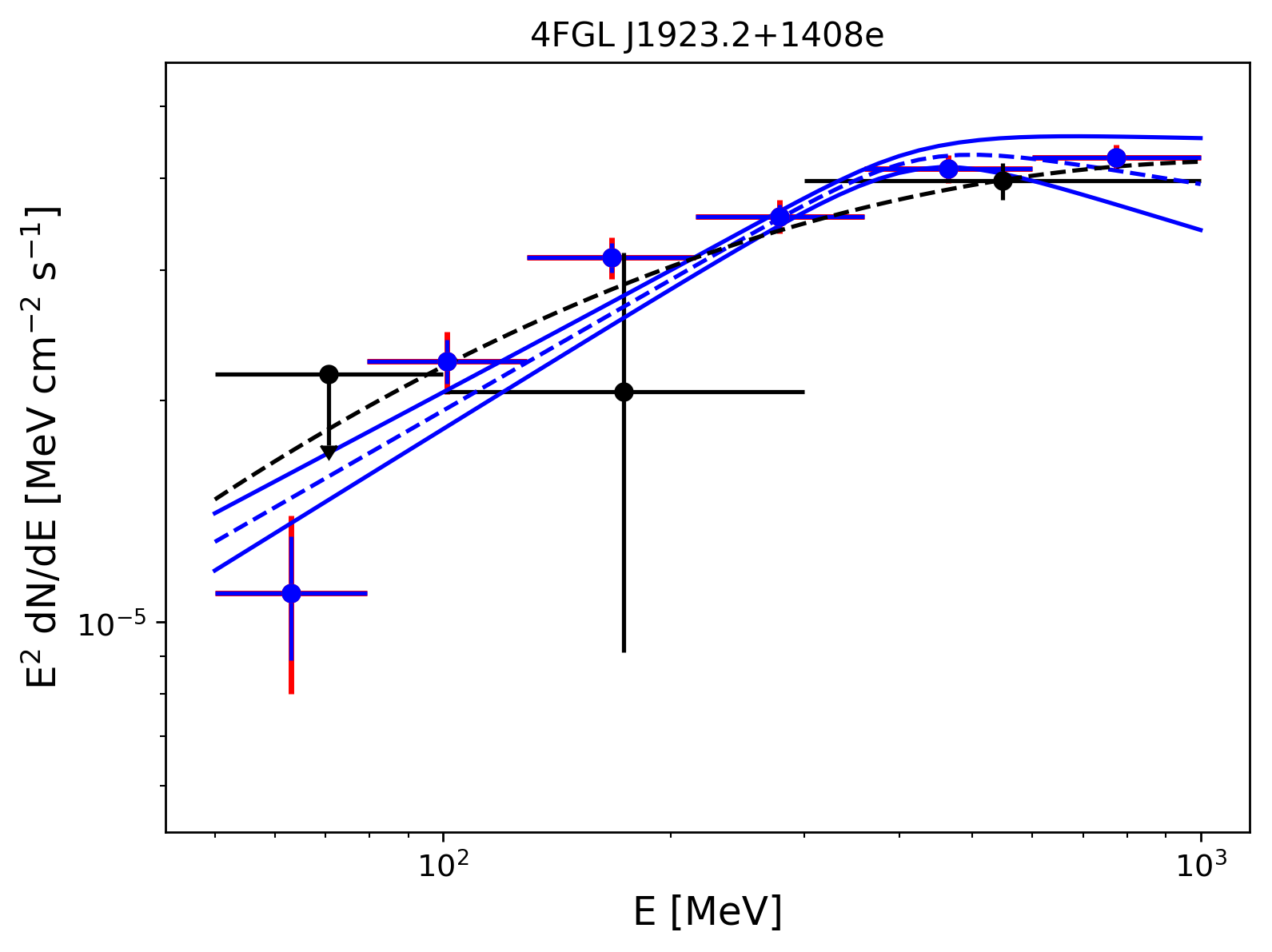}
\includegraphics[width=0.44\textwidth]{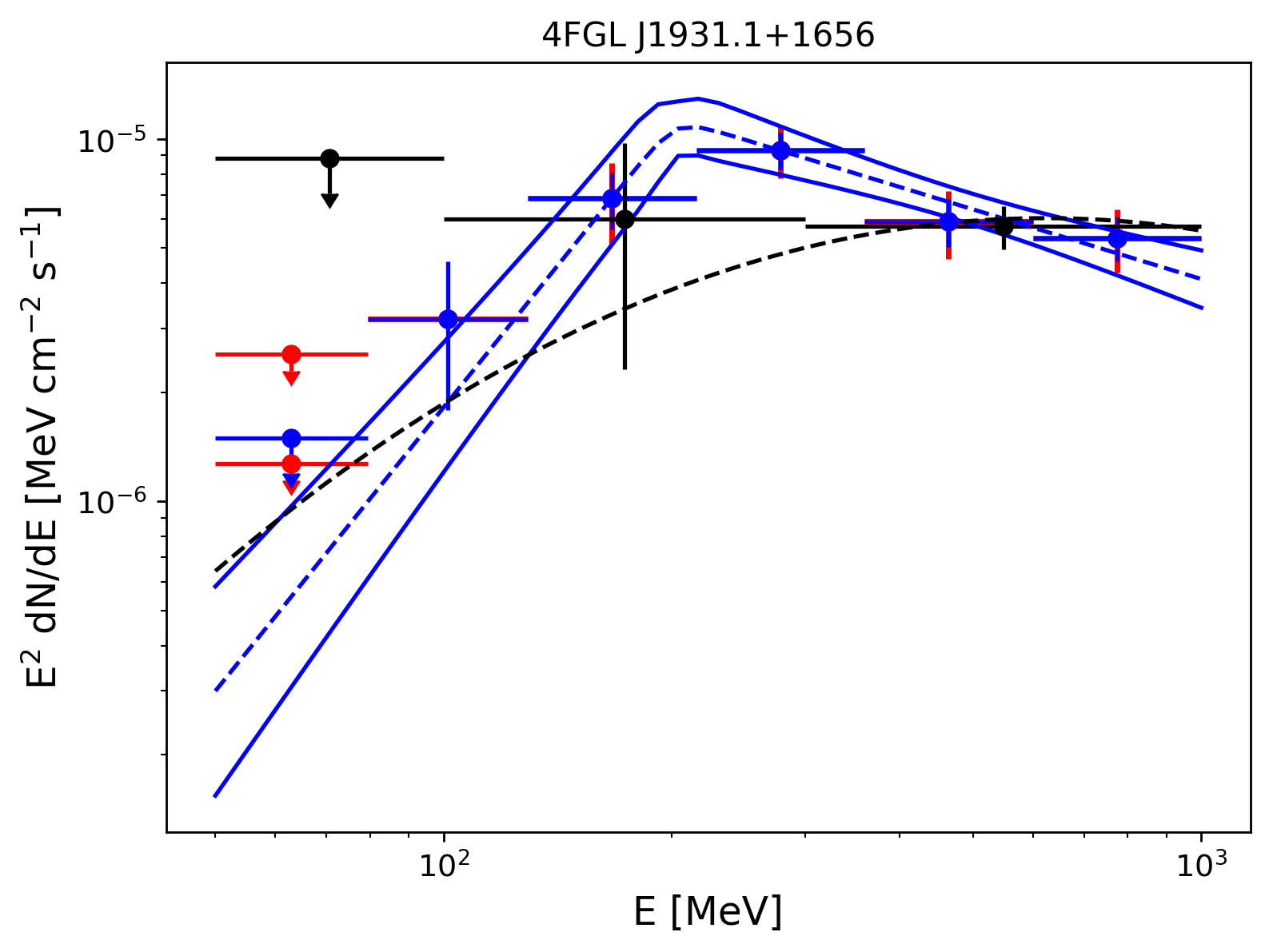}
\end{tabular}
}
\subfigure{
\begin{tabular}{ll}
\centering
\includegraphics[width=0.44\textwidth]{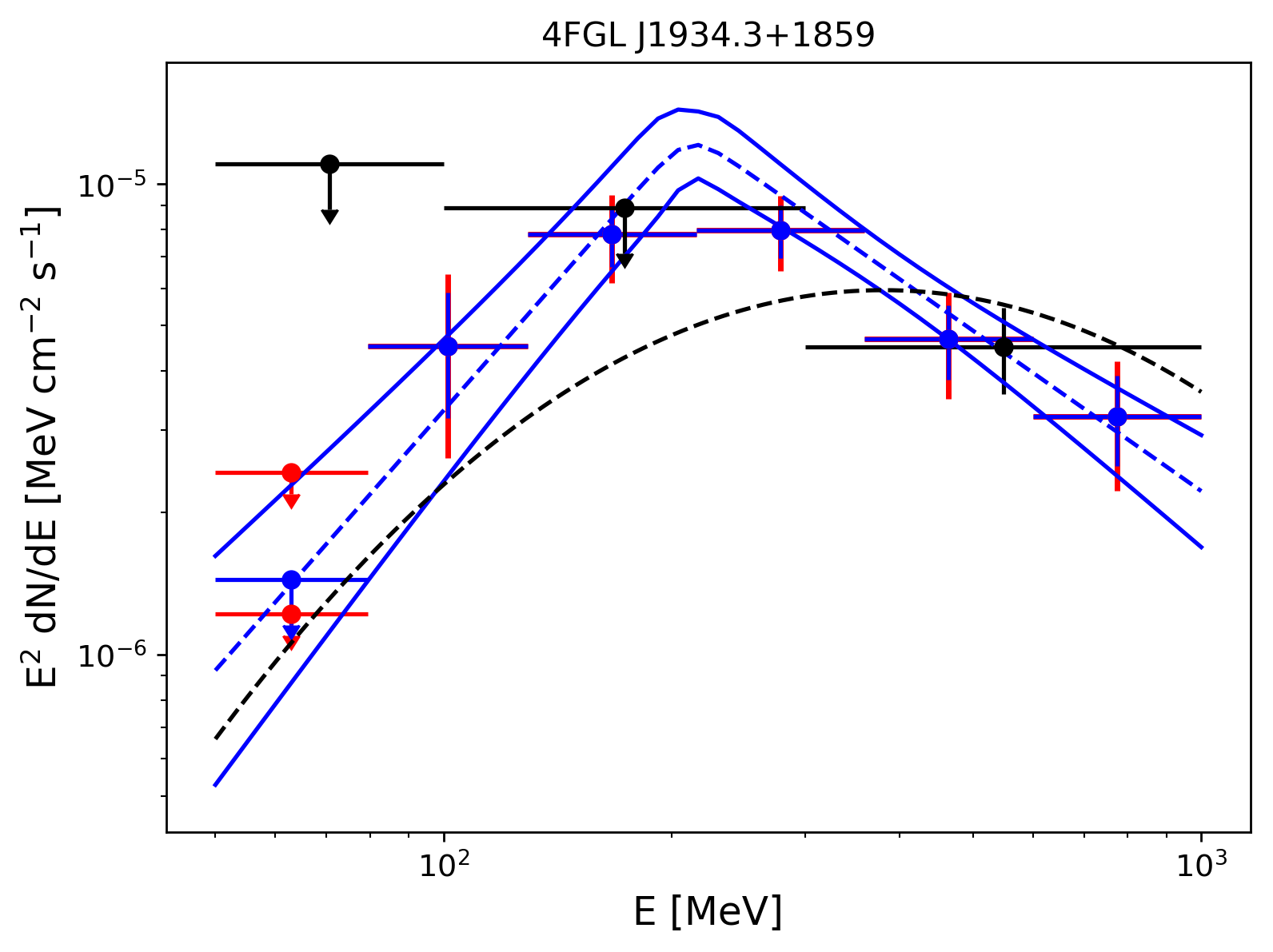}
\includegraphics[width=0.44\textwidth]{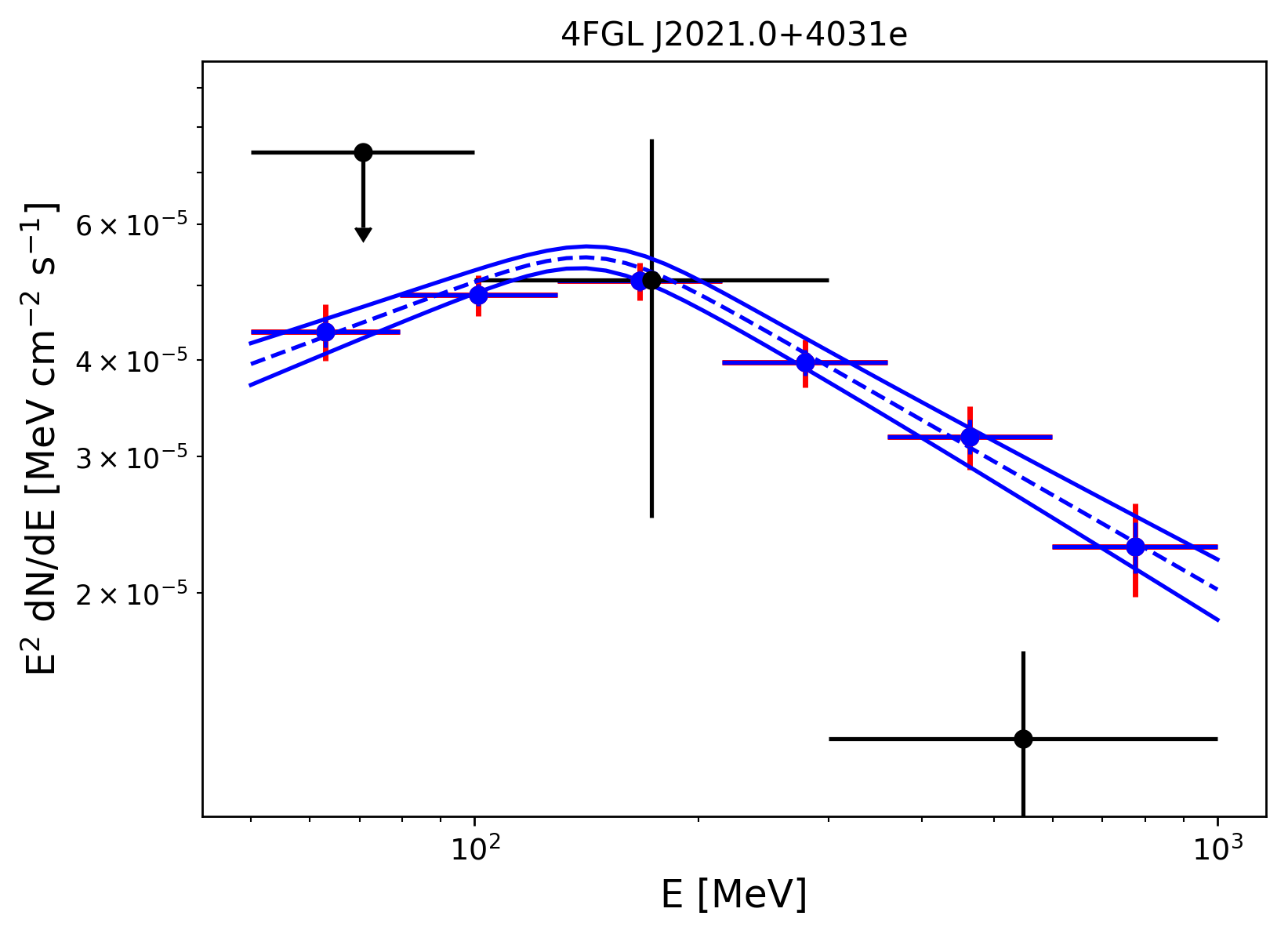}
\end{tabular}
}
\caption{\label{fig:sed8}LAT Spectral energy distributions of 4FGL J1908.7+0812 (top left), 4FGL J1911.0+0905 (top right), 4FGL J1923.2+1408e (middle left), 4FGL J1931.1+1656 (middle right), 4FGL J1934.3+1859 (bottom left), 4FGL J2021.0+4031e (bottom right) with the same conventions used in Figure~\ref{fig:snrsed}.}
\end{figure*}

Among the sources for which a significant spectral break is detected with our pipeline, several are classified as spp, unk or even unassociated as can be seen in Table~\ref{tab:class}. Among these three source classes, spp is the only one for which the fraction of sources with significant break is similar to the analyzed fraction (11.9\% vs 10.7\%), while unk and UNIDs both show a clear decrease between the analyzed fraction and the confirmed one (see Figure~\ref{fig:pop}). The spp are sources of unknown nature but overlapping with known SNRs or PWNe and thus candidates to these classes, while unk are sources associated to counterparts of unknown nature. Unassociated, spp and unk represent $29.7$\% of the 4FGL sources: revealing the mystery of the nature of these unidentified gamma-ray sources might shed new light on the problem of the origin of galactic CRs. In this respect, three sources detected by our pipeline are of special interest since they are coincident with SNRs and/or dense molecular clouds.\\ 
This is the case for 4FGL~J1601.3$-$5224 (Figure~\ref{fig:sed5} middle right) coincident with the SNR G329.7+00.4 which presents a diffuse shell at radio energies~\citep{1996A&AS..118..329W} but is not detected at any other wavelength. Our analysis indicates a soft spectrum $\Gamma_2 = 3.78 \pm 0.32_{\rm stat} \pm 0.89_{\rm syst}$ with large systematics due to the diffuse background. The same systematics affect the value of the energy break showing that our results may suffer from contamination.\\
Similarly, the source 4FGL~J1934.3+1859 (Figure~\ref{fig:sed8} bottom left) is coincident with SNR G054.4$-$00.3 detected as a nearly circular shape and angular diameter of $\sim40$ arcmin at radio energies~\citep{1992A&A...261..289J} while \emph{Swift} and \emph{Suzaku} X-ray observations allowed the detection of the X-ray counterpart~\citep{2017MNRAS.466.1757K} of the gamma-ray pulsar PSR J1932+1916~\citep{2013ApJ...779L..11P} located near the edge of the supernova remnant. \emph{Suzaku} observations also revealed diffuse emission with extent of about 5 arcmin whose spectral properties are compatible with those of PSR+PWN systems. Interestingly, large-scale CO structures across the SNR were observed, indicating the SNR interaction with the ambient molecular gas which is an important ingredient to enhance the gamma-ray emission due to proton-proton interaction. Our analysis reveals a spectral index above the break energy $\Gamma_2 = 3.13 \pm 0.27_{\rm stat} \pm 0.12_{\rm syst}$ which may again indicate that the association with an SNR is spurious or that our low-energy analysis suffers from contamination from other neighboring sources in this crowded region.\\
Finally, the unidentified source 4FGL J1931.1+1656 (Figure~\ref{fig:sed8} middle right) is coincident with the SNR candidate G52.37$-$0.70 detected in a recent THOR+VGPS analysis \citep{2017A&A...605A..58A}. However, the spectral index of $\alpha = 0.3 \pm 0.3$ using VLA observations \citep{2018ApJ...860..133D} seems to indicate that this candidate is unlikely to be an SNR. The $\gamma$-ray spectrum derived by our analysis resembles that of other SNRs and is not affected by large systematics especially the break energy $203 \pm 8 \pm 19$ and the spectral index above the break $\Gamma_2 = 2.64 \pm 0.10 \pm 0.04$. It is the best candidate for proton acceleration among these three potential SNR association.\\
These three regions are extremely complex and would deserve a dedicated analysis at higher energy with \emph{Fermi} to constrain their location and their association with the corresponding SNR, as well as a spectral analysis over a larger energy interval to definitively constrain the type of radiating particles. \\
Even more care should be taken for the extended sources 4FGL~J1633.0$-$4746e (Figure~\ref{fig:sed6} top left) and 4FGL~J1813.1$-$1737e (Figure~\ref{fig:sed6} bottom right) since their disk radii of $0.61^{\circ}$ and $0.6^{\circ}$ respectively in confused Galactic plane regions adds to the complexity of such analysis at low energy.
With its large extension, 4FGL~J1633.0$-$4746e overlaps with both the TeV PWN candidate HESS J1632$-$478 and the unidentified source HESS J1634$-$472, both detected at GeV energies but not included in our list of selected candidates due to their low significance at low energy. This implies that the region contains three sources: a point-like source coincident with HESS J1634$-$472, an extended source coincident with HESS J1632$-$478 but with an extension of $0.256^{\circ}$ almost twice as large as the TeV size, and the very extended source 4FGL~J1633.0$-$4746e overlapping them detected above 10 GeV with a spectral index of $2.25 \pm 0.01_{\rm stat} \pm 0.10_{\rm syst}$ \citep{2017ApJ...843..139A}. Interestingly, our spectral analysis indicates a break at $517 \pm 18_{\rm stat} \pm 252_{\rm syst}$ MeV followed by an index of $\Gamma_2 = 2.11 \pm 0.15_{\rm stat} \pm 0.12_{\rm syst}$ in agreement with the index detected above 10 GeV (though with very large systematics on the break energy due to the diffuse background). The break detected at low energy by our analysis, the hard spectral index $\Gamma_2$ consistent with the one detected at higher energy (which seems to indicate a flat spectrum over a large energy range) and the presence of dense clumps in this region traced NH$_3$(1,1) emission \citep{2017MNRAS.468.2093D} make this source a very interesting proton accelerator. A dedicated analysis would therefore be very valuable in this case.\\
The disk radius of $0.60 \pm 0.06_{\rm stat}^{\circ}$ of the \emph{Fermi} source 4FGL~J1813.1$-$1737e, coincident with the compact TeV PWN candidate HESS J1813$-$178 \citep[Gaussian size of $0.049 \pm 0.04^{\circ}$ in][]{2018A&A...612A...1H}, was first detected by \cite{2018ApJ...859...69A}. The authors reported a hard index of $2.07 \pm 0.09_{\rm stat}$ above 500 MeV compatible with the TeV index. This spectrum is compatible with the spectral index $\Gamma_2 = 2.17 \pm 0.03_{\rm stat} \pm 0.42_{\rm syst}$ derived in our analysis. With such a large extension in the Galactic plane, several sources could contribute to the GeV signal: the PWN powered by PSR J1813$-$1749 thought to emit at TeV energies as seen by H.E.S.S. and HAWC \citep{2017ApJ...843...40A}, the SNR G12.82$-$0.02 whose contribution to the TeV signal was explored by \cite{2007A&A...470..249F} and the giant star-forming region (SFR) W33 that comprises a region of 15' at a distance of 2.4~kpc \citep{2013A&A...553A.117I}. This last hypothesis was considered by \cite{2018ApJ...859...69A}, showing that the energetics, extended morphology and spectrum of the GeV emission are similar to those of the other gamma-ray detected SFR, the Cygnus Cocoon. To firmly establish the presence of protons radiating at gamma-ray energies, such a complex region definitively is worth an individual analysis above 1 GeV to constrain the morphology and a spectral analysis over a larger energy range to model the broad-band emission.\\
Finally, several sources detected by our analysis are completely unassociated and follow-up observations at TeV energies and X-rays would be needed to constrain their nature. They all present values of $\Gamma_2$ much softer than those of the identified SNRs discussed in Section~\ref{subsection:snr}. Similarly, the values of $\Gamma_2 - \Gamma_1$ obtained in our analysis is much larger ($\ge 2.96$) than those of the identified SNRs and dense molecular cloud regions. This tends to indicate that these sources are not associated with SNR shock acceleration.
% In this section we discuss the cases of 4FGL J1601.3-5224 (G329.7+00.4), 4FGLJ1633.0-4746e (HESS J1632-478), 4FGLJ1934.3+1859 (SNR G054.4-00.3), 4FGL J1931.1+1656 (Unid but G52.37−0.70 from THOR), HESSJ1813 (4FGLJ1813.1-1737e, discussed by Araya), 

%Add a line saying that these sources remain unidentified:
%4FGLJ0340.4+5302 (potential pulsar ???), 4FGLJ0709.1-1034 (Unid ?), 4FGLJ0844.1-4330 (Unid), 4FGLJ1547.5-5130 (Unid), 
%4FGLJ1608.8-4803 (Unid), 4FGLJ1626.6-4251 (Unid), 4FGLJ1742.8-2246 (Unid), 4FGLJ1855.2+0456 (Unid), 4FGLJ1856.2+0749 (NVSS J185612+07534), 4FGLJ2108.0+5155 (Unid)

\begin{figure*}[ht]
\centering
\subfigure{
\begin{tabular}{ll}
\centering
\includegraphics[width=0.44\textwidth]{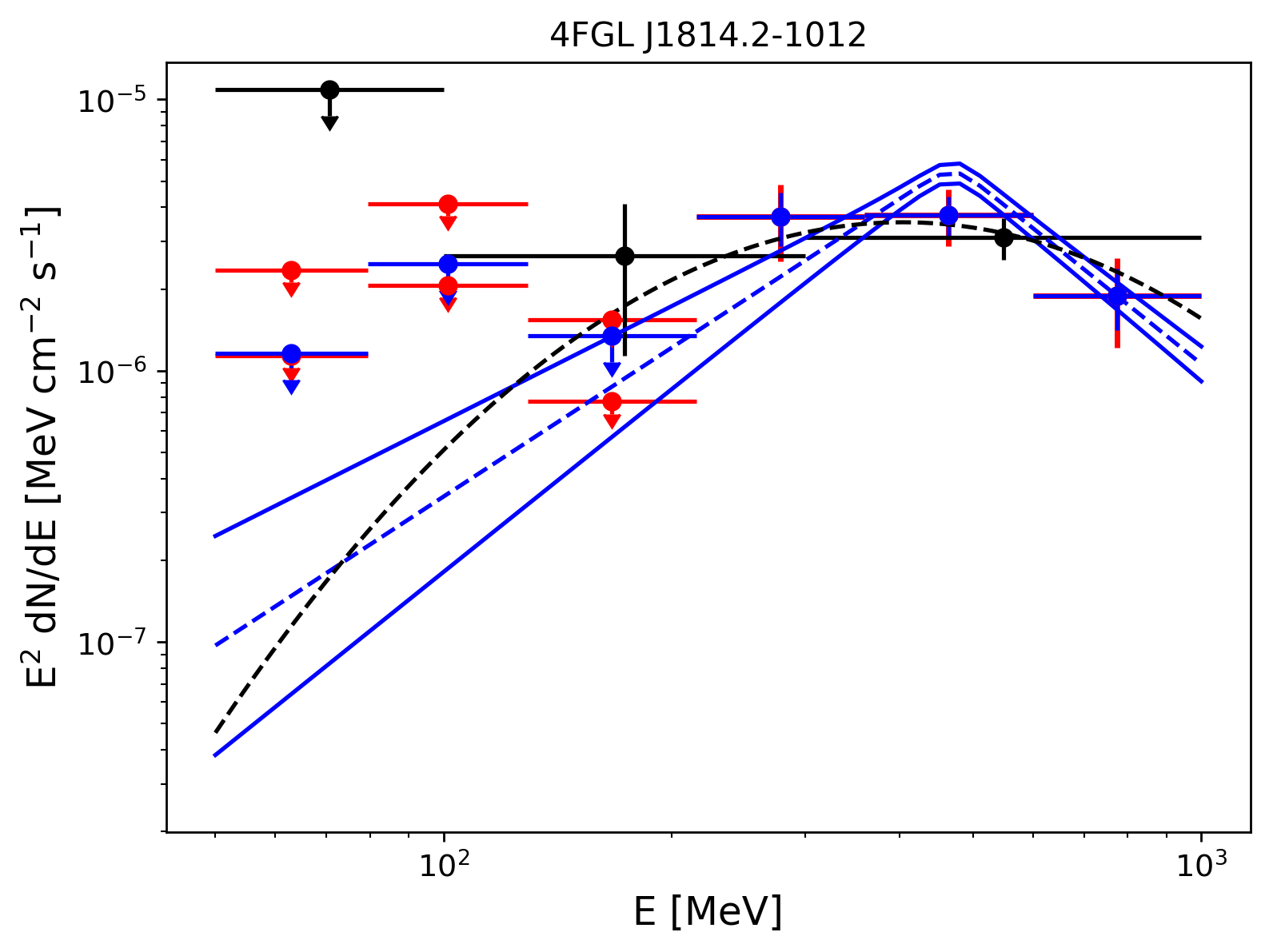}
\includegraphics[width=0.44\textwidth]{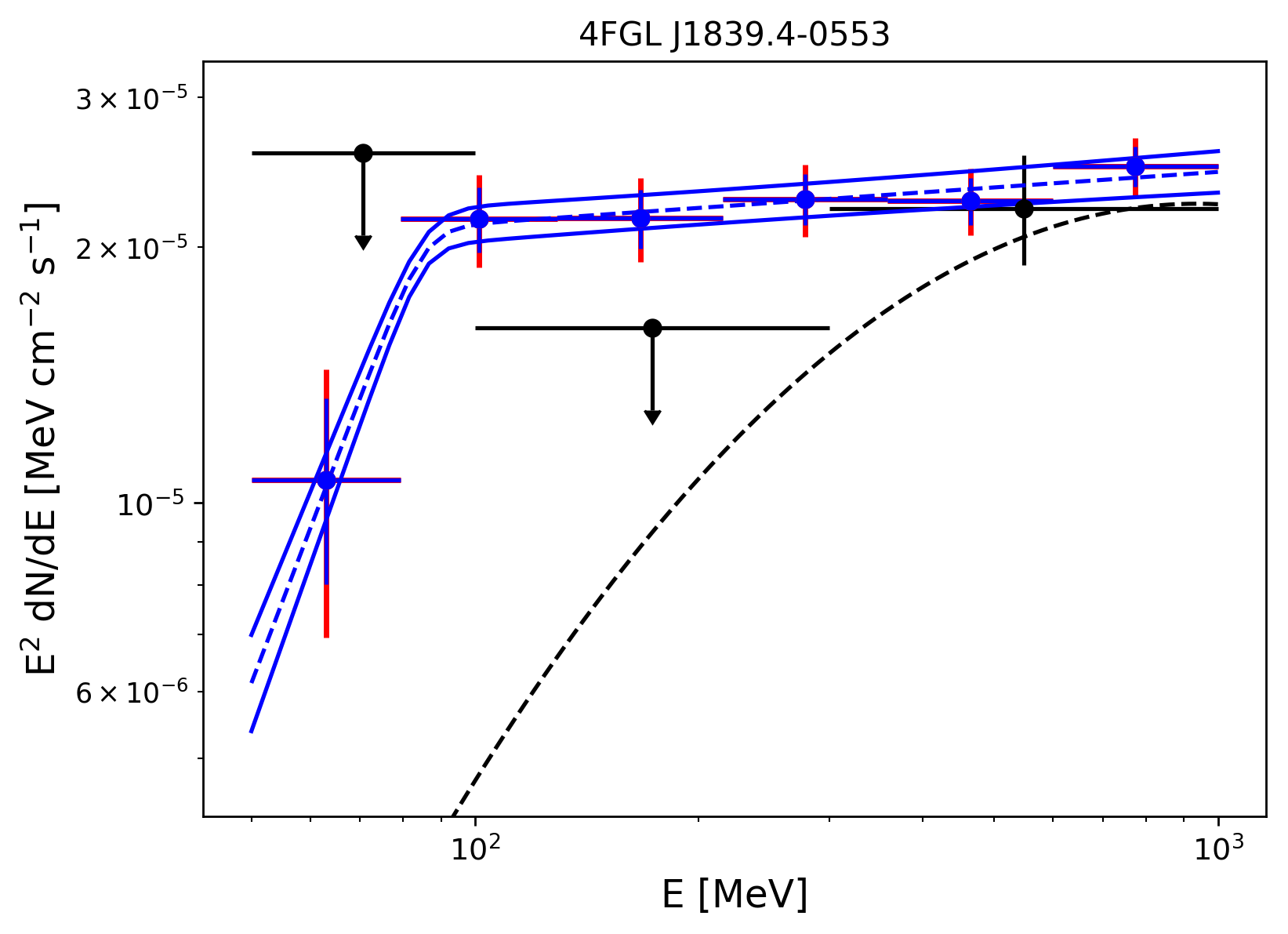}
\end{tabular}
}
\subfigure{
\begin{tabular}{ll}
\centering
\includegraphics[width=0.44\textwidth]{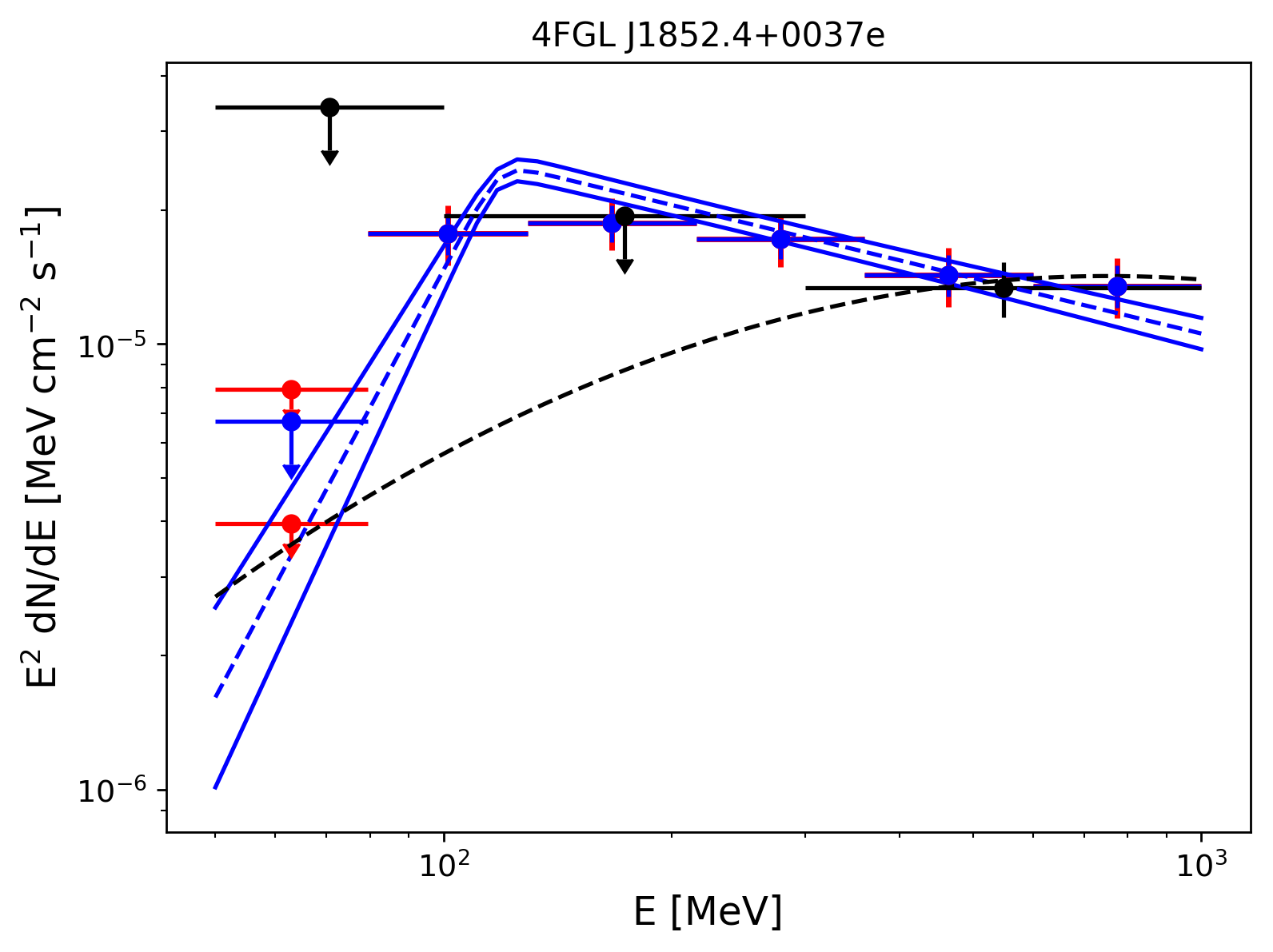}
\includegraphics[width=0.44\textwidth]{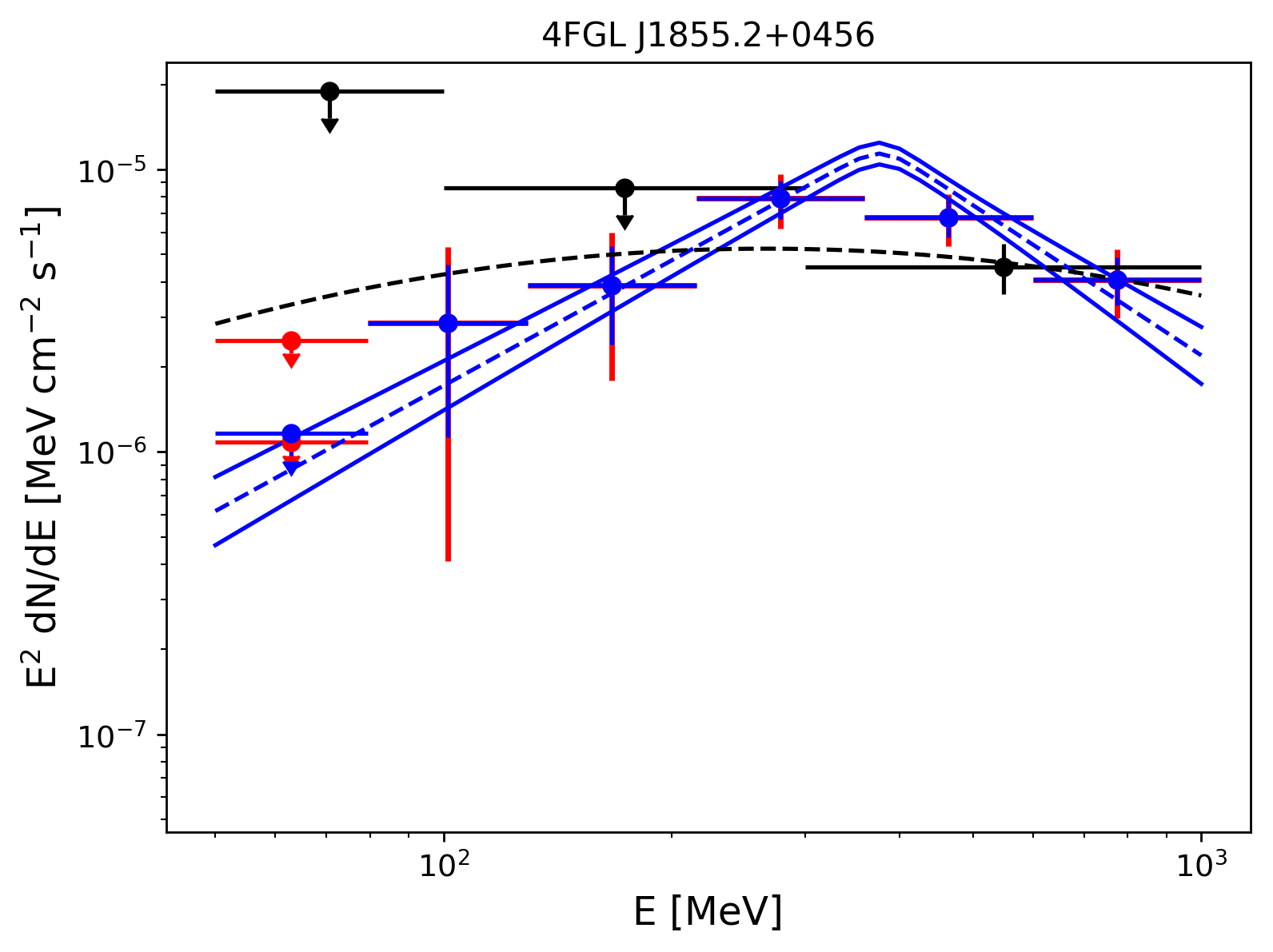}
\end{tabular}
}
\subfigure{
\begin{tabular}{ll}
\centering
\includegraphics[width=0.44\textwidth]{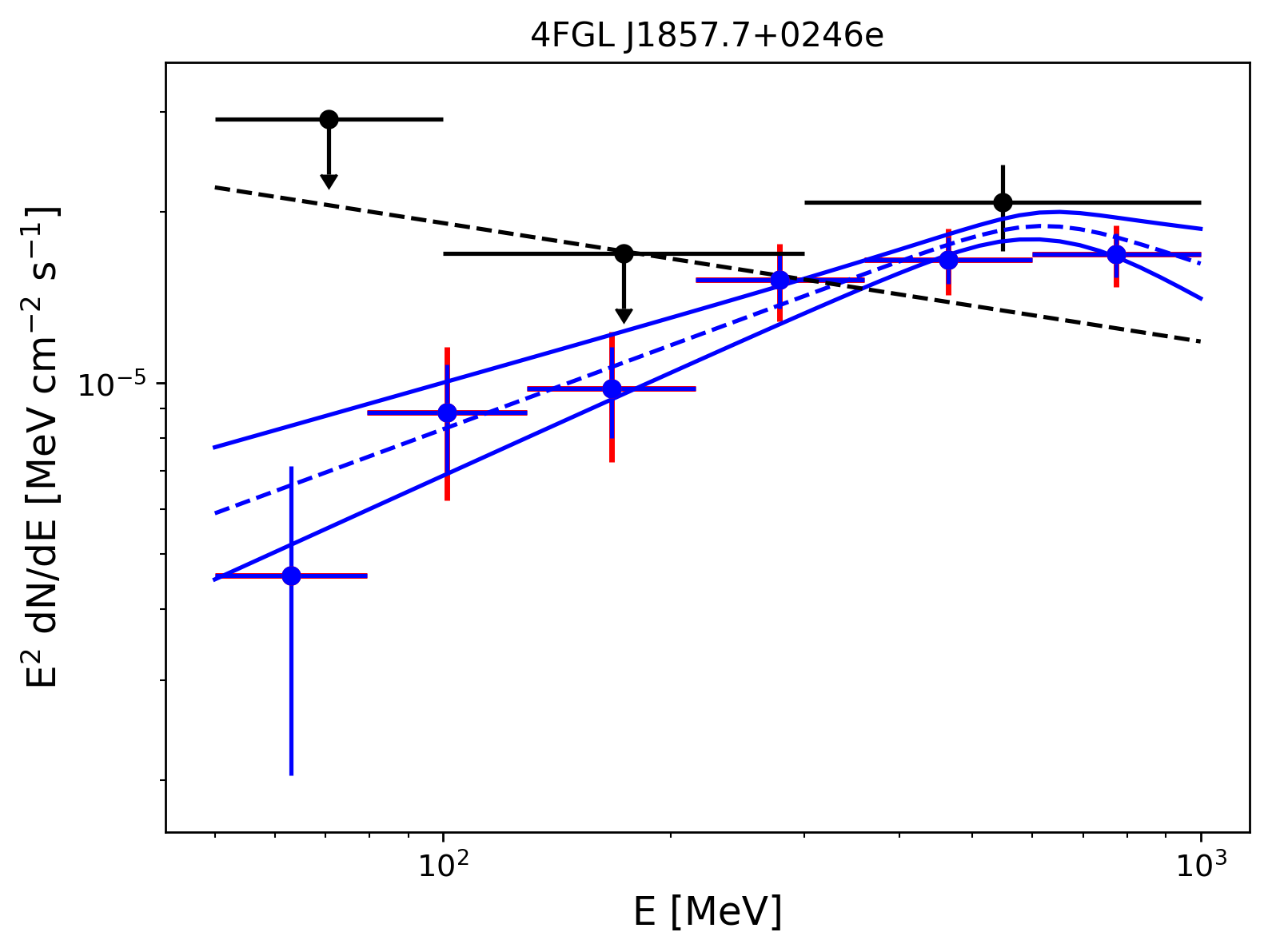}
\includegraphics[width=0.44\textwidth]{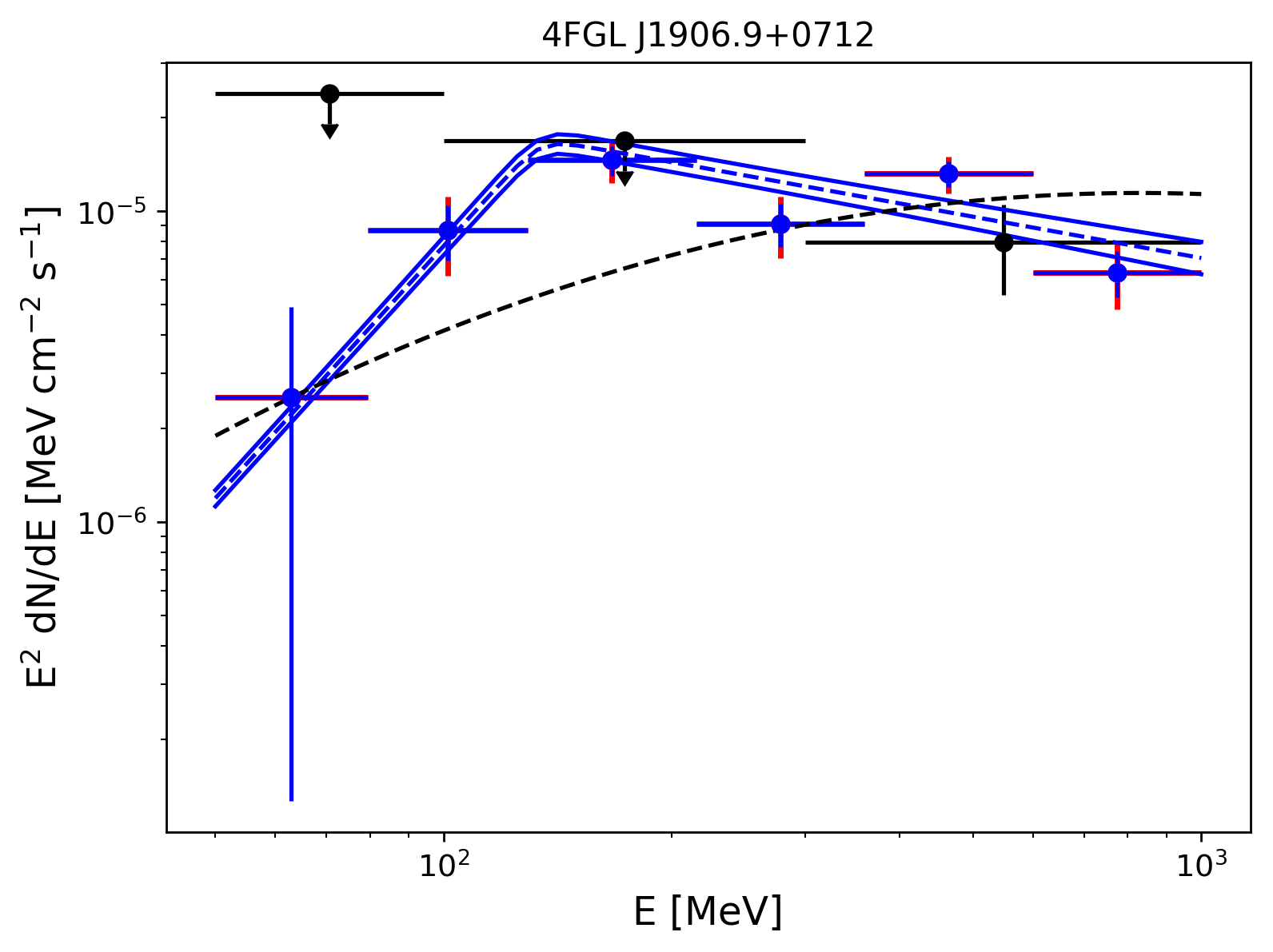}
\end{tabular}
}
\caption{\label{fig:sed7}LAT Spectral energy distributions of 4FGL J1814.2$-$1012 (top left), 4FGL J1839.4$-$0553 (top right), 4FGL J1852.4+0037e (middle left), 4FGL J1855.2+0456 (middle right), 4FGL J1857.7+0246e (bottom left), 4FGL J1906.9+0712 (bottom right) with the same conventions used in Figure~\ref{fig:snrsed}.}
\end{figure*}

\section{Summary}
Using 8 years of Pass~8 LAT data between 50 MeV and 1 GeV, we have analyzed 311 4FGL sources located within $5^{\circ}$ from the Galactic plane and detected 77 sources with significant spectral breaks. We carried out a thorough study of the systematics associated with the diffuse Galactic background and with the effective area for each of them and we confirmed the spectral break for 56 of them. With 13 SNRs identified within this sample of 56 sources, SNRs are the dominant class of sources showing significant breaks at low energy. Only five binaries are included in the sample of 311 sources analyzed but four of them show a significant break at low energies. This seems to indicate that binaries could also have a significant contribution. The spectral characteristics were also evaluated for these 56 sources. The break energy of the sources ranges uniformly between 100 MeV and 550 MeV. However, a clear pattern is detected in the spectral index $\Gamma_2$ of the sources which tends to center at $2.3$ for the population of 13 identified SNRs. Similarly, the value of $\Gamma_2 - \Gamma_1$ tends to center at $\sim 1$ for the same population of sources. This provides an interesting way to constrain the nature of the radiating particles. Our analysis also provides three interesting new proton accelerator candidates: 4FGL J1931.1+1656 is coincident with the SNR candidate G52.37$-$0.70 detected in a recent THOR+VGPS analysis, the extended source 4FGL~J1633.0$-$4746e overlapping the TeV PWN candidate HESS J1632$-$478 and the unidentified source HESS J1634$-$472, and the extended source 4FGL~J1813.1$-$1737e coincident with the compact TeV PWN candidate HESS J1813$-$178 and the star-forming region W33. The current and future observations of the LAT are thus crucial to probe the spectral characteristics of a source at low energy, providing excellent targets of proton acceleration for current and future Cherenkov telescopes such as CTA. 
\label{sec:summary}

% Table 3: Added srcs
%\input{Table_addsrcs_edisp-3_addsrcs.tex}

\acknowledgments
The \emph{Fermi} LAT Collaboration acknowledges generous ongoing support from a number of agencies and institutes that have supported both the development and the operation of the LAT as well as scientific data analysis. These include the National Aeronautics and Space Administration and the Department of Energy in the United States, the Commissariat à l'Energie Atomique and the Centre National de la Recherche Scientifique / Institut National de Physique Nucléaire et de Physique des Particules in France, the Agenzia Spaziale Italiana and the Istituto Nazionale di Fisica Nucleare in Italy, the Ministry of Education, Culture, Sports, Science and Technology (MEXT), High Energy Accelerator Research Organization (KEK) and Japan Aerospace Exploration Agency (JAXA) in Japan, and the K. A. Wallenberg Foundation, the Swedish Research Council and the Swedish National Space Board in Sweden.
Additional support for science analysis during the operations phase is gratefully acknowledged from the Istituto Nazionale di Astrofisica in Italy and the Centre National d'Etudes Spatiales in France.
Work at NRL is supported by NASA.
MLG acknowledges support from Agence Nationale de la Recherche (grant ANR- 17-CE31-0014).

\appendix

\section{The pion-decay bump signature}
\label{appen:pion}

\begin{figure*}[ht]
\begin{center}
\begin{tabular}{ll}
\includegraphics[width=0.49\textwidth]{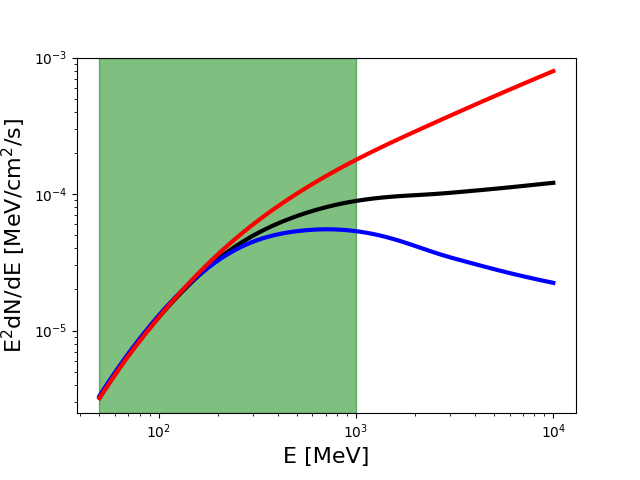}
\includegraphics[width=0.49\textwidth]{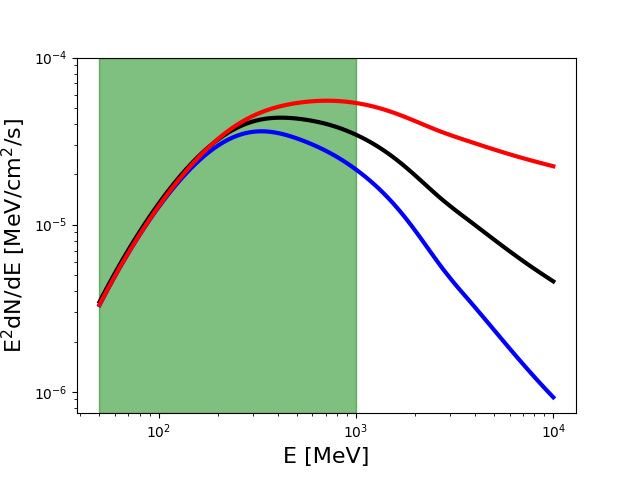}
\end{tabular}
\end{center}
\caption{
\label{fig:pion1} Left: Gamma-ray spectra produced by a power-law distribution of protons with spectral index of 1.5 (red), 2.0 (black), 2.5 (blue) as predicted by $naima$ \citep{naima}. Right: Same figure assuming that an energy break is present in the particle distribution at 1 GeV. The energy interval analyzed in this work is defined by the green area.
}
\end{figure*}

\begin{figure*}[ht]
\begin{center}
\begin{tabular}{ll}
\includegraphics[width=0.49\textwidth]{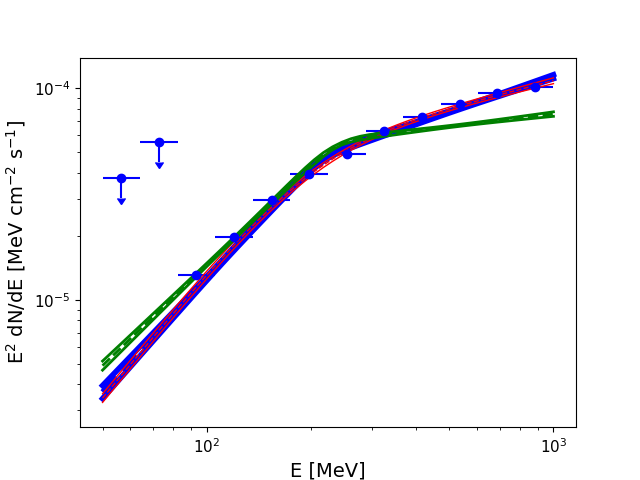}
\includegraphics[width=0.49\textwidth]{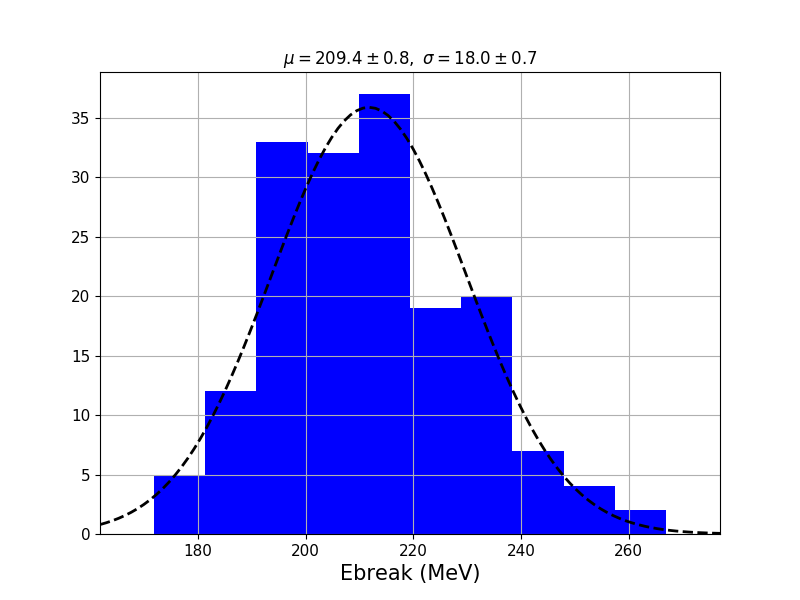}
\end{tabular}
\end{center}
\caption{
\label{fig:gtobssim} Left: Gamma-ray spectrum derived using our analysis pipeline from one simulation of a power-law distribution of protons with spectral index of 2.0 predicted by $naima$ \citep{naima}. The fit assuming a smooth broken power-law model (see Equation \ref{eq:sbpl}) with $\alpha = 0.1$ is presented with the blue curve. The fit assuming $\alpha = 0.5$ is presented with the red curve. The best fit of a similar simulation assuming a proton injection index of 2.5 is presented with the green line. Right: Distribution of the 200 values of energy break fitted for a proton injection spectral index of 2.0. The black line represents our best Gaussian fit.
}
\end{figure*}

As already discussed in the main text, when accelerated protons interact with the interstellar matter, they produce neutral pions which in turn decay into gamma rays. This will create a characteristic signature at low energy in the gamma-ray spectrum called the ``pion-decay bump signature". To better understand how this signature is characterized in our energy interval of interest (50 MeV -- 1 GeV), we have used the python package $naima$ \citep{naima} to derive the gamma-ray emission produced by proton-proton interaction. To do so, $naima$ uses an implementation of the analytical parametrizations of the energy spectra and production rates of gamma rays from \cite{2014PhRvD..90l3014K}, which is accurate within $20$\%. The inclusive $\pi^0$ production cross section is included as a combination of the experimental data cross sections at low energies, the Geant 4.10.0 cross section at intermediate energies and at higher energies the hadronic model Pythia 8.18 as default. We applied $naima$ to three different power-law distributions of protons with spectral index $\Gamma_1$ varying between 1.5 and 2.5. The results presented in Figure~\ref{fig:pion1} (Left) are in perfect agreement with those published in \cite{2013Sci...339..807A} and show that a very steep spectrum is expected below the break energy at $\sim200$~MeV. This Figure also highlights that the pion-decay bump signature might be more difficult to detect for a hard proton distribution (red curve) than for steep injection spectra. This might in turn increase the systematic errors on the derived break energy. Finally, this Figure demonstrates that the restricted energy interval of our analysis does not allow constraining the spectral index of the parent distribution since the gamma-ray spectra trace the energy distribution of parent protons at energies greater than 1 GeV. The upper bound of our energy interval was chosen since middle-aged SNRs commonly exhibit a high energy spectral break at around 1–10 GeV \citep[see the case of W28 with a break at 1 GeV reported by ][]{2010ApJ...718..348A} and a simple broken power-law model would not apply anymore above 1 GeV. To test for this effect, we applied $naima$ to the same power-law distributions of protons adding a break at 1 GeV in their distributions. We assumed that $\Gamma_2 = \Gamma_1 + 1$. Figure~\ref{fig:pion1} (Right) demonstrates that the break energy of the gamma-ray emission detected in our energy interval is not affected. However, this break significantly impacts the spectral index derived assuming a simple broken power-law model. In a second step, we used the gamma-ray emission obtained with $naima$ assuming a power-law distribution of protons with $\Gamma_1 = 2.0$ and $2.5$ to produce 200 Fermi simulated data files for each index over 8 years using the $gtobssim$ tool included in the LAT $fermitools$. We then analyzed these simulations using the $fermitools$ following the same procedure as with the real data, assuming a smooth broken power-law spectral model with $\alpha = 0.1$ (see Equation \ref{eq:sbpl}), except that the SED was produced for 12 energy bins instead of 6 to reflect the high statistics of our simulations (which mimics the flux of IC 443). We did not introduce any diffuse background in our simulations to clearly show how a proton spectrum will be reconstructed at low energies with the LAT in the absence of systematic errors. Figure~\ref{fig:gtobssim} presents the gamma-ray spectrum derived for one of these simulations demonstrating that the spectral index derived above the break does not trace the parent proton distribution. The error bars are extremely small due to the high flux of the simulated source and the absence of diffuse background. This Figure also shows that the smooth broken power-law model used in our analysis with $\alpha = 0.1$ reproduces the gamma-ray spectrum well.  A smoothness parameter $\alpha = 0.5$ was also tested but it does not significantly improves the likelihood of the fit. Finally, one can see that the injection spectral index does not seem to impact the break energy and the index $\Gamma_1$ of our smooth broken power-law fit. The only parameter affected is the index $\Gamma_2$. To confirm this trend, we plotted the distributions of the 200 reconstructed values of the break energy, $\Gamma_1$ and $\Gamma_2$ for each injection spectral index, and fitted a Gaussian on each distribution as can be seen in Figure~\ref{fig:gtobssim} (Right) for the case of the break energy. The results are presented in Table~\ref{tab:gauss} confirming that the only parameter affected by the different injection spectral index is $\Gamma_2$. This study demonstrates that no steep spectrum is predicted for a standard injection spectral index. The only way to produce the steep spectra observed for some of the candidates detected in our analysis would be to include an energy break at (or below) 1 GeV in the injection proton spectrum as shown in Figure~\ref{fig:pion1} (Right).

 \begin{table}[ht!]

\caption{Results of the Gaussian fits of the 200 $naima$ simulations assuming a proton injection spectral index of 2.0 and 2.5.}
\label{tab:gauss}
\begin{tabular}{ c c c}

\multicolumn{3}{c}{} \\
\hline
  &  Proton index = 2.0 & Proton index = 2.5 \\
 \hline
Energy break & $209$ / $18 $ &  $212$ / $15$ \\
$\Gamma_1$ &  $0.24$ / $0.09$ & $0.29$ / $0.09$ \\
$\Gamma_2$ &  $1.43$ / $0.04$ & $1.74$ / $0.04$\\
\hline

\end{tabular}
\tablecomments{The first column indicates the parameter fitted (energy break, $\Gamma_1$ and $\Gamma_2$) , the second and third column presents the mean and sigma of the distribution obtained for a proton injection of 2.0 and 2.5, respectively.}
\end{table}

\section{List of the 311 Galactic plane sources analyzed}
\label{appen:sources}
Table~\ref{tab:candidates} provides the list of all candidates analyzed. Columns 2 and 3 provide the Galactic longitude and latitude of the 311 candidates. Columns 4, 5 and 6 give the curvature
  significance, the significance between 300 MeV and 1 GeV and the source class reported in the 4FGL catalog. Columns 7, 8, 9, 10
  provide the values obtained in our analysis concerning the TS of each source, the improvement of the log-normal representation with respect to the
  power-law model ${\rm TS_{LP}}$ as defined in Section~\ref{subsection:ROI}, the improvement of the smooth broken power-law representation with respect to the power-law model ${\rm TS_{SBPL}}$ and the improvement of the smooth broken power-law representation when fixing $\Gamma_2 = 2$ called ${\rm TS_{SBPL2}}$.\\

% Table 1 : List of candidates
\begin{deluxetable*}{cccccccccc}
\tabletypesize{\footnotesize}
\tablecaption{List of selected Galactic plane candidates\label{tab:candidates}}
%\label{tab:candidates}
\tablehead{
\colhead{4FGL Name} &
\colhead{GLON} &
\colhead{GLAT} &
\colhead{4FGL SigCurv} &
\colhead{4FGL $\sqrt{\rm TS}$} &
\colhead{4FGL Class} &
\colhead{TS} &
\colhead{${\rm TS_{LP}}$} &
\colhead{${\rm TS_{SBPL}}$} &
\colhead{${\rm TS_{SBPL2}}$} \\  
\colhead{} &
\colhead{($^{\circ}$)} &
\colhead{($^{\circ}$)} &
\colhead{} &
\colhead{(0.3 -- 1 GeV)} &
\colhead{} &
\colhead{} &
\colhead{} &
\colhead{} &
\colhead{} \\
}
\startdata
4FGL J0034.6+6438 & 121.13 & 1.83 & 2.6 & 3.9 &  & 7.6 &  &  &  \\ 
4FGL J0039.1+6257 & 121.54 & 0.12 & 7.0 & 5.2 &  & 43.8 & 7.0 &  &  \\ 
4FGL J0129.0+6312 & 127.16 & 0.65 & 3.1 & 5.7 & spp & 55.9 & 5.1 &  &  \\ 
4FGL J0142.5+6650 & 127.93 & 4.46 & 3.4 & 3.6 &  & 6.4 &  &  &  \\ 
4FGL J0144.3+5959 & 129.51 & $-$2.20 & 0.6 & 3.4 &  & 4.9 &  &  &  \\ 
4FGL J0211.5+6219 & 132.09 & 0.89 & 1.1 & 3.5 &  & 0.3 &  &  &  \\ 
4FGL J0221.4+6241e & 133.05 & 1.60 & 5.7 & 8.8 & SNR & 75.9 & 0.5 &  &  \\ 
$\star$4FGL J0222.4+6156e & 133.42 & 0.94 & 14.3 & 30.7 & snr & 1307.1 & 27.2 & 34.8 & 32.4 \\ 
4FGL J0235.3+5650 & 136.82 & $-$3.19 & 4.1 & 7.6 &  & 77.1 & 3.7 &  &  \\ 
$\star$4FGL J0240.5+6113 & 135.68 & 1.09 & 28.2 & 107.3 & HMB & 39495.7 & 127.8 & 127.3 & 100.9 \\ 
4FGL J0302.7+5717 & 139.97 & $-$1.17 & 0.8 & 3.3 &  & 20.3 &  &  &  \\ 
4FGL J0328.5+6114 & 140.71 & 3.92 & 1.4 & 3.9 &  & 12.3 &  &  &  \\ 
4FGL J0330.1+5038 & 146.91 & $-$4.70 & 4.7 & 6.1 &  & 34.8 & 2.7 &  &  \\ 
$\star$4FGL J0330.7+5845 & 142.35 & 2.02 & 4.7 & 6.2 &  & 43.2 & 16.1 & 18.0 & 14.2 \\ 
$\star$4FGL J0340.4+5302 & 146.79 & $-$1.82 & 12.3 & 23.4 &  & 1329.3 & 149.4 & 143.6 & 139.0 \\ 
$\star$4FGL J0426.5+5434 & 150.88 & 3.82 & 9.8 & 18.1 &  & 704.1 & 19.6 & 25.2 & 18.9 \\ 
4FGL J0452.9+5117 & 155.99 & 4.63 & 2.8 & 5.6 &  & 26.4 & 8.0 &  &  \\ 
$\star$4FGL J0500.3+4639e & 160.37 & 2.69 & 5.8 & 12.6 & SNR & 253.0 & 15.2 & 19.8 & 17.1 \\ 
4FGL J0532.6+3358 & 174.26 & 0.36 & 1.3 & 3.5 &  & 11.2 &  &  &  \\ 
4FGL J0533.9+2838 & 178.89 & $-$2.33 & 0.3 & 3.4 & spp & 24.7 &  &  &  \\ 
$\star$4FGL J0540.3+2756e & 180.24 & $-$1.50 & 4.5 & 10.0 & SNR & 202.0 & 10.8 & 12.4 & 9.2 \\ 
4FGL J0608.8+2034 & 189.87 & 0.35 & 2.9 & 3.8 & unk & 15.2 &  &  &  \\ 
4FGL J0609.0+2136 & 188.99 & 0.89 & 4.2 & 3.5 &  & 33.3 & 10.6 & 9.3 & 7.5 \\ 
$\star$4FGL J0609.0+2006 & 190.31 & 0.17 & 4.9 & 3.9 &  & 72.4 & 12.4 & 15.1 & 12.9 \\ 
4FGL J0612.6+1520 & 194.89 & $-$1.37 & 3.9 & 4.4 &  & 18.7 &  &  &  \\ 
$\star$4FGL J0617.2+2234e & 189.05 & 3.03 & 20.8 & 17.0 & SNR & 5456.4 & 84.5 & 99.0 & 79.8 \\ 
4FGL J0618.7+1211 & 198.38 & $-$1.56 & 2.9 & 4.7 &  & 65.4 & 14.6 & 16.3 & 15.0 \\ 
$\star$4FGL J0620.4+1445 & 196.30 & 0.00 & 1.5 & 3.6 &  & 33.5 & 12.0 & 16.2 & 10.4 \\ 
$\star$4FGL J0634.2+0436e & 206.87 & $-$1.70 & 8.4 & 15.5 & snr & 370.6 & 13.1 & 19.7 & 19.4 \\ 
$\star$4FGL J0639.4+0655e & 205.40 & 0.50 & 7.8 & 10.2 & SNR & 170.7 & 44.6 & 42.4 & 39.2 \\ 
4FGL J0642.4+1048 & 202.28 & 2.94 & 2.7 & 3.5 &  & 12.7 &  &  &  \\ 
4FGL J0647.7+0031 & 212.05 & $-$0.57 & 1.1 & 6.5 & spp & 48.7 & 8.5 &  &  \\ 
4FGL J0705.8$-$0004 & 214.64 & 3.17 & 3.1 & 3.2 &  & 15.8 &  &  &  \\ 
$\star$4FGL J0709.1$-$1034 & 224.35 & $-$0.92 & 4.3 & 7.8 &  & 80.8 & 17.0 & 21.9 & 14.9 \\ 
4FGL J0722.7$-$2309 & 237.01 & $-$3.88 & 3.5 & 4.4 &  & 62.2 & 17.3 & 21.5 & 6.9 \\ 
4FGL J0731.5$-$1910 & 234.48 & $-$0.19 & 2.8 & 5.0 &  & 53.8 & 14.4 & 18.4 & 8.0 \\ 
4FGL J0744.0$-$2525 & 241.35 & $-$0.74 & 8.0 & 13.7 &  & 200.6 & 2.4 &  &  \\ 
4FGL J0752.0$-$2931 & 245.78 & $-$1.28 & 4.0 & 5.4 &  & 36.2 & 0.6 &  &  \\ 
4FGL J0822.1$-$4253e & 260.31 & $-$3.37 & 5.9 & 16.7 & SNR & 468.2 & 3.9 &  &  \\ 
4FGL J0833.3$-$4342 & 262.17 & $-$2.20 & 0.0 & 3.7 & spp & 78.7 & 8.4 &  &  \\ 
$\star$4FGL J0844.1$-$4330 & 263.19 & $-$0.53 & 3.2 & 3.4 &  & 65.6 & 18.2 & 23.7 & 12.3 \\ 
4FGL J0844.9$-$4117 & 261.55 & 0.97 & 3.2 & 7.0 &  & 63.3 & 6.7 &  &  \\ 
4FGL J0848.8$-$4328 & 263.69 & 0.16 & 1.8 & 6.8 &  & 87.9 & 6.6 &  &  \\ 
$\star$4FGL J0850.8$-$4239 & 263.30 & 0.96 & 5.0 & 6.9 &  & 83.6 & 19.5 & 28.5 & 19.8 \\ 
4FGL J0851.9$-$4620e & 266.26 & $-$1.23 & 1.1 & 12.0 & SNR & 259.8 & 4.4 &  &  \\ 
4FGL J0853.1$-$4407 & 264.70 & 0.35 & 5.0 & 5.8 &  & 37.8 & 5.2 &  &  \\ 
4FGL J0853.6$-$4306 & 263.98 & 1.07 & 5.6 & 7.3 &  & 43.8 & 0.4 &  &  \\ 
4FGL J0854.8$-$4504 & 265.61 & $-$0.03 & 8.4 & 5.8 &  & 90.6 & 6.0 &  &  \\ 
4FGL J0857.7$-$4507 & 265.98 & 0.33 & 2.1 & 5.8 &  & 23.8 &  &  &  \\ 
4FGL J0859.2$-$4729 & 267.94 & $-$1.03 & 3.9 & 7.0 &  & 135.0 & 6.9 &  &  \\ 
4FGL J0859.3$-$4342 & 265.10 & 1.46 & 5.9 & 11.5 &  & 105.2 & 6.8 &  &  \\ 
4FGL J0900.2$-$4608 & 267.03 & $-$0.01 & 3.1 & 4.4 &  & 43.6 & 9.7 & 11.9 & 6.5 \\ 
$\star$4FGL J0904.7$-$4908c & 269.78 & $-$1.44 & 3.7 & 5.4 & unk & 51.8 & 15.9 & 16.0 & 9.5 \\ 
4FGL J0911.6$-$4738 & 269.46 & 0.43 & 0.9 & 5.4 &  & 72.1 & 11.2 & 15.2 & 8.6 \\ 
4FGL J0917.9$-$4755 & 270.39 & 0.99 & 4.2 & 6.2 &  & 56.8 & 3.7 &  &  \\ 
4FGL J0924.1$-$5202 & 274.01 & $-$1.21 & 1.5 & 4.9 &  & 53.0 & 22.7 & 25.5 & 14.5 \\ 
$\star$4FGL J1008.1$-$5706c & 282.13 & $-$0.98 & 3.3 & 8.6 & unk & 118.4 & 17.8 & 20.5 & 19.7 \\ 
4FGL J1015.5$-$6030 & 284.87 & $-$3.22 & 2.4 & 8.8 &  & 75.3 & 5.1 &  &  \\ 
$\star$4FGL J1018.9$-$5856 & 284.35 & $-$1.68 & 21.2 & 27.7 & HMB & 1800.8 & 10.7 & 22.9 & 20.5 \\ 
4FGL J1020.4$-$5314 & 281.41 & 3.20 & 3.1 & 4.3 &  & 23.0 &  &  &  \\ 
4FGL J1036.3$-$5833e & 286.08 & $-$0.18 & 5.0 & 8.0 &  & 57.9 & 8.0 &  &  \\ 
4FGL J1037.8$-$5810 & 286.06 & 0.24 & 4.2 & 6.5 &  & 55.0 & 0.7 &  &  \\ 
$\star$4FGL J1045.1$-$5940 & 287.60 & $-$0.62 & 12.5 & 26.4 & BIN & 1011.6 & 15.5 & 18.9 & 16.7 \\ 
4FGL J1045.7$-$6414 & 289.79 & $-$4.63 & 1.9 & 3.8 & unk & 10.3 &  &  &  \\ 
4FGL J1046.7$-$6010 & 288.01 & $-$0.97 & 6.2 & 5.8 &  & 55.7 & 2.8 &  &  \\ 
4FGL J1048.5$-$5923 & 287.85 & $-$0.17 & 7.6 & 4.7 &  & 60.1 & 3.4 &  &  \\ 
4FGL J1054.7$-$6008 & 288.88 & $-$0.49 & 1.6 & 3.5 &  & 21.4 &  &  &  \\ 
4FGL J1102.0$-$6054 & 290.02 & $-$0.81 & 6.8 & 7.0 & spp & 77.2 & 4.1 &  &  \\ 
4FGL J1109.4$-$6115e & 290.98 & $-$0.78 & 1.9 & 3.1 &  & 38.9 & 1.9 &  &  \\ 
4FGL J1127.9$-$6158 & 293.30 & $-$0.68 & 4.7 & 4.5 &  & 55.0 & 8.3 &  &  \\ 
4FGL J1139.2$-$6247 & 294.79 & $-$1.07 & 5.2 & 5.9 &  & 27.4 & 3.9 &  &  \\ 
4FGL J1151.4$-$6248 & 296.14 & $-$0.73 & 1.4 & 3.6 & spp & 18.1 &  &  &  \\ 
4FGL J1152.6$-$6207 & 296.12 & $-$0.03 & 2.9 & 3.2 & spp & 15.2 &  &  &  \\ 
4FGL J1154.5$-$5952 & 295.84 & 2.22 & 3.8 & 4.8 &  & 29.4 & 4.9 &  &  \\ 
4FGL J1155.6$-$6547 & 297.25 & $-$3.54 & 2.6 & 4.9 & unk & 25.3 & 8.6 &  &  \\ 
4FGL J1157.7$-$6327 & 296.98 & $-$1.21 & 4.1 & 4.7 &  & 29.5 & 6.5 &  &  \\ 
4FGL J1202.9$-$5717 & 296.40 & 4.96 & 3.3 & 4.4 &  & 8.4 &  &  &  \\ 
4FGL J1203.4$-$6145 & 297.29 & 0.59 & 3.7 & 3.0 &  & 34.2 & 4.3 &  &  \\ 
4FGL J1204.3$-$6111 & 297.28 & 1.17 & 3.8 & 4.3 &  & 21.5 &  &  &  \\ 
4FGL J1205.1$-$5951 & 297.14 & 2.50 & 0.7 & 3.4 &  & 6.7 &  &  &  \\ 
4FGL J1210.4$-$6100 & 297.99 & 1.47 & 1.3 & 3.0 &  & 8.2 &  &  &  \\ 
4FGL J1210.7$-$6005 & 297.87 & 2.38 & 1.9 & 3.1 &  & 14.8 &  &  &  \\ 
4FGL J1213.3$-$6240e & 298.58 & $-$0.13 & 5.3 & 11.0 & snr & 182.4 & 5.8 &  &  \\ 
4FGL J1214.7$-$5858 & 298.21 & 3.56 & 1.3 & 4.2 &  & 4.8 &  &  &  \\ 
4FGL J1216.8$-$5955 & 298.61 & 2.66 & 4.3 & 8.3 &  & 58.1 & 6.1 &  &  \\ 
4FGL J1231.6$-$6511 & 300.86 & $-$2.40 & 5.8 & 8.1 &  & 44.9 & 4.4 &  &  \\ 
4FGL J1244.3$-$6233 & 302.11 & 0.31 & 2.6 & 4.3 &  & 70.3 & 14.7 & 27.7 & 9.1 \\ 
4FGL J1303.0$-$6312e & 304.23 & $-$0.36 & 1.3 & 3.6 & PWN & 57.0 & 6.1 &  &  \\ 
4FGL J1305.5$-$6241 & 304.55 & 0.13 & 1.1 & 3.3 & snr & 44.1 & 6.3 &  &  \\ 
4FGL J1306.3$-$6043 & 304.76 & 2.10 & 11.4 & 6.8 &  & 63.7 & 1.9 &  &  \\ 
4FGL J1309.1$-$6223 & 304.98 & 0.41 & 5.0 & 4.6 &  & 23.6 &  &  &  \\ 
4FGL J1315.9$-$6243 & 305.74 & 0.01 & 4.3 & 3.9 &  & 24.8 &  &  &  \\ 
4FGL J1317.5$-$6316 & 305.86 & $-$0.56 & 6.3 & 4.6 &  & 29.7 & 1.3 &  &  \\ 
4FGL J1320.3$-$6410 & 306.08 & $-$1.48 & 3.7 & 3.0 &  & 19.2 &  &  &  \\ 
4FGL J1328.4$-$6231 & 307.19 & 0.04 & 4.1 & 6.3 &  & 63.8 & 5.2 &  &  \\ 
4FGL J1329.9$-$6108 & 307.56 & 1.38 & 6.9 & 5.8 &  & 44.2 & 3.6 &  &  \\ 
4FGL J1349.1$-$5829 & 310.43 & 3.54 & 1.6 & 4.4 &  & 24.6 &  &  &  \\ 
$\star$4FGL J1351.6$-$6142 & 310.00 & 0.34 & 6.9 & 8.6 &  & 136.4 & 13.6 & 20.9 & 14.6 \\ 
4FGL J1358.2$-$6210 & 310.64 & $-$0.30 & 5.3 & 5.3 & spp & 53.2 & 1.6 &  &  \\ 
$\star$4FGL J1358.3$-$6026 & 311.10 & 1.36 & 6.5 & 7.6 &  & 120.8 & 16.3 & 24.4 & 13.6 \\ 
4FGL J1404.4$-$6159 & 311.40 & $-$0.32 & 5.9 & 3.7 &  & 31.5 & 3.1 &  &  \\ 
$\star$4FGL J1405.1$-$6119 & 311.66 & 0.28 & 9.2 & 7.9 &  & 204.8 & 9.8 & 29.1 & 20.8 \\ 
4FGL J1408.9$-$5845 & 312.86 & 2.62 & 0.6 & 4.6 &  & 43.9 & 9.5 & 13.7 & 4.8 \\ 
4FGL J1409.1$-$6121e & 312.11 & 0.13 & 5.3 & 5.4 &  & 58.3 & 2.7 &  &  \\ 
4FGL J1412.1$-$6631 & 310.86 & $-$4.90 & 4.9 & 5.3 &  & 22.8 &  &  &  \\ 
4FGL J1424.8$-$6536 & 312.38 & $-$4.47 & 0.9 & 4.5 &  & 3.2 &  &  &  \\ 
4FGL J1427.8$-$6051 & 314.40 & $-$0.15 & 4.4 & 10.3 &  & 229.9 & 6.5 &  &  \\ 
4FGL J1435.8$-$6018 & 315.52 & $-$0.02 & 5.7 & 5.0 & spp & 39.2 & 3.8 &  &  \\ 
$\star$4FGL J1442.2$-$6005 & 316.34 & $-$0.13 & 5.2 & 3.5 & spp & 49.3 & 9.6 & 15.8 & 8.0 \\ 
$\star$4FGL J1447.4$-$5757 & 317.84 & 1.52 & 7.0 & 11.7 &  & 185.0 & 17.3 & 19.0 & 14.2 \\ 
4FGL J1449.8$-$5923 & 317.50 & 0.09 & 4.1 & 3.4 &  & 20.3 &  &  &  \\ 
4FGL J1454.3$-$5551 & 319.62 & 2.99 & 3.5 & 4.1 &  & 22.9 &  &  &  \\ 
4FGL J1500.1-5846 & 318.95 & 0.03 & 3.0 & 3.1 &  & 32.2 & 4.2 &  &  \\ 
4FGL J1501.0$-$6310e & 316.95 & $-$3.89 & 4.9 & 5.4 &  & 36.7 & 12.5 & 18.0 & 9.5 \\ 
4FGL J1514.0$-$6240 & 318.49 & $-$4.19 & 0.9 & 3.6 &  & 15.1 &  &  &  \\ 
$\star$4FGL J1514.2$-$5909e & 320.35 & $-$1.21 & 0.1 & 7.0 & PWN & 131.6 & 20.7 & 31.0 & 28.2 \\ 
4FGL J1515.6$-$5817 & 320.95 & $-$0.56 & 1.2 & 3.2 &  & 29.8 & 9.0 &  &  \\ 
4FGL J1517.9$-$5233 & 324.26 & 4.14 & 6.5 & 4.9 &  & 36.0 & 1.6 &  &  \\ 
4FGL J1529.4$-$6027 & 321.24 & $-$3.33 & 4.0 & 4.3 &  & 13.1 &  &  &  \\ 
$\star$4FGL J1534.0$-$5232 & 326.29 & 2.79 & 3.0 & 4.4 &  & 37.1 & 10.9 & 13.7 & 5.8 \\ 
$\star$4FGL J1547.5$-$5130 & 328.58 & 2.37 & 0.3 & 4.8 &  & 59.3 & 34.2 & 32.8 & 22.6 \\ 
4FGL J1551.9$-$6015 & 323.60 & $-$4.84 & 4.5 & 4.6 &  & 23.6 &  &  &  \\ 
$\star$4FGL J1552.9$-$5607e & 326.32 & $-$1.74 & 3.7 & 4.5 & SNR & 159.5 & 10.9 & 11.9 & 9.8 \\ 
4FGL J1553.8$-$5325e & 328.13 & 0.28 & 6.4 & 11.5 &  & 408.9 & 54.8 & 78.2 & 69.0 \\ 
4FGL J1556.0$-$4713 & 332.37 & 4.82 & 0.2 & 6.2 & unk & 67.9 & 13.8 & 16.7 & 7.5 \\ 
$\star$4FGL J1601.3$-$5224 & 329.66 & 0.33 & 6.4 & 7.2 & spp & 187.7 & 57.1 & 48.6 & 35.6 \\ 
$\star$4FGL J1608.8$-$4803 & 333.45 & 2.76 & 3.4 & 4.7 &  & 72.6 & 23.9 & 26.9 & 12.8 \\ 
4FGL J1610.3$-$5154 & 331.01 & $-$0.23 & 4.9 & 3.4 &  & 51.2 & 4.0 &  &  \\ 
4FGL J1616.6$-$5341 & 330.48 & $-$2.17 & 8.1 & 5.3 &  & 28.5 & 6.4 &  &  \\ 
4FGL J1616.6$-$5009 & 332.94 & 0.37 & 6.3 & 4.0 &  & 64.9 & 3.1 &  &  \\ 
4FGL J1618.3$-$4537 & 336.29 & 3.42 & 3.0 & 5.5 &  & 39.2 & 5.3 &  &  \\ 
4FGL J1622.8$-$4454 & 337.37 & 3.36 & 2.4 & 3.1 &  & 14.4 &  &  &  \\ 
4FGL J1623.0$-$4624 & 336.32 & 2.28 & 0.8 & 3.3 &  & 28.4 & 1.4 &  &  \\ 
4FGL J1626.0$-$4917 & 334.63 & $-$0.09 & 0.4 & 3.2 &  & 35.2 & 1.4 &  &  \\ 
4FGL J1626.5$-$4406 & 338.41 & 3.45 & 1.7 & 3.7 &  & 54.6 & 3.8 &  &  \\ 
$\star$4FGL J1626.6$-$4251 & 339.31 & 4.31 & 3.4 & 3.6 &  & 27.4 & 18.4 & 21.8 & 11.2 \\ 
$\star$4FGL J1633.0$-$4746e & 336.52 & 0.12 & 5.7 & 7.6 & spp & 649.8 & 13.3 & 35.6 & 34.2 \\ 
4FGL J1636.3$-$4731e & 337.08 & $-$0.12 & 4.9 & 6.7 & SNR & 133.3 & 5.1 &  &  \\ 
4FGL J1639.3$-$5146 & 334.25 & $-$3.33 & 6.6 & 11.6 &  & 233.4 & 11.0 & 20.8 & 13.2 \\ 
4FGL J1640.3$-$4917 & 336.21 & $-$1.80 & 3.0 & 4.1 &  & 39.3 & 8.2 &  &  \\ 
4FGL J1641.0$-$4619 & 338.51 & 0.08 & 4.4 & 3.6 & spp & 88.4 & 2.5 &  &  \\ 
4FGL J1644.9$-$4921 & 336.64 & $-$2.41 & 2.5 & 4.3 &  & 38.8 & 7.4 &  &  \\ 
4FGL J1645.8$-$4533 & 339.63 & $-$0.05 & 5.0 & 5.3 & unk & 81.3 & 13.5 & 35.5 & 12.7 \\ 
4FGL J1652.2$-$4633e & 339.58 & $-$1.54 & 1.5 & 5.6 &  & 65.9 & 6.7 &  &  \\ 
4FGL J1653.2$-$4349 & 341.80 & 0.06 & 6.8 & 6.1 &  & 141.0 & 5.3 &  &  \\ 
4FGL J1654.2$-$4907 & 337.81 & $-$3.42 & 1.2 & 3.5 &  & 42.7 & 5.9 &  &  \\ 
4FGL J1655.5$-$4737e & 339.10 & $-$2.66 & 0.2 & 6.3 &  & 64.9 & 2.6 &  &  \\ 
4FGL J1700.1$-$4013 & 345.42 & 1.31 & 3.9 & 3.4 & unk & 11.7 &  &  &  \\ 
4FGL J1700.2$-$4237 & 343.54 & $-$0.20 & 4.3 & 3.5 &  & 23.4 &  &  &  \\ 
4FGL J1701.3$-$4924 & 338.30 & $-$4.51 & 2.5 & 3.1 &  & 6.9 &  &  &  \\ 
4FGL J1702.5$-$4803 & 339.50 & $-$3.85 & 1.1 & 3.1 &  & 28.3 & 3.5 &  &  \\ 
4FGL J1705.4$-$4850 & 339.17 & $-$4.70 & 3.7 & 3.0 &  & 27.5 & 7.0 &  &  \\ 
4FGL J1706.4$-$4649 & 340.88 & $-$3.62 & 1.9 & 4.1 &  & 40.6 & 2.2 &  &  \\ 
4FGL J1706.8$-$4540 & 341.85 & $-$2.99 & 2.4 & 3.2 &  & 30.9 & 8.6 &  &  \\ 
4FGL J1708.6$-$4312 & 344.01 & $-$1.76 & 3.2 & 3.4 & unk & 37.9 & 9.9 & 18.5 & 6.7 \\ 
4FGL J1711.1$-$4600 & 342.02 & $-$3.79 & 4.0 & 4.1 & unk & 28.2 & 0.6 &  &  \\ 
4FGL J1712.5$-$4642 & 341.59 & $-$4.39 & 3.4 & 4.2 & spp & 19.5 &  &  &  \\ 
4FGL J1714.8$-$3849 & 348.25 & $-$0.13 & 5.3 & 3.6 &  & 33.2 & 2.7 &  &  \\ 
4FGL J1714.9$-$3324 & 352.65 & 3.01 & 4.4 & 3.4 &  & 25.0 &  &  &  \\ 
4FGL J1718.0$-$3726 & 349.72 & 0.16 & 2.0 & 3.5 & snr & 44.8 & 0.1 &  &  \\ 
4FGL J1719.0$-$4221 & 345.82 & $-$2.83 & 4.1 & 3.4 &  & 28.6 & 6.4 &  &  \\ 
4FGL J1719.4$-$2916 & 356.60 & 4.61 & 3.5 & 3.9 &  & 22.2 &  &  &  \\ 
4FGL J1720.1$-$4358 & 344.60 & $-$3.90 & 1.6 & 3.7 &  & 14.5 &  &  &  \\ 
4FGL J1720.6$-$3706 & 350.30 & $-$0.07 & 4.4 & 3.4 & unk & 93.6 & 8.5 &  &  \\ 
4FGL J1721.7$-$3917 & 348.62 & $-$1.49 & 3.5 & 4.6 &  & 44.0 & 1.7 &  &  \\ 
4FGL J1723.1$-$2859 & 357.29 & 4.09 & 2.7 & 3.0 &  & 17.7 &  &  &  \\ 
4FGL J1724.5$-$3008 & 356.51 & 3.20 & 4.5 & 4.3 &  & 22.8 &  &  &  \\ 
4FGL J1729.1$-$3503 & 352.96 & $-$0.33 & 3.6 & 3.1 &  & 39.5 & 6.5 &  &  \\ 
4FGL J1729.7$-$4242 & 346.64 & $-$4.65 & 3.8 & 3.6 &  & 26.0 & 2.9 &  &  \\ 
4FGL J1729.9$-$4148 & 347.41 & $-$4.18 & 1.4 & 3.2 &  & 8.1 &  &  &  \\ 
4FGL J1730.1$-$3422 & 353.65 & $-$0.14 & 3.5 & 3.8 & unk & 86.7 & 11.9 & 35.5 & 13.4 \\ 
4FGL J1732.7$-$2559 & 0.98 & 3.98 & 3.0 & 3.7 &  & 21.8 &  &  &  \\ 
4FGL J1732.8$-$3725 & 351.40 & $-$2.26 & 2.0 & 3.0 & unk & 13.7 &  &  &  \\ 
4FGL J1734.5$-$2818 & 359.25 & 2.39 & 4.6 & 5.8 &  & 92.9 & 9.9 & 33.5 & 7.7 \\ 
4FGL J1736.1$-$3422 & 354.31 & $-$1.17 & 5.7 & 4.1 &  & 47.7 & 3.3 &  &  \\ 
4FGL J1736.3$-$2929 & 358.46 & 1.42 & 2.6 & 3.8 &  & 64.0 & 0.0 &  &  \\ 
4FGL J1737.6$-$2350 & 3.39 & 4.19 & 2.3 & 3.0 &  & 26.6 & 1.7 &  &  \\ 
4FGL J1738.1$-$2453 & 2.56 & 3.54 & 7.0 & 6.2 &  & 61.3 & 0.9 &  &  \\ 
4FGL J1739.2$-$2717 & 0.66 & 2.05 & 3.2 & 4.2 &  & 25.7 & 1.2 &  &  \\ 
4FGL J1739.3$-$2531 & 2.17 & 2.98 & 4.5 & 4.3 &  & 38.7 & 4.6 &  &  \\ 
4FGL J1741.6$-$3917e & 350.73 & $-$4.72 & 0.3 & 13.4 &  & 224.3 & 6.9 &  &  \\ 
4FGL J1741.6$-$2730 & 0.76 & 1.49 & 3.8 & 3.4 &  & 48.8 & 0.3 &  &  \\ 
$\star$4FGL J1742.8$-$2246 & 4.94 & 3.74 & 4.2 & 6.7 &  & 81.6 & 15.2 & 19.2 & 7.3 \\ 
4FGL J1743.4$-$2406 & 3.87 & 2.93 & 2.9 & 4.0 &  & 29.0 & 12.2 & 14.2 & 8.6 \\ 
4FGL J1743.9$-$3539 & 354.07 & $-$3.21 & 3.7 & 4.9 &  & 31.8 & 0.3 &  &  \\ 
4FGL J1745.6$-$2859 & 359.95 & $-$0.04 & 15.8 & 10.3 & spp & 666.0 & 5.4 &  &  \\ 
4FGL J1745.8$-$3028e & 358.71 & $-$0.84 & 1.2 & 7.6 &  & 261.0 & 4.0 &  &  \\ 
4FGL J1746.2$-$3342 & 355.99 & $-$2.59 & 2.1 & 3.3 & spp & 29.8 & 3.1 &  &  \\ 
4FGL J1746.5$-$2019 & 7.47 & 4.30 & 2.8 & 5.1 &  & 37.0 & 0.1 &  &  \\ 
4FGL J1747.0$-$3505 & 354.89 & $-$3.46 & 5.8 & 7.9 &  & 95.1 & 8.2 &  &  \\ 
4FGL J1747.7$-$2141 & 6.45 & 3.34 & 2.2 & 4.8 &  & 61.5 & 5.3 &  &  \\ 
4FGL J1750.6$-$1906 & 9.02 & 4.08 & 2.7 & 4.4 &  & 25.7 & 7.1 &  &  \\ 
4FGL J1752.3$-$2914 & 0.48 & $-$1.41 & 2.8 & 3.4 &  & 42.9 & 3.7 &  &  \\ 
4FGL J1753.8-2538 & 3.77 & 0.13 & 8.1 & 9.3 &  & 169.6 & 2.8 &  &  \\ 
4FGL J1754.6$-$2933 & 0.48 & $-$2.01 & 5.1 & 3.7 &  & 57.2 & 7.4 &  &  \\ 
4FGL J1757.4$-$3125 & 359.16 & $-$3.46 & 3.9 & 5.1 &  & 68.7 & 6.5 &  &  \\ 
4FGL J1758.3$-$1920 & 9.74 & 2.39 & 3.5 & 4.7 &  & 29.9 & 1.6 &  &  \\ 
4FGL J1758.6$-$2404 & 5.66 & $-$0.04 & 2.5 & 3.8 & unk & 56.5 & 8.3 &  &  \\ 
4FGL J1759.7$-$2141 & 7.85 & 0.94 & 3.7 & 3.4 & unk & 32.4 & 12.0 & 18.2 & 8.5 \\ 
$\star$4FGL J1801.3$-$2326e & 6.53 & $-$0.25 & 9.1 & 5.0 & SNR & 6451.6 & 193.4 & 176.0 & 167.6 \\ 
4FGL J1803.1$-$2724 & 3.26 & $-$2.56 & 3.5 & 4.2 &  & 26.3 & 0.4 &  &  \\ 
4FGL J1803.7$-$3207 & 359.21 & $-$4.97 & 4.2 & 3.4 &  & 21.7 &  &  &  \\ 
4FGL J1803.8$-$2908 & 1.83 & $-$3.54 & 6.2 & 5.2 &  & 57.5 & 4.8 &  &  \\ 
4FGL J1804.9$-$1745 & 11.89 & 1.82 & 2.8 & 3.3 &  & 30.0 & 7.4 &  &  \\ 
4FGL J1806.2$-$1347 & 15.50 & 3.48 & 1.8 & 4.0 &  & 31.8 & 1.7 &  &  \\ 
4FGL J1806.9$-$2824 & 2.80 & $-$3.79 & 1.0 & 3.8 &  & 29.7 & 7.8 &  &  \\ 
4FGL J1808.1$-$1234 & 16.80 & 3.65 & 4.1 & 5.3 &  & 66.5 & 3.5 &  &  \\ 
$\star$4FGL J1808.2$-$1055 & 18.26 & 4.44 & 4.1 & 5.7 &  & 64.8 & 11.3 & 15.6 & 13.7 \\ 
4FGL J1809.2$-$2726 & 3.90 & $-$3.76 & 2.4 & 3.5 &  & 47.3 & 4.6 &  &  \\ 
4FGL J1811.5$-$1844 & 11.79 & $-$0.04 & 4.9 & 5.4 & spp & 113.7 & 5.8 &  &  \\ 
$\star$4FGL J1812.2$-$0856 & 20.48 & 4.50 & 5.0 & 6.4 &  & 99.6 & 16.4 & 19.7 & 9.0 \\ 
$\star$4FGL J1813.1$-$1737e & 12.95 & 0.17 & 5.9 & 8.2 & spp & 240.9 & 32.3 & 30.7 & 25.1 \\ 
4FGL J1813.2-1128 & 18.35 & 3.10 & 3.79 & 3.11 & & 28.9 & 8.9 & &  \\
4FGL J1813.7$-$1152 & 18.08 & 2.79 & 2.9 & 3.3 &  & 15.4 &  &  &  \\ 
4FGL J1814.1$-$1710 & 13.45 & 0.18 & 1.9 & 3.9 & spp & 29.8 & 4.0 &  &  \\ 
$\star$4FGL J1814.2$-$1012 & 19.59 & 3.48 & 3.7 & 6.5 &  & 66.0 & 16.7 & 18.1 & 12.1 \\ 
4FGL J1814.6$-$2537 & 6.09 & $-$3.96 & 3.6 & 3.2 &  & 20.6 &  &  &  \\ 
4FGL J1815.8$-$1416 & 16.21 & 1.19 & 3.5 & 4.3 &  & 53.1 & 7.9 &  &  \\ 
4FGL J1817.9$-$1135 & 18.80 & 2.02 & 3.5 & 4.8 &  & 78.2 & 8.0 &  &  \\ 
4FGL J1818.6$-$1533 & 15.39 & $-$0.00 & 6.9 & 8.3 &  & 272.1 & 8.1 &  &  \\ 
4FGL J1819.9$-$1300 & 17.79 & 0.93 & 2.1 & 4.4 &  & 34.6 & 6.1 &  &  \\ 
4FGL J1820.3$-$1009 & 20.35 & 2.18 & 2.6 & 4.8 & unk & 52.1 & 5.5 &  &  \\ 
4FGL J1820.4$-$1609 & 15.07 & $-$0.66 & 0.8 & 4.8 & unk & 61.0 & 2.8 &  &  \\ 
4FGL J1821.1$-$1422 & 16.72 & 0.03 & 1.5 & 5.7 & spp & 94.1 & 2.8 &  &  \\ 
4FGL J1821.4$-$1516 & 15.97 & $-$0.46 & 4.7 & 4.6 & spp & 117.7 & 3.1 &  &  \\ 
4FGL J1822.8$-$1118 & 19.62 & 1.10 & 2.0 & 3.7 &  & 28.0 & 8.5 &  &  \\ 
4FGL J1823.3$-$1340 & 17.60 & $-$0.11 & 7.1 & 6.7 &  & 114.5 & 5.9 &  &  \\ 
4FGL J1826.2$-$1450 & 16.88 & $-$1.29 & 8.7 & 21.9 & HMB & 2485.1 & 8.0 &  &  \\ 
4FGL J1826.5$-$1202 & 19.40 & $-$0.06 & 4.8 & 3.6 &  & 41.8 & 8.2 &  &  \\ 
4FGL J1828.0$-$1133 & 20.00 & $-$0.14 & 5.8 & 4.2 & spp & 30.8 & 1.8 &  &  \\ 
4FGL J1830.1$-$0212 & 28.52 & 3.73 & 4.7 & 3.8 &  & 33.2 & 8.0 &  &  \\ 
4FGL J1830.2$-$1005 & 21.54 & 0.05 & 5.0 & 6.3 & spp & 93.4 & 9.0 &  &  \\ 
4FGL J1830.7$-$1634 & 15.85 & $-$3.04 & 3.4 & 6.0 &  & 51.3 & 0.0 &  &  \\ 
4FGL J1834.7$-$0724 & 24.43 & 0.31 & 5.7 & 3.1 &  & 27.0 & 1.4 &  &  \\ 
4FGL J1834.9$-$0800 & 23.94 & $-$0.01 & 5.9 & 4.1 &  & 46.0 & 2.4 &  &  \\ 
4FGL J1836.5$-$0651e & 25.13 & 0.16 & 2.1 & 6.2 & pwn & 124.7 & 4.4 &  &  \\ 
4FGL J1836.8$-$0727 & 24.65 & $-$0.18 & 5.8 & 3.6 & unk & 29.2 & 5.7 &  &  \\ 
4FGL J1839.0$-$1502 & 18.13 & $-$4.13 & 1.9 & 3.8 &  & 26.6 & 0.0 &  &  \\ 
$\star$4FGL J1839.4$-$0553 & 26.33 & $-$0.03 & 11.9 & 6.2 & unk & 174.4 & 13.6 & 27.0 & 24.0 \\ 
4FGL J1840.8$-$0453e & 27.37 & 0.12 & 5.7 & 6.3 & spp & 117.5 & 6.7 &  &  \\ 
4FGL J1842.5$-$0359 & 28.37 & 0.16 & 5.7 & 4.3 & unk & 54.9 & 3.8 &  &  \\ 
4FGL J1844.4$-$0306 & 29.37 & 0.13 & 6.5 & 4.7 & unk & 46.3 & 7.6 &  &  \\ 
4FGL J1845.8$-$0831 & 24.71 & $-$2.63 & 4.2 & 4.4 & spp & 39.6 & 6.8 &  &  \\ 
4FGL J1847.2$-$0141 & 30.95 & 0.16 & 6.4 & 3.1 &  & 30.6 & 7.1 &  &  \\ 
4FGL J1849.4$-$0056 & 31.86 & 0.02 & 5.6 & 3.1 & snr & 17.3 &  &  &  \\ 
4FGL J1850.3$-$0031 & 32.35 & $-$0.00 & 4.7 & 3.9 & spp & 47.4 & 8.2 &  &  \\ 
4FGL J1851.5+0718 & 39.46 & 3.30 & 3.0 & 4.8 &  & 30.7 & 2.1 &  &  \\ 
$\star$4FGL J1852.4+0037e & 33.60 & 0.07 & 5.0 & 7.1 & spp & 199.9 & 15.5 & 31.7 & 20.7 \\ 
4FGL J1852.6+0203 & 34.90 & 0.67 & 2.4 & 3.2 &  & 45.2 & 8.2 &  &  \\ 
4FGL J1853.2$-$0922 & 24.77 & $-$4.67 & 2.8 & 4.7 &  & 17.9 &  &  &  \\ 
4FGL J1853.6$-$0620 & 27.53 & $-$3.39 & 1.7 & 3.3 &  & 4.1 &  &  &  \\ 
$\star$4FGL J1855.2+0456 & 37.77 & 1.40 & 4.1 & 5.3 &  & 116.3 & 27.5 & 31.6 & 14.6 \\ 
4FGL J1855.3$-$0740 & 26.53 & $-$4.35 & 5.3 & 4.6 &  & 34.3 & 2.8 &  &  \\ 
$\star$4FGL J1855.9+0121e & 34.65 & $-$0.39 & 28.9 & 24.7 & SNR & 4383.9 & 100.3 & 99.6 & 96.1 \\ 
4FGL J1856.2+0749 & 40.46 & 2.50 & 5.1 & 3.2 & unk & 119.0 & 17.5 & 20.6 & 17.3 \\ 
$\star$4FGL J1857.7+0246e & 36.13 & $-$0.15 & 1.1 & 5.7 & PWN & 247.3 & 15.2 & 24.6 & 21.9 \\ 
4FGL J1858.0+0354 & 37.17 & 0.31 & 8.5 & 5.5 &  & 114.6 & 8.0 &  &  \\ 
4FGL J1859.2$-$0706 & 27.47 & $-$4.96 & 2.7 & 3.0 &  & 21.2 &  &  &  \\ 
4FGL J1900.4+0339 & 37.22 & $-$0.34 & 6.3 & 5.5 &  & 96.7 & 7.3 &  &  \\ 
4FGL J1900.7+0426 & 37.95 & $-$0.05 & 7.0 & 5.1 &  & 72.5 & 7.3 &  &  \\ 
4FGL J1900.9+0538 & 39.03 & 0.47 & 5.4 & 5.2 &  & 63.0 & 8.9 &  &  \\ 
4FGL J1901.1+0730 & 40.72 & 1.26 & 4.5 & 3.6 &  & 28.8 & 2.2 &  &  \\ 
4FGL J1902.2+0448 & 38.45 & $-$0.22 & 6.3 & 3.5 &  & 29.5 & 0.3 &  &  \\ 
4FGL J1903.8+0531 & 39.26 & $-$0.24 & 4.7 & 5.8 & spp & 120.5 & 7.3 &  &  \\ 
4FGL J1906.2+0631 & 40.43 & $-$0.30 & 5.3 & 3.6 & spp & 46.4 & 0.8 &  &  \\ 
$\star$4FGL J1906.9+0712 & 41.12 & $-$0.15 & 7.9 & 3.1 &  & 62.7 & 12.9 & 23.4 & 17.0 \\ 
$\star$4FGL J1908.7+0812 & 42.20 & $-$0.08 & 6.4 & 5.6 &  & 150.8 & 54.3 & 66.6 & 40.0 \\ 
4FGL J1908.8$-$0131 & 33.55 & $-$4.57 & 7.5 & 8.6 &  & 93.1 & 1.3 &  &  \\ 
$\star$4FGL J1911.0+0905 & 43.25 & $-$0.18 & 9.1 & 6.9 & snr & 513.3 & 29.7 & 26.6 & 24.6 \\ 
4FGL J1912.5+1320 & 47.19 & 1.47 & 4.8 & 6.0 &  & 85.3 & 13.5 & 23.2 & 11.5 \\ 
4FGL J1912.7+0957 & 44.21 & $-$0.15 & 7.0 & 4.1 &  & 29.4 & 2.5 &  &  \\ 
4FGL J1915.3+1149 & 46.15 & 0.15 & 6.6 & 5.3 &  & 60.1 & 8.9 &  &  \\ 
4FGL J1916.3+1108 & 45.67 & $-$0.37 & 4.6 & 4.1 & spp & 32.6 & 8.8 &  &  \\ 
4FGL J1918.1+1215 & 46.87 & $-$0.25 & 6.0 & 3.6 & spp & 31.5 & 8.8 &  &  \\ 
4FGL J1920.9+1408 & 48.85 & 0.04 & 4.8 & 6.5 & unk & 89.5 & 4.6 &  &  \\ 
$\star$4FGL J1923.2+1408e & 49.12 & $-$0.46 & 11.4 & 19.0 & SNR & 939.5 & 25.3 & 29.3 & 25.2 \\ 
4FGL J1925.3+1522 & 50.43 & $-$0.32 & 4.1 & 3.4 &  & 21.9 &  &  &  \\ 
4FGL J1928.4+1801 & 53.12 & 0.29 & 2.4 & 4.2 & unk & 54.2 & 7.6 &  &  \\ 
4FGL J1929.0+1729 & 52.72 & $-$0.09 & 6.8 & 5.2 &  & 51.9 & 5.3 &  &  \\ 
$\star$4FGL J1931.1+1656 & 52.48 & $-$0.79 & 7.0 & 7.8 &  & 117.6 & 17.2 & 25.8 & 18.4 \\ 
$\star$4FGL J1934.3+1859 & 54.65 & $-$0.47 & 5.4 & 5.2 & spp & 72.1 & 25.6 & 34.1 & 13.0 \\ 
4FGL J1946.1+2436 & 60.87 & $-$0.09 & 3.9 & 3.7 & unk & 17.0 &  &  &  \\ 
4FGL J1951.0+2523 & 62.12 & $-$0.65 & 4.6 & 4.0 &  & 23.6 &  &  &  \\ 
4FGL J1951.6+2621 & 63.01 & $-$0.27 & 3.7 & 3.8 &  & 15.6 &  &  &  \\ 
4FGL J1952.8+2924 & 65.77 & 1.06 & 5.0 & 4.6 & spp & 52.2 & 13.6 & 25.7 & 10.8 \\ 
4FGL J1954.8+2543 & 62.84 & $-$1.22 & 3.3 & 5.3 &  & 38.6 & 4.8 &  &  \\ 
4FGL J2002.3+3246 & 69.71 & 1.08 & 0.2 & 3.4 & spp & 30.1 & 13.1 & 14.6 & 11.8 \\ 
4FGL J2004.1+2517 & 63.57 & $-$3.23 & 3.3 & 3.1 & unk & 9.0 &  &  &  \\ 
4FGL J2004.3+3339 & 70.68 & 1.20 & 4.0 & 5.8 & unk & 66.9 & 8.4 &  &  \\ 
4FGL J2005.8+3357 & 71.10 & 1.09 & 1.1 & 3.1 &  & 11.2 &  &  &  \\ 
4FGL J2013.5+3613 & 73.85 & 1.02 & 3.3 & 3.6 & spp & 29.4 & 2.4 &  &  \\ 
$\star$4FGL J2021.0+4031e & 78.24 & 2.20 & 2.8 & 3.2 & SNR & 131.1 & 14.2 & 20.5 & 9.4 \\ 
$\star$4FGL J2028.6+4110e & 79.60 & 1.40 & 5.2 & 34.2 & SFR & 965.2 & 119.9 & 105.7 & 98.8 \\ 
$\star$4FGL J2032.6+4053 & 79.81 & 0.63 & 3.8 & 5.7 & HMB & 74.8 & 20.4 & 15.2 & 15.2 \\ 
4FGL J2035.0+3632 & 76.60 & $-$2.34 & 10.7 & 3.9 &  & 12.9 &  &  &  \\ 
$\star$4FGL J2038.4+4212 & 81.53 & 0.54 & 7.2 & 10.2 &  & 134.6 & 10.4 & 14.2 & 9.6 \\ 
4FGL J2041.1+4736 & 86.10 & 3.45 & 7.6 & 11.0 &  & 184.2 & 2.0 &  &  \\ 
4FGL J2043.1+4256 & 82.63 & 0.32 & 3.7 & 4.8 & unk & 40.3 & 0.3 &  &  \\ 
$\star$4FGL J2045.2+5026e & 88.75 & 4.67 & 11.5 & 28.6 & SNR & 412.6 & 42.4 & 42.1 & 34.3 \\ 
4FGL J2047.5+4356 & 83.91 & 0.32 & 3.7 & 5.3 &  & 58.9 & 8.3 &  &  \\ 
4FGL J2052.3+4437 & 84.99 & 0.09 & 4.6 & 6.5 &  & 30.0 & 0.2 &  &  \\ 
$\star$4FGL J2056.4+4351c & 84.87 & $-$0.98 & 3.5 & 5.2 & unk & 81.5 & 17.7 & 18.0 & 10.3 \\ 
4FGL J2058.7+4454 & 85.93 & $-$0.59 & 1.67 & 3.43 & spp & 14.6 &  &  &  \\ 
$\star$4FGL J2108.0+5155 & 92.21 & 2.91 & 3.6 & 8.0 &  & 120.7 & 18.5 & 20.2 & 15.3 \\ 
4FGL J2114.3+5023 & 91.76 & 1.15 & 2.5 & 4.7 &  & 40.0 & 5.8 &  &  \\ 
4FGL J2249.4+6222 & 109.30 & 2.78 & 2.7 & 5.5 &  & 21.8 &  &  &  \\ 
4FGL J2254.9+5802 & 108.01 & $-$1.41 & 1.2 & 3.1 & unk & 26.5 & 4.3 &  &  \\ 
4FGL J2301.9+5855e & 109.20 & $-$1.00 & 1.6 & 5.7 & SNR & 33.1 & 0.8 &  &  \\ 
4FGL J2315.9+5955 & 111.23 & $-$0.76 & 2.6 & 5.1 & unk & 61.4 & 8.5 &  &  \\ 
4FGL J2323.4+5849 & 111.74 & $-$2.13 & 5.6 & 13.8 & snr & 251.2 & 1.1 &  &  \\ 
4FGL J2357.8+6231 & 116.80 & 0.30 & 2.4 & 4.8 & spp & 19.2 &  &  &  \\ 
\enddata
%\tablecomments{Columns 2 and 3 provide the Galactic longitude and
  %latitude of the 311 candidates. Columns 4, 5 and 6 give the curvature
  %significance, the significance between 300 MeV and 1 GeV and the
  %source class reported in the 4FGL catalog. Columns 7, 8, 9, 10
  %provide the values obtained in our analysis concerning the TS of
  %each source, the improvement of the lognormal representation with respect to the
  %power-law model ${\rm TS_{LP}}$ as defined in
  %Section~\ref{subsection:ROI}, the improvement of the smooth broken
  %power-law representation with respect to the power-law model ${\rm TS_{SBPL}}$ and the improvement of the smooth broken power-law
  %representation when fixing $\Gamma_2 = 2$ called ${\rm TS_{SBPL2}}$.\\}
\end{deluxetable*}

%\section{Results of the systematic studies}
%\label{appen:syst}
% Table 2 : Systematics
%\input{Table_syst_edisp-3_addsrcs.tex}

%
\bibliographystyle{hapj}
\bibliography{bibliography_pion.bib}

\end{document}